\documentclass[10pt,reqno]{amsart}
\usepackage[margin = 1.1in]{geometry}
\usepackage{amssymb,mathtools}
\usepackage{bm}
\usepackage{xcolor}
\usepackage{graphicx}
\usepackage{tikz}
\usepackage{booktabs}
\usepackage{algorithm}
\usepackage{algpseudocode}
\usepackage{placeins}
\usepackage[colorlinks=true,linkcolor=blue,citecolor=blue,urlcolor=blue]{hyperref}
\usepackage{aliascnt}
\usepackage{cleveref}
\usepackage[backend=bibtex,
            style=numeric,
            giveninits=true,
            doi=false,
            url=false,
            sortcites=true]{biblatex}

\DeclareFieldFormat[article,inbook,incollection,inproceedings,patent,thesis,unpublished]{title}{#1\isdot}
\renewbibmacro*{in:}{%
  \ifentrytype{article}{}{\bibstring{in}\printunit{\intitlepunct}}}
\DeclareFieldFormat{eprint:arxiv}{%
  arXiv\addcolon\space\nolinkurl{#1}%
  \iffieldundef{eprintclass}{}{\addspace\texttt{\mkbibbrackets{\thefield{eprintclass}}}}}


\newtheorem{assumption}{Assumption}
\newtheorem{remark}{Remark}

\newtheorem{theorem}{Theorem}[section]
\newaliascnt{lemma}{theorem}
\newtheorem{lemma}[lemma]{Lemma}
\aliascntresetthe{lemma}
\newaliascnt{corollary}{theorem}
\newtheorem{corollary}[corollary]{Corollary}
\aliascntresetthe{corollary}
\newaliascnt{proposition}{theorem}
\newtheorem{proposition}[proposition]{Proposition}
\aliascntresetthe{proposition}

\crefname{theorem}{theorem}{theorems}
\Crefname{theorem}{Theorem}{Theorems}
\crefname{lemma}{lemma}{lemmas}
\Crefname{lemma}{Lemma}{Lemmas}
\crefname{corollary}{corollary}{corollaries}
\Crefname{corollary}{Corollary}{Corollaries}
\crefname{proposition}{proposition}{propositions}
\Crefname{proposition}{Proposition}{Propositions}

\newcommand{\R}{\mathbb{R}}
\newcommand{\C}{\mathbb{C}}
\newcommand{\E}{\mathbb{E}}
\newcommand{\dd}{\mathrm{d}}

\newcommand{\cL}{\mathcal{L}}
\newcommand{\cP}{\mathcal{P}}
\newcommand{\cX}{\mathcal{X}}
\newcommand{\cB}{\mathcal{B}}

\newcommand{\cN}{\mathcal{N}}
\newcommand{\cR}{\mathcal{R}}

\newcommand{\inner}[2]{\left\langle #1,#2 \right\rangle}

\newcommand{\supp}{\operatorname{supp}}
\newcommand{\dist}{\operatorname{dist}}
\newcommand{\diam}{\operatorname{diam}}

\newcommand{\varpsi}{\psi}

\title[Learning observable propagation with sparse local moments]{Learning observable propagation with sparse local moments}
\author[Yuehaw Khoo]{Yuehaw Khoo\texorpdfstring{\textsuperscript{*}}{}}
\author[Bowen Li]{Bowen Li\texorpdfstring{\textsuperscript{\textdagger}}{}}
\author[Yuguan Wang]{Yuguan Wang\texorpdfstring{\textsuperscript{\textdaggerdbl}}{}}
\author[Siyao Yang]{Siyao Yang\texorpdfstring{\textsuperscript{\textsection}}{}}
\thanks{\rightskip=0pt plus 1fil\relax\textsuperscript{*}\,Department of Statistics and CCAM, University of Chicago, Chicago, IL 60637, USA \mbox{(ykhoo@uchicago.edu)}}
\thanks{\textsuperscript{\textdagger}\,Department of Mathematics, City University of Hong Kong, Hong Kong SAR, China \mbox{(bowen.li@cityu.edu.hk)}}
\thanks{\textsuperscript{\textdaggerdbl}\,Program in Applied and Computational Mathematics, Princeton University, Princeton, NJ 08544, USA \mbox{(yuguanw@princeton.edu)}}
\thanks{\textsuperscript{\textsection}\,Institute of Computational Mathematics and Scientific/Engineering Computing, Academy of Mathematics and Systems Science, Chinese Academy of Sciences, Beijing 100190, China \mbox{(yangsiyao@amss.ac.cn)}}
\date{November 27, 2026}

\begin{document}

\begin{abstract}
We propose a unified framework for predicting local moment dynamics in high-dimensional many-body systems from short-time trajectories. Thanks to the locality of the dynamics, we construct the transition moment matrices by projecting the Markov operator onto low-order connected-cluster basis functions supported near the target observable. Exploiting the spatial decay of time-lagged correlations, {these matrices} can be approximated in sparse-plus-low-rank form using trajectory samples, with computational cost scaling linearly in the system dimension and the number of trajectories. The learning error can be further reduced by correcting the memory effect induced by the Galerkin projection. 
We numerically demonstrate the effectiveness of the framework in extrapolating the dynamics of various many-body systems, including Langevin and Glauber dynamics for classical systems as well as Lindblad dynamics for open quantum systems. 
For reversible local binary-spin dynamics, we establish volume-uniform $\mathcal{O}(\varepsilon)$-$L^2(\pi)$ approximation bounds for local observable dynamics using connected-cluster spaces with neighborhood radius $\mathcal O(\log(1/\varepsilon))$ and cluster order $\mathcal O(\log^2(1/\varepsilon))$.
We further establish error bounds for repeated one-step propagation using exact moment matrices and their sparse-plus-rank-one compressions, uniformly in the time step and number of iterations.

\end{abstract}
\maketitle

\section{Introduction}
In many-body stochastic dynamics, such as Langevin dynamics of interacting particles~\cite{Pavliotis2014}, Glauber dynamics of spin systems~\cite{glauber1963time}, and Lindblad dynamics of open quantum systems~\cite{lindblad1976generators}, the quantities of interest are usually local observables, including one-site magnetizations, neighboring correlations, and low-order marginals. Predicting the moments of such observables, that is, their expectations, is challenging because the probability density or density operator of a system with $d$ sites has space complexity exponential in $d$. In some settings, one can avoid representing the full probability law by estimating local moments directly from simulated trajectories.  However, long trajectories can still be costly to simulate, and changing the initial distribution generally requires a fresh ensemble. In this work, we use short-time trajectories to construct a reduced model for the moment dynamics of a given local observable. Once constructed, the model can be reused across initial distributions and can predict moments beyond the time horizon of the training trajectories.

Specifically, we consider a time-homogeneous Markov process $\{X_t\}_{t\geq0}$ with probability law $\rho_t$. The full probability law evolves according to
\begin{equation*}
    \rho_t=\cP_t^*\rho_0, 
\end{equation*}
where $\cP_t$ is the Markov (semigroup) operator acting on observables and $\cP_t^*$ is its adjoint acting on probability laws. This is the propagator form of the corresponding forward equation, such as the Fokker--Planck equation or the master equation for a density operator. 
Keeping only a vector $\boldsymbol\mu(t)$ of moments of local observables amounts to a projection of the full dynamics, which gives the variation-of-constants form of the Nakajima--Zwanzig equation \cite{nakajima1958quantum,zwanzig1960ensemble}:
\begin{equation*}
    \boldsymbol\mu(t) = M(t)\boldsymbol\mu(0) +\int_0^t M(t-s)\boldsymbol\eta(s)\,\dd s. 
\end{equation*}
Here $M(t)$ is the reduced moment propagator, the transpose of the Gram-normalized \emph{transition moment matrix}, and $\boldsymbol\eta$ accounts for information in the initial law not captured by the selected moments.
Because of the projection, $M(t)$ need not have the semigroup property, leading to non-Markovian memory effects.  Therefore, two problems arise in determining the reduced model:
\begin{itemize}
    \item[(1)] Which observables should we include in $\boldsymbol{\mu}(t)$ to keep the complexity  manageable? 
    \item[(2)] How do we obtain $M(t)$ from data and correct for the memory effects?
\end{itemize}

We address them as follows:
\begin{itemize}
\item \textbf{Efficient discretization of the Markov operator.} 
For the first problem, we use the locality of the dynamics. An observable evolved over a short time period $\tau$ remains concentrated near its support, so we approximate it in a connected-cluster basis whose functions each depend on only a few sites near the target observable. We take $\boldsymbol\mu(t)$ to be the moments of these basis functions, which can keep $\boldsymbol{\eta}(t)$ small. On graphs of uniformly bounded degree, the transition moment matrix has at most $\mathcal O(d)$ rows and columns for fixed neighborhood radius, cluster order, and one-site basis size.

\item \textbf{Linear-scaling estimation and compression.} 
For the second problem, we estimate the transition moment matrix from short-time trajectories. We develop an algorithm that constructs it in sparse-plus-low-rank form with $\mathcal O(d)$ cost. The resulting matrix is applied iteratively to predict moment dynamics over longer times. The transfer tensor method (TTM) \cite{cerrillo2014non,pollock2018tomographically} is used to account for the remaining memory effects.
\end{itemize}

Beyond the method itself, we demonstrate its  wide range of applications and justify it theoretically:
\begin{itemize}
\item \textbf{Unified framework for predicting classical and quantum dynamics.} Our framework predicts local moment dynamics across a broad class of Markov processes, including Langevin dynamics for classical many-body systems, Glauber dynamics for classical spin systems, and stochastic unravelings of Lindblad dynamics for open quantum systems.
\item \textbf{Theoretical guarantees.}
For reversible local binary-spin dynamics with a spectral gap, we show that a connected-cluster space with neighborhood radius $\mathcal O(\log(1/\varepsilon))$ and cluster order $\mathcal O(\log^2(1/\varepsilon))$ achieves $L^2(\pi)$ error at most $\varepsilon$ for both local observable approximation and repeated one-step propagation using exact moment matrices under the stationary law $\pi$. Our numerical experiments illustrate that the method can also yield accurate predictions with a uniform reference measure, as in the two-dimensional Glauber example and beyond. 
We further establish error bounds for sparse-plus-rank-one compression of the Gram and transition moment matrices and the resulting repeated propagation of local observables. 
\end{itemize}

Among related works, extended dynamic mode decomposition~\cite{williams2015data} and trajectory-based Galerkin methods~\cite{thiede2019galerkin} estimate the matrix of a Markov or Koopman operator in a given set of {basis functions} from trajectories. They usually require a tensor-product basis, whose size is exponential in $d$ and thus prohibitively expensive in high dimensions. To avoid this cost, one can represent the full operator by a tensor network~\cite{Lucke2022tgEDMD,doi:10.1137/090752286,TANG2024113110,doi:10.1137/24M1654075} or a neural network~\cite{osti_2281727,li2020fourier,JMLR:v24:21-1524}. Such representations are not easily iterated to reach longer times. Repeated application of a tensor-network operator increases the  ranks and requires re-compression. Composing a neural-network-based operator with itself accumulates error without control. In contrast, our reduced model consists of only a few matrices of fixed size acting on coefficient vectors, so each iteration costs a few matrix--vector products. Another classical way to address high dimensionality is to approximate a function of many variables by a sum of terms, each depending on a small group of variables called a \textit{cluster}, as in the Mayer cluster expansion~\cite{brydges1984short,meneghelli2014mayer} and the atomic cluster expansion~\cite{drautz2019atomic,dusson2022atomic}. This idea has been applied to Markov operators in our earlier work~\cite{khoo2025fastoperator}, using clusters of a few sites anywhere in the system together with a global low-rank compression. Similarly, the more recent work of Facelli et al.~\cite{facelli2026fast} also considers low-degree basis truncation for observable propagation in open quantum systems. Unlike these previous approaches, in this work, we further explore the locality of the dynamics of local observables~\cite{PhysRevLett.104.190401,barthel2012quasilocality,wild2023classical}, which allows us to keep a sparse set of {basis functions supported on sites near the target observable}, and to replace the global low-rank compression in~\cite{khoo2025fastoperator} by a more general sparse-plus-low-rank representation.

The rest of the paper is organized as follows. Section~\ref{sec:problem} introduces the Markov operator, the transition moment matrix, and the interaction graph. Section~\ref{sec:method} discusses the convenient basis representation, Monte Carlo estimation, matrix compression, the transfer tensor method, and the selection of the neighborhood for {the basis functions}. Section~\ref{sec:applications} describes moment prediction and its classical and quantum applications. Section~\ref{sec:glauber_guarantees} establishes error bounds for reversible local binary-spin dynamics, with proofs in Appendix~\ref{app:gb-proofs}. Section~\ref{sec:numerics}  presents the numerical results. In Section~\ref{sec:conclusion}, we conclude the paper.

\section{Preliminaries}
\label{sec:problem}
Let $V=[d]=\{1,2,\ldots,d\}$ be the index set of the sites (or coordinates). We consider a time-homogeneous Markov process $\{X_t\}_{t\geq0}$ on a state space $\Omega$ whose components are indexed by $V$.  For example, $\Omega$ can be a subset of $\C^d.$ Let $\rho_t$ denote the law of $X_t$. For an observable $g:\Omega\to\C$, we are interested in computing the moment
\begin{equation}
    \mu_g(t)  := \E_{X_t\sim\rho_t}[g(X_t)] = \int_\Omega(\cP_tg)(x)\,\rho_0(\dd x). \label{eq:moment_backward_relation}
\end{equation}
Here $\cP_t$ is the Markov operator, defined by
\begin{equation*}
    (\cP_t g)(x):=\E[g(X_t)\mid X_0=x], 
\end{equation*}
which satisfies the semigroup property $\cP_{t+s}=\cP_t\circ\cP_s$ for any $t,s\geq0$.

We use $t$ for continuous time and $\tau$ for a fixed estimation or propagation time period.
To obtain a finite-dimensional weak representation of $\cP_{\tau}$, we choose basis functions $\{\psi_i\}_{i=1}^n$, collected in the row vector $\Psi=[\psi_1,\ldots,\psi_n]$, and fix a reference probability measure $\rho_\mathrm{ref}$. For a time step ${\tau}$, the \textit{transition moment matrix} $M_{\rho_\mathrm{ref}}^\tau \in \C^{n\times n}$ is defined by
\begin{equation}\label{eq:transition_moment_matrix}
M_{\rho_{\text{ref}}}^{\tau}(i,j) = \langle \psi_{i}, \cP_{\tau}  \psi_{j}\rangle_{\rho_{\text{ref}}},
\end{equation}
where the inner product is
\begin{displaymath}
    \langle f, g\rangle_{\rho_{\text{ref}}} := \int_{\Omega} f^*(x)g(x)\,\rho_{\text{ref}}(\dd x).
\end{displaymath}
This matrix serves as a finite-dimensional surrogate of the operator $\cP_{\tau}$.
In particular, when $\rho_{\text{ref}}$ is  uniform, we write $\langle f,g\rangle_{\rho_{\text{ref}}}$ simply as $\langle f,g\rangle$. We also write $\langle f \rangle_{\rho_{\text{ref}}} = \int_\Omega f(x)\,\rho_{\text{ref}}(\dd x)$ for future use. At ${\tau}=0$, the transition moment matrix is the Gram matrix.

The local dependence among the indexed components is encoded by the interaction graph
\begin{equation*}
    G=(V,E), 
\end{equation*}
whose edge $(i,j)\in E$ represents a direct interaction between sites $i$ and $j$. Let $\dist_G(i,j)$ be the number of edges in a shortest path from $i$ to $j$. For a non-empty set $S\subseteq V$, we define the distance $\dist_G(j,S):=\min_{i\in S}\dist_G(j,i)$ and the graph neighborhood of $S$ with radius $\xi\geq0$ as
\begin{equation*}
    \cN_\xi(S) := \{j\in V:\dist_G(j,S)\leq \xi\}. 
\end{equation*}
We write $G[S]$ for the subgraph of $G$ induced by $S$. We say $G[S]$ is \emph{connected} if there is a path between any two vertices in $S$ within the subgraph.

We consider an observable depending on a small connected target set  $C\subseteq V$ 
\begin{equation}
    g(x)=g_C(x_C), \label{eq:cluster_observable}
\end{equation}
where $x_C:=(x_i)_{i\in C}$ and $|C|\ll d$. This includes local moments used to reconstruct low-order marginals of a high-dimensional probability density and one-site observables in quantum many-body systems. Representative applications are discussed in Section~\ref{sec:applications}. 

\section{Approximating the Markov operator}
\label{sec:method}

Our framework is motivated by the locality of the Markov process on the interaction graph $G$.  Specifically, we assume that the generator of the Markov process,
\begin{equation*}
    \cL g:=\lim_{h\rightarrow 0^+}\frac{\cP_hg-g}{h},
\end{equation*}
satisfies the following assumption.
\begin{assumption}
    \label{assump:local_generator}
The generator admits a decomposition
\begin{displaymath}
    \mathcal L=\sum_{Z\subseteq V}\mathcal L_Z, \qquad G[Z]\text{ connected},\quad |Z|\le R_{\mathrm{loc}}.
\end{displaymath}
Each $\mathcal L_Z$ acts locally on $Z$: for every observable $g_C$ supported on $C$,
\begin{equation*}
    \mathcal L_Zg_C=0\quad\text{if }Z\cap C=\varnothing,\qquad \supp(\mathcal L_Zg_C)\subseteq C\cup Z.
\end{equation*}
Here $R_{\mathrm{loc}}$ bounds the size of an interaction set, independently of the system dimension $d$.
\end{assumption}
This assumption is satisfied by many Gibbs samplers for local Hamiltonians and can be relaxed to suitably decaying quasi-local interactions.  We focus on the finite-range case above for simplicity. It leads to the following observation about the short-time propagation of an observable of the form  \eqref{eq:cluster_observable}. First, for a fixed short time period ${\tau}$, the evolved observable after ${\tau}$ has the formal expansion
\begin{equation}
    \cP_{\tau}g_C = e^{{\tau}\cL}g_C = \sum_{\ell=0}^{\infty} \frac{{\tau}^\ell}{\ell!}\cL^\ell g_C. \label{eq:local_semigroup_taylor}
\end{equation}
Assumption \ref{assump:local_generator} implies that the $\ell$th term in the expansion is supported within $\cN_{\ell R_{\mathrm{loc}}}(C)$, so distant sites enter only through high powers of $\cL$. Provided that $\|\cL^\ell g_C\|$ does not grow too rapidly with $\ell$, the factor ${\tau}^\ell/\ell!$ suppresses these high-order contributions. Thus, $\cP_{\tau}g_C$ can be well approximated by basis functions supported within a local window near $C$. 
In each nonzero term of \eqref{eq:local_semigroup_taylor}, every successive block intersects the previously reached sites, preserving connectivity and adding at most $R_{\mathrm{loc}}-1$ new sites. 
For reversible local binary-spin dynamics, Section~\ref{sec:glauber_guarantees} combines this observation with the spectral gap assumption to approximate $\cP_tg_C$ by small connected clusters within a fixed neighborhood of $C$, uniformly in $t$.
Related connected-cluster arguments for short-time quantum dynamics are developed in \cite{wild2023classical}, which show that only interaction clusters connected to the observable's support contribute and establish bounds on the short-time truncation error. Complementary spatial-locality results for local Markovian quantum dynamics are given in \cite{PhysRevLett.104.190401} and \cite{barthel2012quasilocality}.

\subsection{Choice of basis functions}
\label{sec:target_adapted_basis}
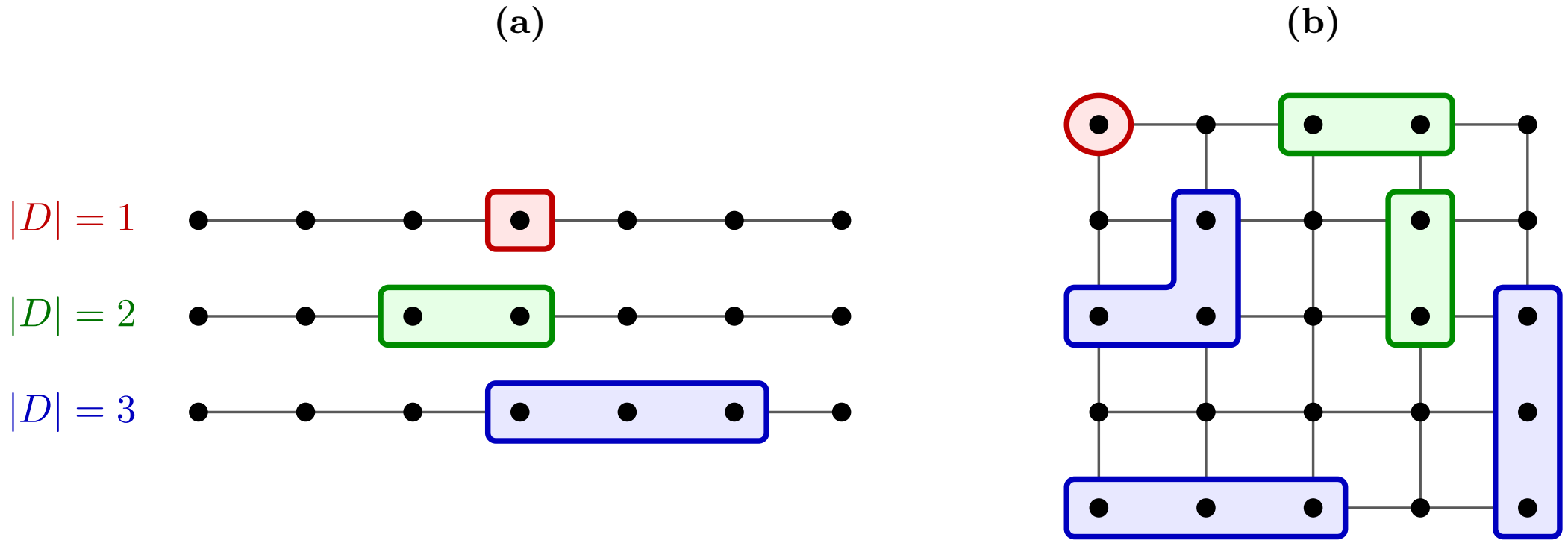
\begin{figure}[!htbp]
    \centering
    \begin{tikzpicture}[
        x=0.95cm,
        y=0.85cm,
        site/.style={circle,fill=black,inner sep=1.7pt},
        edge/.style={draw=black!65,line width=0.65pt},
        cluster/.style={rounded corners=2pt,line width=1.4pt}
    ]
        \begin{scope}[shift={(0,-1)}]
            \node[font=\small\bfseries] at (3,4.05)
                {(a)};

            \foreach \y in {0,1,2} {
                \foreach \x in {0,...,5}
                    \draw[edge] (\x,\y)--(\x+1,\y);
            }

            \draw[cluster,draw=red!75!black,fill=red!10]
                (2.70,1.70) rectangle (3.30,2.30);
            \draw[cluster,draw=green!55!black,fill=green!10]
                (1.70,0.70) rectangle (3.30,1.30);
            \draw[cluster,draw=blue!75!black,fill=blue!9]
                (2.70,-0.30) rectangle (5.30,0.30);

            \foreach \y in {0,1,2} {
                \foreach \x in {0,...,6}
                    \node[site] at (\x,\y) {};
            }

            \node[anchor=east,red!75!black] at (-0.45,2)
                {$|D|=1$};
            \node[anchor=east,green!45!black] at (-0.45,1)
                {$|D|=2$};
            \node[anchor=east,blue!75!black] at (-0.45,0)
                {$|D|=3$};
        \end{scope}

        \begin{scope}[shift={(8.4,0)}]
            \node[font=\small\bfseries] at (2,3.05)
                {(b)};

            \foreach \y in {-2,...,2} {
                \foreach \x in {0,...,3}
                    \draw[edge] (\x,\y)--(\x+1,\y);
            }
            \foreach \x in {0,...,4} {
                \foreach \y in {-2,...,1}
                    \draw[edge] (\x,\y)--(\x,\y+1);
            }

            \draw[cluster,draw=red!75!black,fill=red!10]
                (0,2) circle[radius=0.30];
            \draw[cluster,draw=green!55!black,fill=green!10]
                (1.70,1.70) rectangle (3.30,2.30);
            \draw[cluster,draw=green!55!black,fill=green!10]
                (2.70,-0.30) rectangle (3.30,1.30);
            \draw[cluster,draw=blue!75!black,fill=blue!9]
                (-0.30,-0.30)--(1.30,-0.30)--(1.30,1.30)--
                (0.70,1.30)--(0.70,0.30)--(-0.30,0.30)--cycle;
            \draw[cluster,draw=blue!75!black,fill=blue!9]
                (-0.30,-2.30) rectangle (2.30,-1.70);
            \draw[cluster,draw=blue!75!black,fill=blue!9]
                (3.70,-2.30) rectangle (4.30,0.30);

            \foreach \x in {0,...,4} {
                \foreach \y in {-2,...,2}
                    \node[site] at (\x,\y) {};
            }
        \end{scope}
    \end{tikzpicture}
\caption{Examples of connected supports $D$ on (a) a one-dimensional chain and (b) a two-dimensional nearest-neighbor grid. Each outlined connected support generates the family $\{\prod_{i\in D}\varphi_{\alpha_i}^{(i)}(x_i):\bm\alpha\in[p]^D\}$. The basis $\cB_k(S)$ collects the distinct functions generated by all connected supports $D\subseteq S$ with $1\leq |D|\leq k$.}
    \label{fig:connected_cluster_basis}
\end{figure}

In this subsection, we elaborate on the choice of basis functions.
Let $S\subset V$ be a set of sites. For each  site $i\in S$, we choose one-site functions $\{\varphi_a^{(i)}\}_{a=1}^p$ with $\varphi_1^{(i)}\equiv 1$. For a cluster order $k\ge 1$, we define the \textit{connected-cluster basis} on $S$
\begin{equation}
    \label{connected_cluster_basis}
    \cB_k(S)= \left\{ \prod_{i\in D}\varphi_{\alpha_i}^{(i)}(x_i): D\subseteq S,\ G[D]\text{ is connected},\ 1\leq |D|\leq k,\ \bm\alpha\in[p]^D \right\} 
\end{equation}
where $D$ is the \textit{connected support} used to generate the basis functions and $\bm\alpha=(\alpha_i)_{i\in D}\in[p]^D$ is a multi-index, with $[p]:=\{1,\ldots,p\}$.   \Cref{fig:connected_cluster_basis} illustrates connected supports $D$ on a chain and a two-dimensional lattice. On a graph of uniformly bounded degree, the number of basis functions in $\mathcal B_k(S)$ is $n=\mathcal O(|S|\,p^k)$ for fixed $k$, compared with $\mathcal O(p^{|S|})$ for the full tensor-product basis. Similar cluster ansätze appear in atomic cluster expansions \cite{drautz2019atomic,dusson2022atomic} and Mayer cluster expansions \cite{brydges1984short,meneghelli2014mayer}.

\subsection{Monte Carlo approximation of moment matrices}
\label{sec:monte_carlo_weak}

We now construct Monte Carlo estimators of the transition moment matrices from source states sampled from $\rho_{\text{ref}}$ and trajectories evolved over {a time period} ${\tau}$. We first draw independent source states
\begin{equation*}
    X_0^{(s)}\sim\rho_{\text{ref}}, \qquad s=1,\ldots,N_{\mathrm{src}},
\end{equation*}
and from each source state we simulate $N_{\mathrm{traj}}$ independent trajectories up to time ${\tau}$. Denoting the corresponding endpoints by $X_{\tau}^{(j|s)}$, $j=1,\ldots,N_{\mathrm{traj}}$, we define the feature matrices $\Phi_0,\Phi_{\tau} \in \C^{n\times N_{\mathrm{src}}}$ whose columns are given by
\begin{equation*}
    \Phi_0(:,s) := \Psi\bigl(X_0^{(s)}\bigr)^*, \qquad \Phi_{\tau}(:,s) := \frac{1}{N_{\mathrm{traj}}} \sum_{j=1}^{N_{\mathrm{traj}}} \Psi\bigl(X_{\tau}^{(j|s)}\bigr)^*.
\end{equation*}
Here $\Psi(x)$ is the row vector of the basis functions in $\mathcal{B}_k(S)$ evaluated at $x$. The transition moment matrices are estimated by
\begin{equation}
    \label{eqn:empirical_moment_matrices} \widehat M_{\rho_{\text{ref}}}^0 = \frac{1}{N_{\mathrm{src}}}\Phi_0\Phi_0^*, \qquad \widehat M_{\rho_{\text{ref}}}^{\tau} = \frac{1}{N_{\mathrm{src}}}\Phi_0\Phi_{\tau}^*.
\end{equation}
Conditioned on $X_0^{(s)}$, the vector $\Phi_{\tau}(:,s)$ is a Monte Carlo estimator of $(\cP_{\tau}\Psi^*)(X_0^{(s)})$. Thus, this construction requires only sampling the initial reference measure, evolving the source states over a short time period, and evaluating the chosen basis functions.

\subsection{Fast construction of transition moment matrices}
\label{sec:compression}
We approximate $\widehat M_{\rho_{\text{ref}}}^{\tau}$ in \eqref{eqn:empirical_moment_matrices} by the sum of a sparse near-field component and a low-rank far-field component. This splitting is analogous to the near-field and far-field decomposition in the fast multipole method~\cite{GREENGARD1987325} and in rank-structured matrices~\cite{hackbusch1999sparse}, where interactions between well-separated groups are compressed by low-rank approximations. In our case, the key motivation is the approximate factorization for basis functions with well-separated supports
 \begin{equation}\label{zero_correlation}
    M_{\rho_{\text{ref}}}^{\tau}(i,j) =  \langle \psi_{i} ,\cP_{\tau} \psi_{j} \rangle_{\rho_{\text{ref}}} \approx \langle  \overline{\psi_{i}} \rangle_{\rho_{\text{ref}}}  \langle \cP_{\tau} \psi_{j} \rangle_{\rho_{\text{ref}}} =: H_{\rho_{\text{ref}}}^{\tau}(i,j)
\end{equation}
where $H_{\rho_{\text{ref}}}^{\tau}$ has rank at most one. For a product reference measure and a fully uncoupled many-body system, \eqref{zero_correlation} is exact when the supports of $\psi_i$ and $\psi_j$ are disjoint. Local interactions create a connected correction, but locality suggests that this correction should decrease with the distance between the supports. For two nonconstant basis functions, we define
\begin{equation*}
    \dist_G(\psi_i,\psi_j) := \min_{u\in \supp(\psi_i),\,v\in \supp(\psi_j)}\dist_G(u,v). 
\end{equation*}
We set $\dist_G(\psi_i,\psi_j)=0$ if either function is constant. For a moderate time period, the locality argument in~\eqref{eq:local_semigroup_taylor} suggests that $\cP_{\tau} \psi_j$ remains concentrated near $\supp (\psi_j)$. Consequently, for pairs satisfying
\begin{equation*}
    \dist_G(\psi_i,\psi_j) \geq \delta,
\end{equation*}
we expect~\eqref{zero_correlation} to be accurate when $\delta$ is sufficiently large relative to the time period. The related spatial decay estimates are available for finite-range interacting particle systems~\cite[Proposition~3.3]{doukhan2008functional} and for local quantum dynamics~\cite{chen2023speed,PhysRevLett.104.190401}.

We therefore approximate the empirical transition moment matrix $\widehat M_{\rho_{\text{ref}}}^{\tau}$ by the decomposition
\begin{equation*}
    \widetilde M_{\rho_{\text{ref}}}^{\tau} = P + Q.
\end{equation*}
The sparse component $P$ captures local dynamical correlations between basis functions with nearby supports, while the low-rank component $Q$ captures the approximately factorized contributions of basis functions with well-separated supports. They are obtained by solving the optimization problem
\begin{equation}\label{eq:compression_optimization}
     \begin{split}
    &\min_{P,Q} \| \widehat M_{\rho_{\text{ref}}}^{\tau} - (P + Q)\|_F^2,\\
    &\text{subject to }  P(i,j) = 0 \text{ if } \dist_G(\psi_i,\psi_j) \geq \delta,\quad  \text{rank}(Q) \leq r
     \end{split}
\end{equation}
with given near-field radius $\delta$ and rank $r$. We initialize the two components by
\begin{equation}
    \label{compression 1} Q_0 = \widehat H_{\rho_{\text{ref}}}^{\tau},\qquad P_0 := ( \widehat M_{\rho_{\text{ref}}}^{\tau} - \widehat H_{\rho_{\text{ref}}}^{\tau} )\odot \mathbf{1}_\delta \quad \text{where} \quad  \mathbf{1}_\delta(i,j) =
    \begin{cases}
    1, & \dist_G(\psi_i,\psi_j)<\delta\\
    0, & \text{otherwise}
    \end{cases} 
\end{equation}
where $\widehat H_{\rho_{\text{ref}}}^{\tau}$ is the Monte Carlo estimate of $H_{\rho_{\text{ref}}}^{\tau}$. The population counterpart of this initialization is analyzed in Theorem~\ref{thm:st-matrix-compression}.

Starting from $(P_0,Q_0)$ in~\eqref{compression 1}, we alternate the low-rank and sparse updates in~\eqref{eq:compression_optimization}, following the update structure of GoDec~\cite{zhou2011godec}.
The sparse update uses the prescribed graph-distance mask $\mathbf{1}_\delta$, and the randomized SVD~\cite{halko2011finding} uses matrix-free products with the empirical moment matrix through the feature matrices $\Phi_0$ and $\Phi_\tau$.
We summarize the procedure in Algorithm~\ref{alg:matrix_free_compression}.

\begin{algorithm}[htbp!]
\caption{Fast construction of $\widetilde M_{\rho_{\text{ref}}}^{\tau}$}
\label{alg:matrix_free_compression}
\begin{algorithmic}[1]
\Require Initial sparse component $P_0$ and low-rank component $Q_0$ as in \eqref{compression 1}; near-field radius $\delta$;  rank $r$; maximal iteration number $\nu_{\max}$.
\Ensure Sparse component $P_{\nu_{\max}}$ and low-rank component $Q_{\nu_{\max}}$ such that $\widehat M_{\rho_{\text{ref}}}^{\tau} \approx P_{\nu_{\max}}+Q_{\nu_{\max}}$.
\For{$\nu=1,\ldots,\nu_{\max}$}
\State Update the low-rank component $Q_\nu$ as a rank-$r$ approximation of the residual $\widehat  M_{\rho_{\text{ref}}}^{\tau} - P_{\nu-1}$ using randomized SVD: first form the sketched residual
\begin{displaymath}
    Y = (\widehat  M_{\rho_{\text{ref}}}^{\tau} - P_{\nu-1})\Gamma =  \frac{1}{N_{\rm src}}\Phi_0(\Phi_{\tau}^*\Gamma) - P_{\nu-1}\Gamma \in\C^{n\times r}
\end{displaymath}
for a Gaussian random matrix $\Gamma\in\C^{n\times r}$; compute the QR factorization $Y = UR$ and form
\begin{displaymath}
    Z = (\widehat  M_{\rho_{\text{ref}}}^{\tau} - P_{\nu-1})^*U = \frac{1}{N_{\rm src}}\Phi_{\tau}(\Phi_0^* U) -P_{\nu-1}^* U  \in\C^{n\times r}.
\end{displaymath}
Finally, compute {the SVD} $Z^*=W\Sigma V^*$; set $Q_\nu = UW\Sigma V^* $.
\State Update the sparse component $P_\nu$ by evaluating only the entries in the near field:
\[
    P_\nu = ( \widehat M_{\rho_{\text{ref}}}^{\tau} - Q_\nu )\odot \mathbf{1}_\delta = P_{\nu-1}+(Q_{\nu-1}-Q_\nu)\odot\mathbf{1}_\delta.
\]
\EndFor
\end{algorithmic}
\end{algorithm}

Given the feature matrices, directly forming~\eqref{eqn:empirical_moment_matrices}  costs $\mathcal O(N_\mathrm{src}n^2).$ Let $n_\delta$ be the number of nonzero entries of $\mathbf{1}_\delta$ in~\eqref{compression 1}. The initialization of the sparse component costs $\mathcal O(N_\mathrm{src}n_\delta).$  Each iteration costs $\mathcal{O}(N_{\rm src}n r+n_\delta r+nr^2)$ including randomized sketching, computing QR factorization and SVD, and sparse update. For fixed cluster order, one-site basis size, and uniformly bounded graph degree, a fixed radius $\delta$ gives $n_\delta=\mathcal O(n)$, and, for fixed rank $r$ and iteration count $\nu_{\max}$, the construction of the compressed matrix costs $\mathcal{O}(N_\mathrm{src}n)$. Since $n=\mathcal O(d)$,  this stage scales linearly rather than quadratically in the number of sites $d$.

Before ending this subsection, we remark that in many cases when {the} learning time period $\tau$ is small and {the} reference measure is separable, the near-field component dominates the transition moment matrix. In such scenarios, the low-rank background is near zero and the refinement step has little effect, and one can simply use the initial sparse-plus-rank-one approximation to construct the transition moment matrix.

\subsection{Closure by the transfer tensor method}
\label{sec:ttm}

The transition moment matrix approximates the Markov operator over a single time period $\tau.$ We now approximate $\cP_{m\tau}$ for $m>1.$
By linearity, the evolution of any observable in $\mathrm{span}(\Psi)$ is determined by the action of $\cP_{m\tau}$ on the basis functions, namely, $\cP_{m\tau} \Psi$. However, the evolved basis functions can have components outside $\mathrm{span}(\Psi)$. We need to project them back onto $\mathrm{span}(\Psi)$ and encode the results in the \emph{dynamical maps} $\mathcal{E}_m \in \mathbb{C}^{n \times n}$, which satisfy
\begin{equation}
    \label{eq:dynamical_map} \Pi\cP_{m{\tau}}\Psi=\Psi\mathcal E_m, \qquad m\geq0.
\end{equation}
Here $\Pi$ denotes the $L^2_{\rho_\mathrm{ref}}$ orthogonal projection onto $\text{span}(\Psi)$. By the definition of the transition moment matrix in~\eqref{eq:transition_moment_matrix}, we have 
\begin{equation}
    \mathcal E_m = ( M_{\rho_{\text{ref}}}^0 )^{-1} M_{\rho_{\text{ref}}}^{m{\tau}} \label{eq:lag_map_new}
\end{equation}
assuming $M_{\rho_{\text{ref}}}^0$ is nonsingular. In particular, $\mathcal E_0=I_n$. Although the Markov operators satisfy the semigroup property, $\mathrm{span}(\Psi)$ is not generally invariant under $\cP_{\tau}$, and hence $\mathcal E_m$ need not equal $\mathcal E_1^m$. The \emph{transfer tensors} $\mathcal T_m\in\C^{n\times n}$ are defined by the exact discrete convolution
\begin{equation}
    \label{eq:TTM_dynamical_map} \mathcal E_m = \sum_{\ell=1}^{m} \mathcal T_\ell \mathcal E_{m-\ell},\qquad m\geq 1.
\end{equation} 
Given the dynamical maps, $\mathcal T_m$ can be obtained recursively from \eqref{eq:TTM_dynamical_map} as
\begin{equation}
    \mathcal T_1=\mathcal E_1, \qquad \mathcal T_m = \mathcal E_m - \sum_{\ell=1}^{m-1} \mathcal T_\ell\mathcal E_{m-\ell}, \qquad m\geq 2. \label{eq:TTM_recursion_new}
\end{equation}
The transfer tensor $\mathcal T_m$ with $m\ge 2$ accounts for the memory effects induced by the projection, which the Markovian approximation $\mathcal E_m\approx \mathcal E_1^m$ neglects.
For example, $\mathcal T_2=\mathcal E_2-\mathcal E_1^2$ is the first such correction, and the higher-order transfer tensors describe the departure of the projected dynamics from the Markovian relation $\mathcal E_m\approx\mathcal E_1^m$.
In practice, computing all $\mathcal E_m$ directly from trajectories up to every prediction time is expensive. When the transfer tensors decay sufficiently rapidly, we choose a memory length $K_{\mathrm{mem}}$ and neglect $\mathcal T_m$ for $m>K_{\mathrm{mem}}$, so that only the first $K_{\rm mem}$ dynamical maps are needed to approximate the longer-time projected evolution.

In summary, our implementation of TTM consists of three steps:
\begin{itemize}
\item \textbf{Data generation:} Estimate the transition moment matrices $M_{\rho_{\text{ref}}}^{m{\tau}}$ from trajectories for $m=0,\ldots,K_{\rm mem}$ for the chosen basis functions.
\item \textbf{Forming transfer tensors:} Compute the dynamical maps $\mathcal E_m$ for $m=1,\ldots,K_{\rm mem}$ by \eqref{eq:lag_map_new}, and then compute the transfer tensors $\mathcal T_m$ by \eqref{eq:TTM_recursion_new}.
\item \textbf{Memory truncation:} Use the first $K_{\rm mem}$ transfer tensors in \eqref{eq:TTM_dynamical_map} to approximate the longer-time projected evolution.
\end{itemize}
When compression is combined with TTM, each required transition moment matrix $M_{\rho_{\text{ref}}}^{m{\tau}}$ is compressed separately. Since dynamical correlations may spread over larger distances as the {time period} increases, the near-field radius $\delta$ and rank $r$ needed to achieve a target accuracy may depend on $m.$

\subsection{Selection of the neighborhood radius}
\label{sec:radius_detection}
Given the interaction graph $G$, the effective neighborhood radius $\xi$ needed to predict a target observable may still be unknown.
We propose a simple sensitivity-based rule to choose it. 

For continuous-state dynamics, fix $g_C$ and $t>0$, assume $u_t=\cP_tg_C$ is differentiable, and define
\begin{equation}
    S_j(t) :=\E_{\rho_{\text{ref}}}\!\left[|\partial_j u_t|^2\right],  \quad u_t = \mathcal P_t g_C, \quad j\in V. \label{eq:radius_site_sensitivity}
\end{equation}
This quantity measures the sensitivity of the evolved observable to the initial coordinate $x_j$. Since $u_t$ is expected to remain concentrated near the target set $C$, $S_j(t)$ should be small when site $j$ is far from $C$. We estimate that by the Monte Carlo method. We draw $x^{(s)}\overset{i.i.d.}{\sim}\rho_{\text{ref}}$, $s=1,\ldots,N'_{\rm src}$, and for each source state, we approximate $\partial_j u_t$ by a central finite difference:
\begin{equation}\label{eq:partial u_t finite difference}
    \partial_j u_t(x^{(s)}) \approx \frac{u_t(x^{(s)}+\Delta x e_j) -u_t(x^{(s)}-\Delta x e_j)}{2\Delta x}
\end{equation}
with $\Delta x>0$ and $e_j$ being the $j$-th unit vector. To evaluate $u_t(x^{(s)}\pm\Delta x e_j)$, we simulate $N'_{\rm traj}$ independent trajectory pairs starting from $x^{(s)}\pm \Delta x e_j$:
\begin{equation}\label{eq:u_t MC approx}
    u_t(x^{(s)}\pm\Delta x e_j) =   \E\left[g_C\left(X_{t,j,\pm}^{(s)}\right)\mid X_{0,j,\pm}^{(s)} = x^{(s)}\pm \Delta x e_j \right] \approx \frac{1}{N'_{\rm traj}}\sum_{\ell=1}^{N'_{\rm traj}} g_C\left (X_{t,j,\pm}^{(\ell|s)} \right ).
\end{equation}
The two ``$\pm$'' trajectories within each pair are driven by the same Brownian increments, while different pairs are independent.
This common-noise coupling is used to reduce the variance of the finite-difference estimate. Inserting \eqref{eq:partial u_t finite difference} and \eqref{eq:u_t MC approx} into \eqref{eq:radius_site_sensitivity}, we obtain a Monte Carlo estimator of $S_j(t)$:
\begin{equation}
    \widehat S_j(t) :=\frac{1}{N'_{\rm src}}\sum_{s=1}^{N'_{\rm src}} \left| \frac{1}{N'_{\rm traj}}\sum_{\ell=1}^{N'_{\rm traj}} \frac{g_C\left(X_{t,j,+}^{(\ell|s)}\right) - g_C\left(X_{t,j,-}^{(\ell|s)}\right)}{2\Delta x}\right|^2. \label{eq:radius_sensitivity_estimator}
\end{equation}
We then choose $\xi$ as the largest graph distance from $C$ to a site whose relative sensitivity $\frac{\widehat S_j(t)}{\max_{a \in V} \widehat S_a(t)}$ exceeds a prescribed threshold $0<\varepsilon<1$, provided $\max_a\widehat S_a(t)>0$:
\begin{equation*}
    \xi =\max \left\{\dist_G(j,C):j\in V,\; \widehat S_j(t)/\max_{a \in V} \widehat S_a(t) > \varepsilon \right\}. 
\end{equation*}
For discrete-state dynamics, we similarly compute the sensitivity by changing the initial state at one site while keeping all other sites fixed. We compare the expected future observable for the original and perturbed configurations, then average the squared differences over sampled initial configurations and admissible alternative local states.
These expectations can similarly be estimated from paired short trajectories. The neighborhood radius is then selected using the same relative-sensitivity criterion as in the continuous-state case.

\section{Predicting moment dynamics}
\label{sec:applications}
We now explain how the estimated transition moment matrices predict the moment dynamics of a given observable. We first introduce an iterative Galerkin scheme for a general observable and then apply it to the target local observable $g_C$ (\Cref{sec:predicting_local_observable}), with applications to classical (Section \ref{sec:marginal_prediction} and Section \ref{sec:glauber_application}) and quantum (\Cref{sec:lindblad_application}) many-body systems. 

Consider a general $d$-dimensional observable $f$ on $\Omega$. For a given time period $\tau$, let $u_{\tau}=\cP_{\tau} f$ and approximate it by
\begin{equation*}
    u_{\tau}(x)\approx\Psi(x)c_{\tau} 
\end{equation*}
where $c_{\tau}$ is the corresponding column coefficient vector. The choice of $\Psi(x)$ for a target local observable $f = g_C$ will be discussed later in \Cref{sec:predicting_local_observable}. Assuming $f$ admits the expansion $f(x) = \Psi(x)c_0$, testing against the basis functions gives
\begin{equation*}
    \inner{\Psi}{u_{\tau}}_{\rho_{\text{ref}}} =\inner{\Psi}{\cP_{\tau} f}_{\rho_{\text{ref}}}, \qquad M_{\rho_{\text{ref}}}^{0}c_{\tau} =M_{\rho_{\text{ref}}}^{\tau}c_0{.}
\end{equation*}
Provided that $M^0_{\rho_\mathrm{ref}}$ is nonsingular, we obtain
\begin{equation*}
    c_{\tau} = (M_{\rho_{\text{ref}}}^{0})^{-1} M_{\rho_{\text{ref}}}^{\tau} c_0.
\end{equation*}
More generally, \eqref{eq:dynamical_map} gives $\Pi\cP_{m{\tau}}f=\Psi\mathcal E_m c_0$.
In practice, we plug in the estimators $\widetilde M_{\rho_{\text{ref}}}^{0}$ and $\widetilde M_{\rho_{\text{ref}}}^{\tau}$ obtained by the compression algorithm in \Cref{sec:compression}.
Accordingly, we use $\widetilde c$ to denote the approximate coefficients.
We assume that $\widetilde M_{\rho_{\text{ref}}}^{0}$ is invertible and set $\widetilde{\mathcal E}_0=I$.
We form the estimated dynamical maps $\widetilde{\mathcal E}_m=(\widetilde M_{\rho_{\text{ref}}}^{0})^{-1}\widetilde M_{\rho_{\text{ref}}}^{m{\tau}}$ for $m=1,\cdots,K_{\text{mem}}$ and form the transfer tensors $\widetilde{\mathcal T}_\ell$ by \eqref{eq:TTM_recursion_new}.
The coefficients are then updated iteratively without estimating further transition moment matrices, by
\begin{equation}
    \widetilde c_m =
    \begin{cases}
    \widetilde{\mathcal E}_m c_0, & 0\leq m\leq K_{\rm mem},\\[2mm]
    \displaystyle\sum_{\ell=1}^{K_{\rm mem}}\widetilde{\mathcal T}_\ell\widetilde c_{m-\ell}, & m>K_{\rm mem}.
    \end{cases}
    \label{eq:TTM_prediction_new}
\end{equation}
Setting $K_{\rm mem}=1$ neglects higher-order memory corrections and reduces the scheme to repeated one-step propagation:
\begin{equation}
    \widetilde c_{m+1} =(\widetilde M_{\rho_{\text{ref}}}^{0})^{-1} \widetilde M_{\rho_{\text{ref}}}^{\tau}\widetilde c_m, \qquad \widetilde c_0=c_0. \label{eq:coefficient_iteration}
\end{equation}
For exact transition moment matrices, this iteration is exact when $\mathrm{span}(\Psi)$ is invariant under $\mathcal{P}_{\tau}$. In Theorem~\ref{thm:st-log-radius}, we control the $L^2(\pi)$ error of the above repeated propagation with exact moment matrices in the case when $\mathrm{span}(\Psi)$ is not invariant, for reversible local binary-spin dynamics with $\rho_{\text{ref}}=\pi$. The bound is uniform in the time step $\tau$ and the iteration number $m$.
Given the coefficients, the moment of the observable $f$ at time $t_m=m{\tau}$ follows from \eqref{eq:moment_backward_relation} as
\begin{equation}
    \mu_f(t_m)=\E_{\rho_0}[\cP_{t_m}f] \approx\E_{\rho_0}[\Psi]\,\widetilde c_m, \label{eq:reconstructed_dynamics}
\end{equation}
where the row vector $\E_{\rho_0}[\Psi]$ is computed once.

\subsection{Predicting dynamics of local observables}
\label{sec:predicting_local_observable}
We now specify the choice of basis functions for predicting the moment of a local observable $g_C$ with connected target set $C$. For simplicity, we assume that $g_C$ is a product of the one-site functions from Section~\ref{sec:target_adapted_basis}:
\begin{equation}
    g_C(x_C) = \prod_{i\in C}\varphi_{\alpha_i}^{(i)}(x_i). \label{eq:target_tensor_expansion}
\end{equation}
A general $g_C$ supported on $C$ can be approximated by a linear combination of such products.

To predict the dynamics of $g_C$, we choose the basis functions $\cB_k\left(\mathcal{N}_\xi(C)\right),$ where $\xi\geq0$ is the neighborhood radius. The neighborhood $\mathcal{N}_\xi(C)$ includes nearby sites to capture the local spreading of the dynamics. We choose $k\geq |C|$, so that the basis contains $g_C$ in \eqref{eq:target_tensor_expansion}. We then construct the corresponding transition moment matrices and apply the TTM scheme in~\eqref{eq:TTM_prediction_new}, or the repeated one-step propagation in~\eqref{eq:coefficient_iteration}, starting from the coefficient vector $c_0$ of $g_C$. The desired moment then follows from~\eqref{eq:reconstructed_dynamics}.

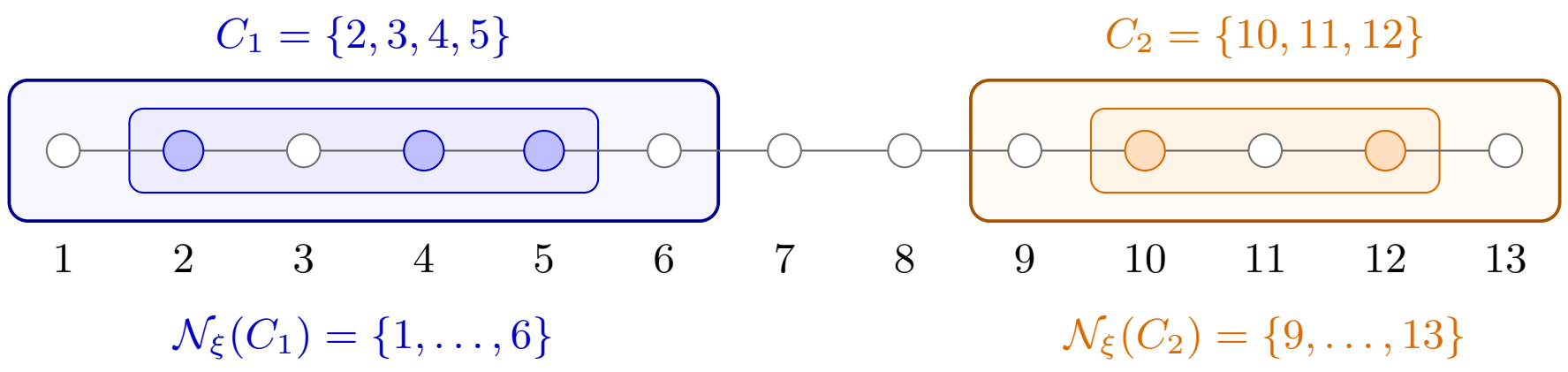
\begin{figure}[!htbp]
    \centering
    \begin{tikzpicture}[
        x=0.9cm,y=0.85cm,
        site/.style={circle,draw=black!55,fill=white,inner sep=2.5pt},
        first/.style={circle,draw=blue!75!black,fill=blue!25,inner sep=3pt},
        second/.style={circle,draw=orange!85!black,fill=orange!25,inner sep=3pt}
    ]
        \draw[rounded corners=4pt,draw=blue!55!black,fill=blue!3,line width=0.7pt]
            (0.55,-0.62) rectangle (6.45,0.62);
        \draw[rounded corners=4pt,draw=orange!65!black,fill=orange!3,line width=0.7pt]
            (8.55,-0.62) rectangle (13.45,0.62);
        \draw[rounded corners=3pt,draw=blue!75!black,fill=blue!7]
            (1.55,-0.37) rectangle (5.45,0.37);
        \draw[rounded corners=3pt,draw=orange!85!black,fill=orange!9]
            (9.55,-0.37) rectangle (12.45,0.37);
        \foreach \i in {1,...,12} \draw[black!60] (\i,0)--(\the\numexpr\i+1\relax,0);
        \foreach \i in {1,...,13} {
            \node[site] at (\i,0) {};
            \node[font=\small] at (\i,-0.95) {\i};
        }
        \foreach \i in {2,4,5} \node[first] at (\i,0) {};
        \foreach \i in {10,12} \node[second] at (\i,0) {};
        \node[blue!75!black,font=\small] at (3.5,1.00) {$C_1=\{2,3,4,5\}$};
        \node[orange!85!black,font=\small] at (11,1.00) {$C_2=\{10,11,12\}$};
        \node[blue!75!black,font=\small] at (3.5,-1.65) {$\mathcal N_\xi(C_1)=\{1,\ldots,6\}$};
        \node[orange!85!black,font=\small] at (11,-1.65) {$\mathcal N_\xi(C_2)=\{9,\ldots,13\}$};
    \end{tikzpicture}
\caption{Connected sets and neighborhoods for the disconnected target set $C=\{2,4,5,10,12\}$ on a chain. Colored sites mark the target sites. Adding intermediate sites $3$ and $11$, which carry the constant one-site function, gives the connected sets $C_1$ and $C_2$ shown by the inner boxes. The outer boxes show the separated neighborhoods $\mathcal N_\xi(C_\ell)$ with neighborhood radius $\xi=1$.}
    \label{fig:grouped_support_readout}
\end{figure}

\begin{remark}[Disconnected target observables]
\label{rem:disconnected_targets}
If $C$ is not connected, nearby components can be joined by adding intermediate sites to form disjoint connected sets $C_1,\ldots,C_q$ covering $C$, as illustrated in \Cref{fig:grouped_support_readout}. The added sites carry the constant one-site function, so the target observable is unchanged. For a product target, write $g_C=\prod_{\ell=1}^q g_\ell$, where $g_\ell$ is supported on $C_\ell$, and predict each factor using $\cB_k(\mathcal N_\xi(C_\ell))$ with $k\geq\max_\ell |C_\ell|$. For well-separated sets ${C_\ell}$, choose the neighborhood radius $\xi$ so that the neighborhoods $\mathcal{N}_\xi({C_\ell})$ remain well separated. If the factors ${g_\ell}$ remain weakly correlated over the prediction interval, the target moment can be approximated by
\[
    \mu_{g_C}(t) \approx \prod_{\ell=1}^q\mu_{g_\ell}(t).
\]
\end{remark}

\subsection{Langevin dynamics for classical many-body systems}
\label{sec:marginal_prediction}
For applications, we first consider a classical many-body system on the interaction graph $G=(V,E)$. Let $\cX_i=\R$ be the continuous local state space at site $i$, so that the full state space is $$\Omega=\prod_{i\in V}\cX_i.$$ We assume that the many-body potential is a sum of local interaction terms,
\begin{equation}
    U(x) = \sum_{Z\subseteq V} U_Z(x_Z), \label{eq:graph_many_body_potential}
\end{equation}
where the sum is over connected interaction sets $Z$ of uniformly bounded size. The corresponding Markov process is the overdamped Langevin dynamics on $\R^d$:
\begin{equation*}
    \dd X_t = -\nabla U(X_t)\,\dd t + \sqrt{2\beta^{-1}}\,\dd W_t, 
\end{equation*}
where $\beta>0$ is the inverse temperature and $W_t$ is a standard Brownian motion.
Its generator acts on an observable $f$ as
\begin{equation*}
    \cL f := -\nabla U\cdot\nabla f+\beta^{-1}\Delta f. 
\end{equation*}
The local decomposition of $U$ makes $\cL$ compatible with the locality condition in Assumption~\ref{assump:local_generator}. To generate the short-time trajectories required by Section~\ref{sec:monte_carlo_weak}, we use the Euler--Maruyama method: for a time step $\Delta t$, one trajectory is updated by
\begin{equation}
    \overline X_{\ell+1} = \overline X_\ell -\Delta t\,\nabla U(\overline X_\ell) +\sqrt{2\beta^{-1}\Delta t}\,\zeta_\ell, \qquad \zeta_\ell\overset{\mathrm{i.i.d.}}{\sim}\mathcal N(0,I_d). \label{eq:euler_maruyama_langevin}
\end{equation}

An important task for such systems is to predict the time evolution of low-order marginals. Let $\rho_t:\R^d\to[0,\infty)$ be the probability density of $X_t$ satisfying the Fokker-Planck equation
\begin{equation}
    \label{eqn:fokker-planck} \partial_t\rho_t = \mathcal L^*\rho_t = \beta^{-1}\Delta\rho_t + \nabla\!\cdot(\rho_t\nabla U)
\end{equation}
where $\mathcal L^*$ is the adjoint operator of $\mathcal L$. The marginal of $\rho_t$ on $S\subseteq V$ is given by
\begin{equation*}
    \rho_t^S(x_S) = \int \rho_t(x) \,\dd x_{V\setminus S}. 
\end{equation*}
To reconstruct the marginal on a bounded domain $D_S\subset\R^{|S|}$, let $\Psi^S=[\psi_{\bm\alpha}^S]_{\bm\alpha\in[p]^S}$ collect the products of one-site functions $\psi_{\bm\alpha}^S(x_S)=\prod_{i\in S}\varphi_{\alpha_i}^{(i)}(x_i)$.
We use the moment-matching approximation
\begin{equation*}
    \rho_t^{S,\mathrm{rec}}(x_S)=\Psi^S(x_S)c_t^S,\qquad x_S\in D_S.
\end{equation*}
Define the moment vector and the Lebesgue Gram matrix by
\begin{equation*}
    \boldsymbol\mu_S(t):=\E_{\rho_t}[(\Psi^S(X_S))^*],\qquad M_S:=\int_{D_S}(\Psi^S(x_S))^*\Psi^S(x_S)\,\dd x_S.
\end{equation*}
Assuming these moments are finite and $M_S$ is finite and invertible, choose $c_t^S$ to satisfy
\begin{equation*}
    M_Sc_t^S=\boldsymbol\mu_S(t),\qquad \rho_t^{S,\mathrm{rec}}(x_S)=\Psi^S(x_S)M_S^{-1}\boldsymbol\mu_S(t).
\end{equation*}
The components of $\boldsymbol\mu_S(t)$ are moments of local observables and can be predicted as in Section~\ref{sec:predicting_local_observable}.
This construction matches those moments by integration over $D_S$; the Gram matrix here uses Lebesgue measure, rather than the reference probability measure used to estimate the transition moment matrices.
If $\rho_t^S|_{D_S}\in L^2(D_S)$ and the marginal is supported in $D_S$, the reconstruction from exact moments coincides with its $L^2(D_S)$-orthogonal projection onto $\operatorname{span}(\Psi^S)$.
Otherwise, the choice of $D_S$ introduces an additional approximation error. Moreover, moment matching alone does not guarantee the nonnegativity of the reconstructed density.

\subsection{Glauber dynamics for classical spin systems}
\label{sec:glauber_application}
We next consider classical spins on the interaction graph $G=(V,E)$. Each site carries a spin $s_i\in\{-1,1\}$, and a configuration $s=(s_i)_{i\in V}$ belongs to $\Omega=\{-1,1\}^V$. The energy $U(s)$ is a sum of local interaction terms, as in \eqref{eq:graph_many_body_potential}, with connected interaction sets of uniformly bounded size, and the graph has uniformly bounded degree.

Glauber dynamics is a continuous-time Markov jump process that updates one spin at a time. We denote its state at time $t$ by $s_t=(s_{i,t})_{i\in V}$. Let $s^{(i)}$ denote the configuration obtained by reversing $s_i$ and leaving all other spins unchanged. The generator acts on an observable $f:\Omega\to\R$ as
\begin{equation}
    \cL f(s) = \sum_{i\in V}c_i(s)\bigl[f(s^{(i)})-f(s)\bigr] \label{eq:glauber_generator}
\end{equation}
where the heat-bath flip rate is
\begin{equation}
    c_i(s)=\frac{1}{1+\exp \bigl(\beta\Delta_iU(s)\bigr)}, \qquad i\in V, \label{eq:glauber_flip_rate}
\end{equation}
where $\Delta_iU(s):=U(s^{(i)})-U(s)$ is the energy difference. Since $\Delta_i U$ depends only on interaction terms containing $i$, each summand in \eqref{eq:glauber_generator} acts on a connected set of uniformly bounded size and hence satisfies Assumption~\ref{assump:local_generator}. 
Analogous to the Fokker-Planck equation \eqref{eqn:fokker-planck} for continuous-state dynamics, the master equation evolving the probability mass function $\rho_t(s)$ of the spin configuration reads
\begin{equation*}
    \frac{\dd\rho_t(s)}{\dd t} = \mathcal L^* \rho_t(s) =  \sum_{i\in V} \bigl[c_i(s^{(i)})\rho_t(s^{(i)})-c_i(s)\rho_t(s)\bigr] 
\end{equation*}
Again, directly evolving $\rho_t$ requires storing $2^d$ probabilities. Instead, we generate trajectories using the exact Gillespie algorithm \cite{gillespie1977exact}, which updates one spin at a time as summarized in \Cref{alg:glauber}. 
At configuration $s$, the waiting time to the next flip is exponentially distributed with parameter $c_{\mathrm{tot}}(s) := \sum_i c_i(s)$. A site $i$ is then selected with probability $c_i(s)/c_{\mathrm{tot}}(s)$, and the state is updated to $s^{(i)}$. If the total rate is zero, the state remains unchanged. The configuration is constant between jumps, so it can be read at the prescribed {observation times} without time-discretization error.

With the stochastic trajectories, we can again use our proposed moment-based approach to predict the dynamics of local observables of interest, such as single-spin means $\mu_i(t)=\E_{\rho_t}[s_i]$ and two-spin moments $\mu_{ij}(t)=\E_{\rho_t}[s_i s_j]$.
Under such discrete-state dynamics, integration over $\Omega$ is replaced by summation over spin configurations. The normalized one-site functions used to construct the cluster bases in Section~\ref{sec:target_adapted_basis} are
\begin{equation*}
\varphi_1^{(i)}(s_i)=1, \qquad \varphi_2^{(i)}(s_i)=s_i,\qquad s_i\in\{-1,1\}.
\end{equation*}
Their distinct products on connected supports form the connected-cluster basis with $p=2$. These basis functions are orthonormal under the uniform reference measure, so in this case the population Gram matrix is $M_{\rho_{\text{ref}}}^0=I_n$.

For a target observable with connected target set $C$, we use $\cB_k(\mathcal N_\xi(C))$ as in \Cref{sec:predicting_local_observable}. Source states sampled from the chosen reference measure and their short-time jump trajectories supply the feature matrices and transition moment matrices in \Cref{sec:monte_carlo_weak}. The moments are then predicted via~\eqref{eq:reconstructed_dynamics}, using either repeated one-step propagation or TTM.
When predicting disconnected observables, such as a two-spin moment $\mu_{ij}(t)=\E_{\rho_t}[s_i s_j]$ where $i$ and $j$ are far apart, we predict the two single-spin means in separate neighborhoods and approximate $\mu_{ij}(t) \approx \mu_i(t)\mu_j(t)$ as discussed in Remark~\ref{rem:disconnected_targets}, provided that the spins remain weakly correlated over the prediction interval.

\begin{algorithm}[htbp!]
\caption{Gillespie simulation of Glauber dynamics} \label{alg:glauber}
\begin{algorithmic}[1]
\Require Initial configuration $s_0\in\Omega$; energy $U$; inverse temperature $\beta>0$; final time $T>0$.
\Ensure Final configuration $s_T$.

\State Initialize $t=0$ and $s=s_0$.

\While{$t<T$}
    \State Compute the flip rates $c_i(s)$ from \eqref{eq:glauber_flip_rate} and set ${c_{\mathrm{tot}}}(s)=\sum_{i\in V}c_i(s)$. 

    \State Sample the waiting time $\Delta t$ from an
    exponential distribution with rate ${c_{\mathrm{tot}}}(s)$.

    \If{$t+\Delta t\geq T$}
        \State \textbf{break}.
    \EndIf

    \State Sample a site $i\in V$ with probability $ \Pr(i=j\mid s)=\frac{c_j(s)}{{c_{\mathrm{tot}}}(s)}$, $j\in V$.

    \State Update $t\gets t+\Delta t$ and $s\gets s^{(i)}$.
\EndWhile

\State Set $s_T=s$.
\end{algorithmic}
\end{algorithm}

\subsection{Lindblad dynamics for open quantum systems}
\label{sec:lindblad_application}
Our framework also applies to local observable prediction in open quantum systems.  Consider a finite-dimensional quantum many-body system defined on the interaction graph $G=(V,E)$. Associated with each site $i\in V$ is a Hilbert space $\mathcal H_i$, and the Hilbert space of the full system is
\begin{displaymath}
    \mathcal H=\bigotimes_{i\in V}\mathcal H_i.
\end{displaymath}
At time $t$, the  quantum state is described   by a density operator  $\rho_t\in  \C^{\mathrm{dim}(\mathcal H) \times \mathrm{dim}(\mathcal H)}$  satisfying
\begin{equation*}
    \rho_t= \rho_t^*,\qquad \rho_t\succeq 0,\qquad \operatorname{Tr}(\rho_t)=1.
\end{equation*}
These conditions are the quantum counterparts of the nonnegativity and normalization of a classical probability distribution.

In an open quantum system, the system of interest interacts with an external environment with many degrees of freedom. After tracing out the environment and under suitable approximations, the density operator $\rho_t$ of the system is governed by the  Lindblad master equation~\cite{gorini1976completely,lindblad1976generators}:
\begin{equation}
    \label{eqn:linblad} \frac{\dd\rho_t}{\dd t} = \mathcal{L}^* \rho_t = -\mathrm{i}[H,\rho_t] + \sum_{j} \left( L_{j}\rho_tL_{j}^* - \frac12\{L_{j}^* L_{j},\rho_t\} \right),
\end{equation}
with initial density operator $\rho_0$, where the Lindblad generator acting on a given observable $A$ is defined by
\begin{equation*}
     \cL A = \mathrm{i}[H,A] + \sum_{j} \left( L_j^*A L_j - \frac12\{L_j^*L_j,A\} \right). 
\end{equation*}
Here $H=H^*\in\C^{\dim(\mathcal H)\times\dim(\mathcal H)}$ is the system Hamiltonian which satisfies the structure \eqref{eq:graph_many_body_potential}. $L_j$ are the jump operators, and each $L_j$ is assumed to have bounded local support. The commutator and anticommutator are defined by $[A,B]:=AB-BA$ and  $\{A,B\}:=AB+BA$, respectively. Equation~\eqref{eqn:linblad} plays the role of a forward evolution equation for the density operator, analogous to the Fokker–Planck equation \eqref{eqn:fokker-planck}.

Directly evolving the density operator again requires storing a matrix whose size grows exponentially with the number of sites.  Alternatively, one can represent the solution by an ensemble of wave-function trajectories. We write these trajectories as $\{\varpsi_t\}_{t\ge0}\subset\mathcal H$, with $\|\varpsi_t\| = 1$, such that
\begin{equation*}
     \rho_t = \mathbb E[\varpsi_t \varpsi_t^*].
\end{equation*}
In quantum dynamics,  this stochastic trajectory representation is called an \textit{unraveling}~\cite{PhysRevLett.68.580,RevModPhys.70.101}.  The representation is not unique. A standard quantum-jump unraveling is given by
\begin{equation*}
\begin{aligned}
    \dd\varpsi_t =\left[ -\mathrm{i}H-\frac12\sum_j \left( L_j^* L_j-\lambda_j(t)I \right) \right]\varpsi_{t-} \dd t +\sum_j\left(\frac{L_j}{\sqrt{\lambda_j(t)}}-I \right)\varpsi_{t-} \dd N_j(t),
\end{aligned}
\end{equation*}
where \(\varpsi_{t-}\) denotes the left limit of the state process at time \(t\). The counting process $N_j(t)$ counts jumps in channel $j$ and has predictable intensity $\lambda_j(t)=\|L_j\varpsi_{t-}\|^2$, so that $\E[\dd N_j(t)\mid\mathcal F_{t-}]=\lambda_j(t)\,\dd t$ (see~\cite{PhysRevA.70.012106}).
When $\lambda_j(t)=0$, channel $j$ has no jump and its normalized jump increment is taken to be zero. To generate these trajectories, we use the Monte Carlo {wave-function} (MCWF) method~\cite{PhysRevLett.68.580,RevModPhys.70.101}, which is a first-order time-discrete jump process that allows at most one jump per step,  analogous to the Euler--Maruyama method in~\eqref{eq:euler_maruyama_langevin}. 
The details of MCWF are summarized in \Cref{alg:mcwf}.
Here $\Delta t$ is the internal integration step, distinct from the propagation time period $\tau$.

A quantum observable is represented by a Hermitian operator $A$ on $\mathcal H$. Its moment in the state $\rho_t$ is
\begin{equation*}
     \mu_A(t) :=\operatorname{Tr}(A\rho_t)=\E[\langle \varpsi_t, A \varpsi_t \rangle],
\end{equation*}
analogous to the classical moment in~\eqref{eq:moment_backward_relation}. Writing $\cP_t = e^{t\mathcal L}$, we have, similarly to~\eqref{eq:moment_backward_relation}, 
\begin{equation*}
    \mathrm{Tr}(A\rho_t)=\mathrm{Tr}((\cP_t A) \rho_0).
\end{equation*}
Given \(N_{\rm traj}\) independent trajectories generated by MCWF with initial states sampled from a pure-state ensemble representing \(\rho_0\), we approximate 
\begin{equation*}
    \rho_t \approx \frac{1}{N_\mathrm{traj}} \sum_{i=1}^{N_\mathrm{traj}} \varpsi_t^{(i)}  (\varpsi_t^{(i)} )^*, \qquad \mathrm{Tr}(A\rho_t) \approx \frac{1}{N_\mathrm{traj}} \sum_{i=1}^{N_\mathrm{traj}}\langle \varpsi_t^{(i)}  , A \varpsi_t^{(i)}  \rangle .
\end{equation*}

\begin{algorithm}[htbp!]
\caption{MCWF}
\label{alg:mcwf}
\begin{algorithmic}[1]
\Require Normalized initial state $\varpsi_0\in\mathcal H$; Hamiltonian $H$ and jump operators $L_j$; time step $\Delta t>0$; final time $T>0$.
\Ensure Normalized approximation $\varpsi_T$ at time $T$.

\State Initialize $t=0$ and $\varpsi=\varpsi_0$, and form {the} effective Hamiltonian
\begin{equation*}
 H_{\mathrm{eff}} := H-\frac{\mathrm{i}}{2}\sum_j L_j^*L_j.
\end{equation*}
\While{$t<T$}
    \State Set $h=\min\{\Delta t,T-t\}$ and compute the evolved state before {a} jump
    \[
        \widetilde\varpsi=\exp(-\mathrm{i}H_{\mathrm{eff}}h)\varpsi.
    \]
    \State Compute the channel weights $q_j=\|L_j\varpsi\|^2$ and set $q_{\mathrm{tot}}=\sum_jq_j$. Set the jump probability as 
\begin{equation*}
p =
\begin{cases}
1-\|\widetilde\varpsi\|^2
\approx h \sum_j
\inner{\varpsi}{L_j^*L_j\varpsi},
& \text{if } q_{\mathrm{tot}}>0, \\[2pt]
0,
& \text{if } q_{\mathrm{tot}}=0.
\end{cases}
\end{equation*}
 \State With probability $p$, select channel $j$ with probability $q_j/q_{\mathrm{tot}}$ and perform the jump:
     \[
        \varpsi\gets\frac{L_j\varpsi}{\|L_j\varpsi\|}.
    \]
 Otherwise, directly perform the normalization $\varpsi\gets\frac{\widetilde\varpsi}{\|\widetilde\varpsi\|}$. Update $t\gets t+h${.}
\EndWhile
\State Set $\varpsi_T=\varpsi$.
\end{algorithmic}
\end{algorithm}

In this work, we consider a system of $d$ spin-$1/2$ particles as our numerical example (see \Cref{sec:quantum_results_new}). For such a system, $\mathcal H_i=\mathbb C^2$ and $\dim(\mathcal H)=2^d$, so $\rho$ is represented by a $2^d\times 2^d$ matrix. A local observable $A$ supported on sites $S\subset V$ is given by
\begin{equation*}
     A=A_S\otimes I_{S^c},
\end{equation*}
where $A_S$ is a Hermitian operator on $\mathcal H_S:=\bigotimes_{i\in S}\mathcal H_i$, and $I_{S^c}$ is the identity on the remaining sites.

The one-site operators are spanned by the identity and the three \textit{Pauli matrices}
\begin{equation*}
     \sigma_0 =
    \begin{pmatrix}
    1&0\\
    0&1
    \end{pmatrix},
    \qquad \sigma_x=
    \begin{pmatrix}
    0&1\\
    1&0
    \end{pmatrix},
    \qquad \sigma_y=
    \begin{pmatrix}
    0&-\mathrm{i}\\
    \mathrm{i}&0
    \end{pmatrix},
    \qquad \sigma_z=
    \begin{pmatrix}
    1&0\\
    0&-1
    \end{pmatrix}.
\end{equation*}
For a connected support $D=\{i_1,\ldots,i_{|D|}\}$,  the corresponding basis operators are the \textit{Pauli strings}
\begin{equation*}
     A_{D,\bm\alpha} := \sigma_{\alpha_1}^{(i_1)} \cdots \sigma_{\alpha_{|D|}}^{(i_{|D|})}, \qquad \alpha_1,\ldots,\alpha_{|D|}\in\{0,x,y,z\},
\end{equation*}
where $\sigma_\alpha^{(i)}$ acts as   $\sigma_\alpha$ on site $i$ and as the identity elsewhere. The distinct Pauli strings form the quantum counterpart of the connected-cluster basis in Section~\ref{sec:target_adapted_basis}.
Their moments $\mathrm{Tr}(A_{D,\bm\alpha}\rho_t)$ are the corresponding local moments.

In classical Glauber dynamics, storing a single spin configuration requires memory proportional to the number of sites, whereas the quantum MCWF method stores wavefunctions whose dimension grows exponentially with that number.
For spin-$\frac{1}{2}$ systems, each walker is represented as a vector in a $2^d$-dimensional Hilbert space, where $d$ is the number of spins.
In practice, for larger one-dimensional chain systems, we use the tensor-jump method (TJM)~\cite{sander2025tjm}, which represents each trajectory as a matrix product state~\cite{RevModPhys.77.259}. For fixed local and bond dimensions, this representation reduces the storage and the cost of local operations from exponential to linear scaling in the number of sites.

\section{Error bounds for reversible spin dynamics}
\label{sec:glauber_guarantees}

We bound the errors of connected-cluster approximation and repeated one-step propagation for reversible local binary-spin dynamics, including heat-bath Glauber dynamics.
Throughout this section, the reference measure is the stationary law $\rho_{\text{ref}}=\pi$, and observable errors are measured in $L^2(\pi)$.

Let \(G=(V,E)\) be a finite connected graph on \(d\) sites with maximum degree at most \(\Delta\ge 1\). Write $s=(s_i)_{i\in V}\in\Omega=\{-1,1\}^V$ for a spin configuration.
We use the neighborhoods from Section~\ref{sec:problem}, with $\cN_\xi(i)=\cN_\xi(\{i\})$, and write $\dist_G(S,T)=\min_{i\in S,j\in T}\dist_G(i,j)$ for nonempty sets.
We strengthen Assumption~\ref{assump:local_generator} by requiring each local term to be a Markov generator on its block:
\begin{equation}
 \begin{gathered}
 \cL=\sum_{A\in\mathcal I}\cL_A,\qquad \cL_A=\ell_A\otimes I_{A^c},\qquad |A|\le R_{\mathrm{loc}},\qquad \diam_G(A)\le a,\\
 J_{\mathrm{loc}}=\max_{i\in V}\sum_{A:i\in A}\|\cL_A\|_{\infty\to\infty}<\infty.
 \end{gathered}
 \label{eq:st-generator}
\end{equation}
Here $a\ge1$, and $\ell_A$ acts on the configurations of the connected block $A$. 
The blocks $A\in\mathcal I$ may overlap, and a single jump may update several spins.

We assume that the full generator $\cL$ satisfies detailed balance with respect to a strictly positive probability law $\pi$ and that its spectral gap $\lambda$ is bounded below by a positive constant independent of graph volume.
The kernel of $\mathcal L$ consists of constant functions. 
We work with real-valued observables, and write $\pi(h)=\E_\pi[h]$, $\|h\|_\pi^2=\langle h,h\rangle_\pi$.
The spectral gap gives the following contraction, and $\mathcal D$ denotes the Dirichlet form of $\cL$:
\begin{equation}
 \|\cP_t h\|_\pi\le e^{-\lambda t}\|h\|_\pi\quad\text{if }\pi(h)=0,\qquad \mathcal D(h,q):=-\langle h,\cL q\rangle_\pi.
 \label{eq:st-gap}
\end{equation}

\subsection{Connected-cluster approximation and propagation}

For $S\subseteq V$, let $\chi_S(s)=\prod_{i\in S}s_i$ be the Walsh monomial of degree $|S|$, with $\chi_\varnothing=1$.
For a nonempty window $B\subseteq V$ and integer $k\ge1$, the basis is indexed by the following distinct supports:
\begin{equation}
 \label{eq:st-dictionary}
W_k(B)=\operatorname{span}\mathcal B_k(B),
\qquad
\mathcal R=\{S\subseteq B:\chi_S\in\mathcal B_k(B)\}.
\end{equation}
Here the coordinate support $S$ may be disconnected. 
The cluster order $k$ bounds both the size of the generating set $D$ and the degree of $\chi_S$.
Let $\Psi=[\chi_S]_{S\in\cR}$ be the row vector of basis functions and let $\Pi$ be the $L^2(\pi)$-orthogonal projection onto $W_k(B)$.

Let $g_C$ depend on a nonempty connected target set $C$, as in \eqref{eq:cluster_observable}.
We abbreviate the exact moment matrix \eqref{eq:transition_moment_matrix} under $\pi$ by $M^t=M_\pi^t$ for $t\ge0$.
For a time period $\tau>0$, the one-step map from Section~\ref{sec:ttm} is
\begin{equation*}
 M^t(S,T)=\langle\chi_S,\cP_t\chi_T\rangle_\pi,\qquad \mathcal E_1=(M^0)^{-1}M^\tau.
\end{equation*}
The distinct Walsh monomials are linearly independent and $\pi$ has full support, so the Gram matrix $M^0$ is positive definite. 
By \eqref{eq:dynamical_map}--\eqref{eq:lag_map_new}, the exact-matrix version of \eqref{eq:coefficient_iteration} for $g_C=\Psi c_0$ yields
\begin{equation}
 u_m=\Psi\mathcal E_1^m c_0=(\Pi\cP_\tau)^m g_C.
 \label{eq:st-exact-iteration}
\end{equation}
\begin{theorem}[Cluster approximation]
\label{thm:st-log-radius}
For every $\varepsilon>0$ there are a neighborhood radius $\xi\ge0$ and an integer cluster order $k\ge |C|$:
\begin{equation}
 \xi=\mathcal O(\log(1/\varepsilon)),\qquad k=|C|+\mathcal O(\log^2(1/\varepsilon)),
 \label{eq:st-log-radius-orders}
\end{equation}
such that the basis \eqref{eq:st-dictionary} on $B=\cN_\xi(C)$ gives
\begin{equation}
 \sup_{\tau>0,\ m\in\mathbb N_0}\|\cP_{m\tau}g_C-u_m\|_\pi\le\varepsilon,
 \label{eq:st-propagation-accuracy}
\end{equation}
where \(\mathbb N_0=\{0,1,2,\ldots\}\) denotes the nonnegative integers, and $u_m$ is the exact-matrix iteration \eqref{eq:st-exact-iteration}.
\end{theorem}

Taking $m=1$ and $\tau=t>0$ in \eqref{eq:st-propagation-accuracy} readily gives the local cluster approximation bound for $\cP_tg_C$: 
\begin{equation*}
 \sup_{t\ge0}\|(I-\Pi)\cP_tg_C\|_\pi\le\varepsilon.
\end{equation*}
The detailed proof is given in Appendix~\ref{app:st-exact-propagation}, which shows, for some $c > 0$,  
\begin{equation*}
\sup_{\tau > 0,\;m\in\mathbb N_0} \|\mathcal P_{m\tau}g_C-u_m\|_\pi \;\lesssim\; e^{-c\min\{\xi,\sqrt{k}\}}.
\end{equation*}
The proof idea can be summarized as follows.
The spectral gap ensures that both the exact and projected trajectories $\cP_{m\tau}g_C$ and $u_m$ approach the stationary mean exponentially fast, so it suffices to control the evolution up to a relaxation time $T=\mathcal O(\lambda^{-1}\log(1/\varepsilon))$.
By locality, a spin at graph distance $r$ from $C$ can influence $\cP_tg_C$ only through a chain of at least $\lceil r/a\rceil$ overlapping generator blocks, yielding a spatial influence bound of order $e^{vt-r/a}$, with speed $v$ in \eqref{eq:st-local-propagation-constants} below.
This allows us to freeze the exterior dynamics at distance $\ell a$, with an error of order $e^{vT-\ell}$ for $t\le T$.
The resulting localized evolution admits a Krylov approximation in $W_k(B)$.
Finally, an integrated residual bound for this approximation, combined with Duhamel's formula, controls the repeated propagation error uniformly in $\tau$ and $m$.

\subsection{Transition moment matrix compression}
\label{sec:gb-compression}
We next establish the error bound for compressing the transition moment matrix. The analysis is restricted to the sparse-plus-rank-one initialization applied to the exact moment matrices under the stationary law $\pi$. Note that the rank-one term in \eqref{zero_correlation} is independent of time, since $\pi$ is stationary. 
Let $\boldsymbol\mu_\pi=(\pi(\chi_S))_{S\in\cR}$ be the vector of stationary means, with $(\boldsymbol\mu_\pi)_\varnothing=1$, and set the rank-one matrix $H=\boldsymbol\mu_\pi\boldsymbol\mu_\pi^\top$.
For an integer cutoff $\delta\ge1$, let $\mathbf{1}_\delta$ be the near-field mask defined in \eqref{compression 1}, which involves pairs of nonempty supports at distance less than $\delta$, as well as the constant row and column. 
Here and in the corresponding proofs, tildes denote compressed population matrices at cutoff $\delta$ and quantities derived from them.
Applying the same mask to $M^t-H$ for every $t\ge0$, including the Gram matrix at $t=0$, we define 
\begin{equation}
 \widetilde M^t=\underbrace{\mathbf{1}_\delta\odot(M^t-H)}_{\text{sparse}}+\underbrace{H}_{\text{rank one}},\qquad \widetilde{\mathcal E}_1=(\widetilde M^0)^{-1}\widetilde M^\tau,\qquad \widetilde u_m=\Psi\widetilde{\mathcal E}_1^{\,m}c_0.
 \label{eq:st-compression}
\end{equation}
This is the population counterpart of the sparse-plus-rank-one initialization in \eqref{compression 1}, applied to both the Gram and transition moment matrices.
Theorem~\ref{thm:st-compression} provides a sufficient condition for $\widetilde M^0$ to be positive definite, ensuring that the compressed iteration \eqref{eq:st-compression} is well-defined.

Let $\cR_+=\cR\setminus\{\varnothing\}$ and denote the centered covariance blocks by $C^t=(M^t-H)_{\cR_+,\cR_+}$ and $\widetilde C^t=(\widetilde M^t-H)_{\cR_+,\cR_+}$.
Write $G_c=C^0$ and $\widetilde G_c=\widetilde C^0$ for the centered Gram matrices, and define the one-step dissipation matrices by $D_\tau=G_c-C^\tau$ and $\widetilde D_\tau=\widetilde G_c-\widetilde C^\tau$.
Since \eqref{eq:st-compression} uses the same mask at every time, we have
\begin{equation}
 \widetilde C^t=(\mathbf{1}_\delta)_{\cR_+,\cR_+}\odot C^t,\qquad \widetilde D_\tau=(\mathbf{1}_\delta)_{\cR_+,\cR_+}\odot D_\tau.
 \label{eq:st-common-defect}
\end{equation}

\begin{theorem}[Compression error of moment matrices]
\label{thm:st-matrix-compression}
Let $\cR$ be any collection of distinct supports of size at most $k$, with $\varnothing\in\cR$. Let $E_\delta$ be the spatial-tail bound defined in \eqref{eq:st-matrix-tail-bound}. Then the compressed matrices in \eqref{eq:st-compression} satisfy
\begin{align}
 \|\widetilde M^t-M^t\|_2 &\le E_\delta e^{-\lambda t/2}, \qquad t\ge0, \label{eq:st-matrix-error}\\
 \|\widetilde D_\tau-D_\tau\|_2 &\le E_\delta(1-e^{-\lambda\tau/2}), \qquad \tau>0. \label{eq:st-defect-error}
\end{align}
Here \(\|\cdot\|_2\) denotes the spectral norm.
\end{theorem}

Proposition~\ref{prop:st-growth} gives sufficient conditions for $E_\delta$ to decay exponentially in $\delta$, uniformly in the graph volume and the window size $|\cN_\xi(C)|$ for fixed $k$.
The common-mask identity \eqref{eq:st-common-defect} lets us estimate the discarded entries of $D_\tau$ directly, yielding the small-$\tau$ factor in \eqref{eq:st-defect-error} used in Theorem~\ref{thm:st-compression}.
The proof of Theorem~\ref{thm:st-matrix-compression} relies on the observation that $M^t-H$ is the stationary time-lagged covariance matrix, whose entries satisfy
\[
-\frac{\mathrm d}{\mathrm dt}(M^t-H)(S_1,S_2) =\mathcal D(\chi_{S_1},\cP_t\chi_{S_2}),
\]
where $\mathcal D$ is the Dirichlet form defined in \eqref{eq:st-gap}. The locality and the spectral gap assumptions for the generator yield spatial and temporal decay of this Dirichlet kernel. Integrating the resulting bound over $[t,\infty)$ and $[0,\tau]$ gives \eqref{eq:st-matrix-error} and \eqref{eq:st-defect-error}, respectively. We defer the proof to Appendix~\ref{app:st-matrix-compression}.

\subsection{Error bounds for observable propagation}

We now quantify the error introduced by compressing the Gram and transition moment matrices in repeated one-step propagation.

\begin{theorem}
\label{thm:st-compression}
Under the assumptions of Theorem~\ref{thm:st-matrix-compression}, suppose $G_c\succeq g_*I$ for some $g_*>0$. For a relative compression tolerance $0<\varepsilon\le1$, choose $\delta$ such that
\begin{equation}
    E_\delta\le \frac{g_*}{8}\varepsilon.
    \label{eq:st-compression-cutoff}
\end{equation}
Then $\widetilde M^0$ is positive definite, and for every $f=\Psi c_0$, the exact and compressed iterates $u_m=\Psi\mathcal E_1^m c_0$ and $\widetilde u_m=\Psi\widetilde{\mathcal E}_1^{\,m}c_0$ satisfy
\begin{equation}
 \sup_{\tau>0,\ m\in\mathbb N_0}  \|\widetilde u_m-u_m\|_\pi \le \varepsilon\|f-\pi(f)\|_\pi.
 \label{eq:st-compression-propagation}
\end{equation}
\end{theorem}
For each fixed basis, $G_c$ is positive definite because it is the Schur complement of the constant entry in $M^0$.
When $E_\delta$ exponentially decays (Proposition~\ref{prop:st-growth}), one can take $\delta=\mathcal O(\log(1/\varepsilon))$.  The proof uses a telescoping identity for the matrix powers and controls the compression errors in the centered Gram matrix $G_c$ and the one-step dissipation matrix $D_\tau$ for small $\tau$ (see Appendix~\ref{app:st-compressed-propagation}). The following corollary is a direct consequence of Theorems~\ref{thm:st-log-radius} and~\ref{thm:st-compression}.

\begin{corollary}[Total propagation error]
For a local observable $g_C$, choose the connected-cluster space $W_k(B)$ by Theorem~\ref{thm:st-log-radius} with tolerance $\varepsilon_{\rm proj}>0$.
For the resulting basis, take $g_*>0$ such that $G_c\succeq g_*I$ and choose $\delta$ by Theorem~\ref{thm:st-compression} with tolerance $0<\varepsilon_{\rm comp}\le1$. Then, the compressed propagation of the local observable $g_C$ satisfies 
\begin{equation*}
 \sup_{\tau>0,\ m\in\mathbb N_0} \|\cP_{m\tau}g_C-\widetilde u_m\|_\pi \le \varepsilon_{\rm proj} + \varepsilon_{\rm comp}\|g_C-\pi(g_C)\|_\pi.
\end{equation*}
\end{corollary}

\begin{remark}[Graphs of polynomial growth]
As mentioned above, for a fixed cluster order $k$, under the model assumption \eqref{eq:st-generator} and the uniform polynomial volume growth \eqref{eq:st-polynomial-growth} of the graph, the exponential decay of $E_\delta$ is estimated by Proposition~\ref{prop:st-growth}. In addition, suppose that the Gram lower bound $G_c\succeq g_*I$ for some $g_*>0$ is uniform in the family of graphs and the window radius $\xi$. Then, from Theorem~\ref{thm:st-compression}, we can conclude, for every $f=\Psi c_0$ and all sufficiently large $\delta$,
\[
 \sup_{\tau>0,\ m\in\mathbb N_0}\|\widetilde u_m-u_m\|_\pi \lesssim e^{-b\delta}\|f-\pi(f)\|_\pi,
\]
 with constant $b > 0$ independent of graph volume and window size.  In particular, a cutoff $\delta=\mathcal O(\log(1/\varepsilon))$ suffices for relative compression tolerance $\varepsilon$ as $\varepsilon\to0$.
\end{remark}

\section{Numerical experiments}
\label{sec:numerics}

In this section, we present our numerical results on predicting the moments of local observables for high-dimensional Langevin dynamics (\Cref{sec:GL_marginal_results_new}),  Glauber dynamics (\Cref{sec:glauber_results}) and Lindblad dynamics (\Cref{sec:quantum_results_new}). 
In all experiments, the predictions are compared with an empirical reference obtained by averaging over simulated states. 

\subsection{Langevin dynamics for classical many-body systems}
\label{sec:GL_marginal_results_new}
We first consider the Langevin dynamics of Section \ref{sec:marginal_prediction}, and use the estimated transition moment matrices to predict low-order marginals of a high-dimensional density. Specifically, we consider the Ginzburg–Landau model \cite{hoffmann2012GL} on a 1D chain and a 2D lattice.

\begin{figure}[!htbp]
    \centering
\includegraphics[width=0.75\textwidth] {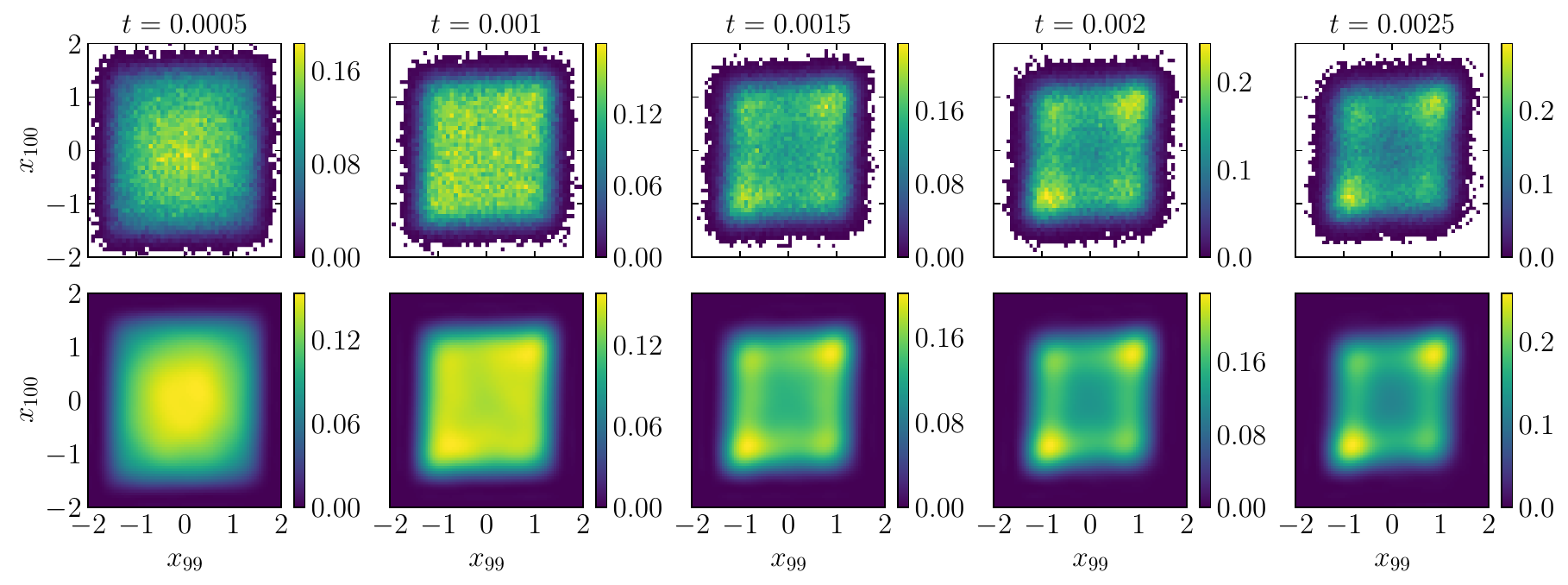} \caption{$(99,100)$-marginal evolution in the $d=100$ 1D Ginzburg--Landau model. Top: empirical reference (histograms); bottom: prediction obtained by iteratively applying the estimated transition moment matrix.}
    \label{fig:GL1d_d100_near_new}
\end{figure}

\subsubsection{1D Ginzburg--Landau model}
The one-dimensional Ginzburg--Landau model on a chain has potential
\begin{equation*}
    U_{\rm 1D}(x) = \frac{\lambda}{2} \sum_{i=0}^{d}\left(\frac{x_{i+1}-x_i}{h}\right)^2 + \frac{1}{4\lambda} \sum_{i=1}^{d}(1-x_i^2)^2, 
\end{equation*}
defined on the hypercube $[-2,2]^d$, with $x_0=x_{d+1}=0$ and $h=1/(d+1)$. We consider a $d=100$ example with $\lambda=0.005$ and $\beta=1/40$. The initial density $\rho_0$ is the product of standard Gaussians truncated to $[-2,2]$. For each pair of target sites, the transition moment matrix is constructed from connected-cluster basis functions on the two neighborhoods of radius $\xi=1$ around the target sites, with total cluster order $k=2$. We use $15$ Legendre functions per site as the one-site functions. The transition moment matrices are estimated from $N_{\rm src}=2\times10^4$ source states and $N_{\rm traj}=100$ trajectories per source, with {time period} ${\tau}=2.5\times10^{-3}$ and Euler--Maruyama step $\Delta t={\tau}/20$. The reference measure ${\rho_{\mathrm{ref}}}$ is the uniform distribution on $[-2,2]^d$. An independent ensemble of $2\times10^5$ trajectories provides the empirical reference.

Figure~\ref{fig:GL1d_d100_near_new} and Figure~\ref{fig:GL1d_d100_far_new} below show the predicted evolution of the adjacent $(99,100)$-marginal and the distant $(50,100)$-marginal, respectively. In both cases, the reconstructed marginal captures the evolution from the initially broad central mass toward four modes near $(\pm1,\pm1)$. It also reproduces the stronger diagonal correlation of the adjacent pair and the nearly symmetric four-mode structure of the distant pair.

\begin{figure}[!htbp]
    \centering
\includegraphics[width=0.75\textwidth] {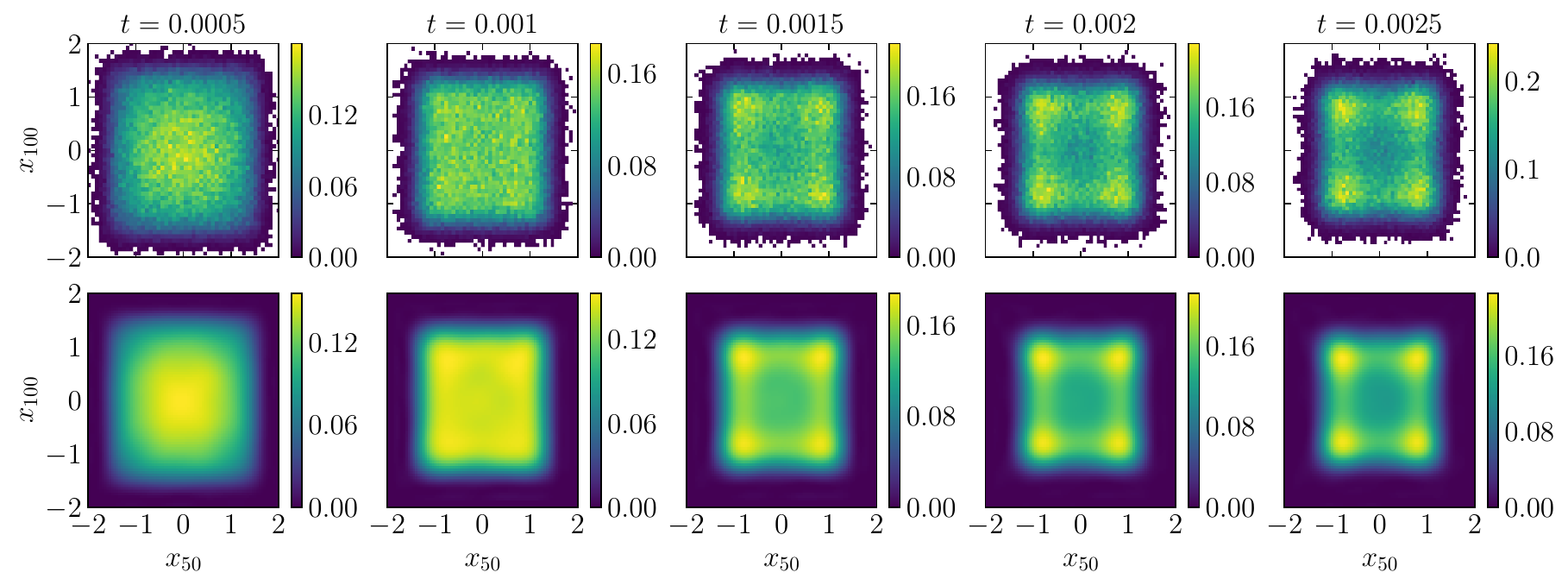} \caption{$(50,100)$-marginal evolution in the $d=100$ 1D Ginzburg--Landau model. Top: empirical reference (histograms); bottom: prediction obtained by iteratively applying the estimated transition moment matrix.}
    \label{fig:GL1d_d100_far_new}
\end{figure}

\begin{figure}[!htbp]
    \centering
    \begin{minipage}[t]{0.49\textwidth}
        \centering
\includegraphics[width=0.75\linewidth] {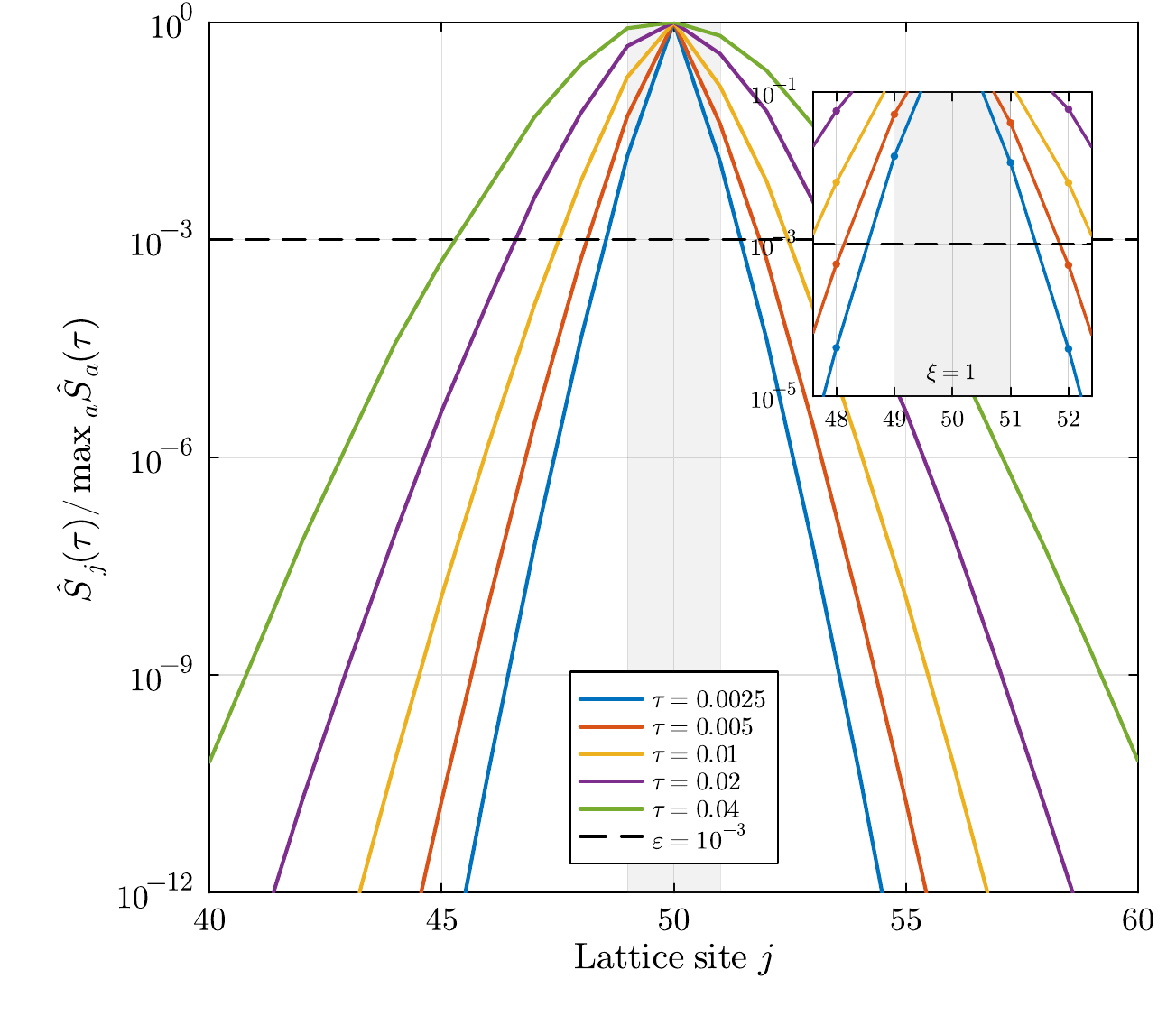} \par\smallskip {\small (a) Variation of {time period}}
    \end{minipage}\hfill
    \begin{minipage}[t]{0.49\textwidth}
        \centering
\includegraphics[width=0.75\linewidth] {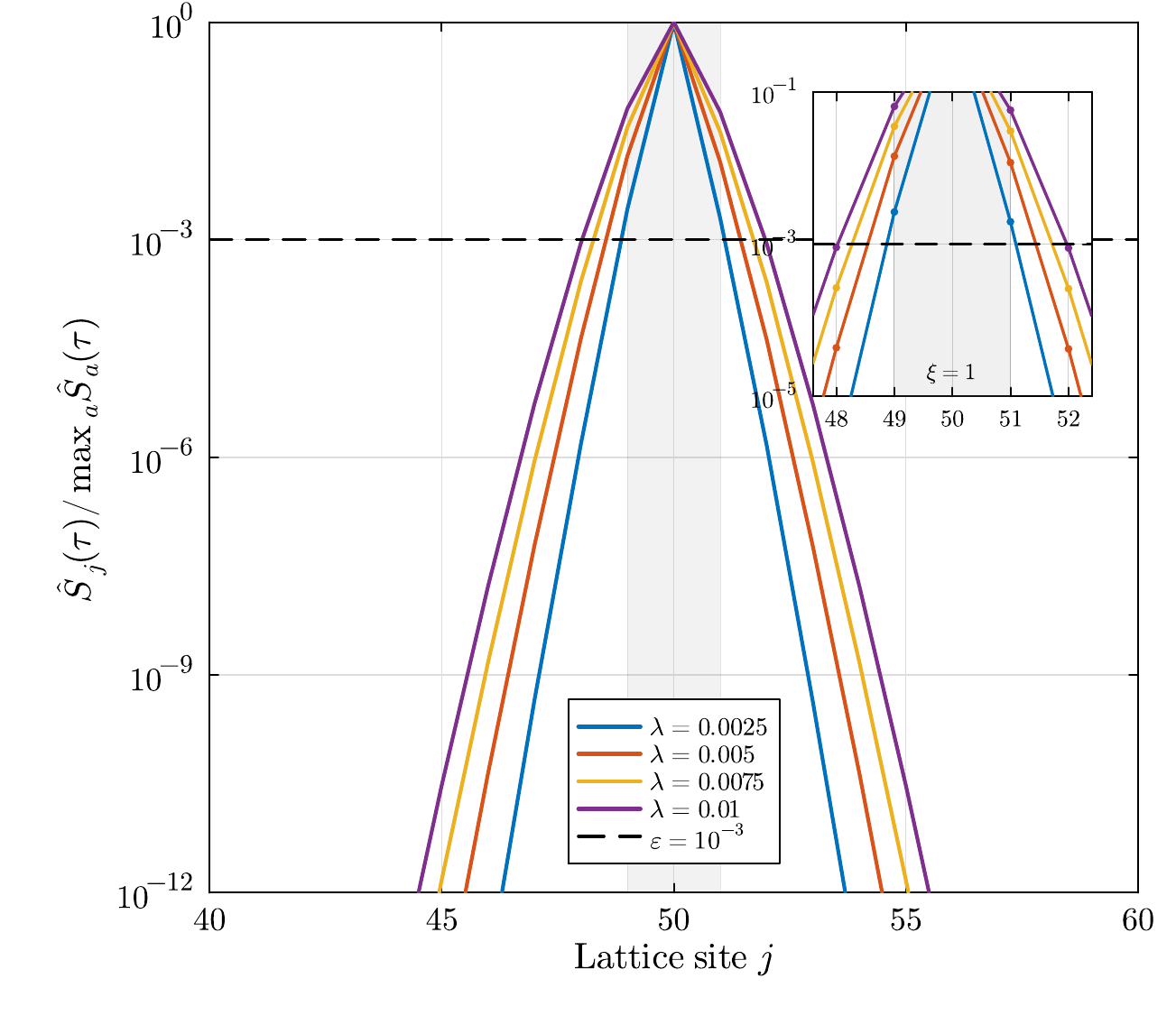} \par\smallskip {\small (b) Variation of coupling intensity}
    \end{minipage}
\caption{Normalized site sensitivities for $g_C(x)=x_{50}$ in the $d=100$ 1D Ginzburg--Landau model with $\beta=1/40$. (a) Time periods $\tau=0.0025,0.005,0.01,0.02,0.04$ at fixed $\lambda=0.005$. (b) Coupling parameters $\lambda=0.0025,0.005,0.0075,0.01$ at fixed $\tau=0.0025$. Each curve is normalized by its maximum over the tested sites $1,\ldots,100$. Values are shown on a logarithmic scale. }
    \label{fig:GL1d_radius}
\end{figure}

We further justify the choice of the neighborhood radius $\xi=1$ by the selection rule introduced in Section~\ref{sec:radius_detection} in a separate experiment with target observable $g_C(x)=x_{50}$, under the same model parameters. We use $N'_{\rm src}=256$ source states, $N'_{\rm traj}=64$ trajectory pairs per source state and perturbed coordinate, the finite-difference perturbation $\Delta x=5\times10^{-3}$ in \eqref{eq:radius_sensitivity_estimator} and an Euler--Maruyama step $\Delta t=1.25\times10^{-4}$. Initial-coordinate perturbations are tested at all 100 sites, while every trajectory evolves on the full 100-site chain. The panels in Figure~\ref{fig:GL1d_radius} display the positive sensitivity estimates $\widehat S_j(\tau)$, normalized separately for each curve by the largest estimate among the tested sites. For readability, only sites $40,\ldots,60$ are shown.

Figure~\ref{fig:GL1d_radius}(a) shows the sensitivity profiles at $\lambda=0.005$ for several {time periods}. Longer {time periods} produce broader relative sensitivity profiles. Figure~\ref{fig:GL1d_radius}(b) fixes $\tau=2.5\times10^{-3}$ and varies $\lambda$. Stronger coupling likewise broadens the relative sensitivity profiles. 
For $\lambda=0.005$ and $\tau=2.5\times10^{-3}$, the plotted relative sensitivities outside $\{49,50,51\}$ lie below the dashed cutoff $\varepsilon=10^{-3}$, supporting the choice $\xi=1$.
Figure~\ref{fig:GL1d_radius}(b) further shows that with higher coupling intensity $\lambda$, the relative sensitivity spreads to more distant sites, reinforcing the need for a larger neighborhood radius if the coupling is stronger.
 
\subsubsection{2D Ginzburg--Landau model}
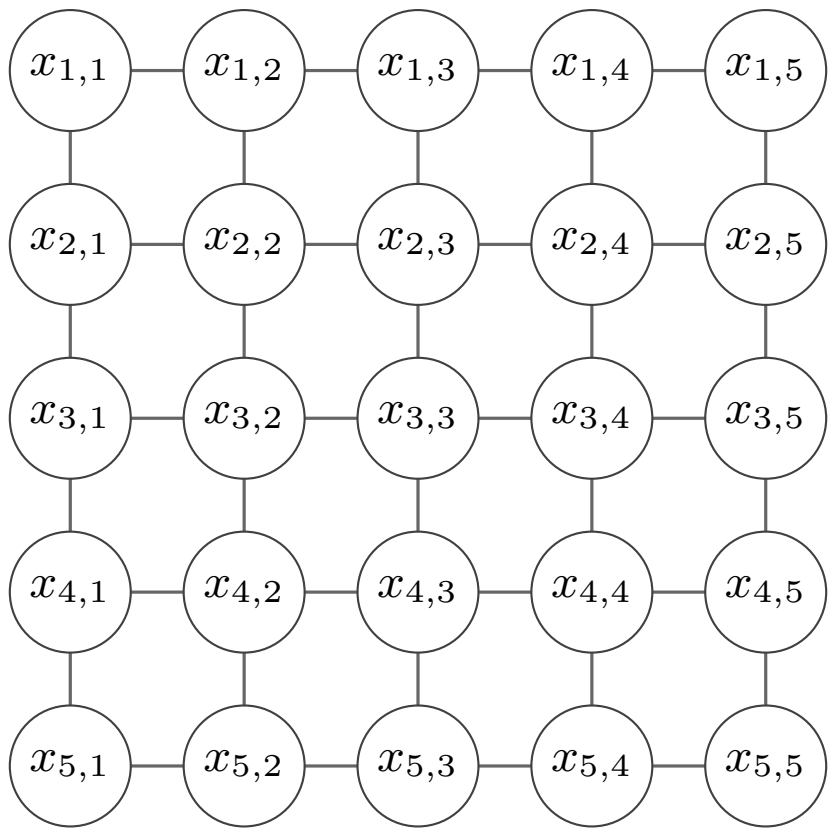
\begin{figure}[!htbp]
    \centering
    \begin{tikzpicture}[
        x=1.15cm,y=1.15cm,
        glsite/.style={circle,draw=black!75,fill=white,
                       minimum size=8mm,inner sep=0pt,font=\small}
    ]
        \foreach \i in {1,...,5}
            \draw[black!60,line width=0.6pt] (1,-\i)--(5,-\i);
        \foreach \j in {1,...,5}
            \draw[black!60,line width=0.6pt] (\j,-1)--(\j,-5);
        \foreach \i in {1,...,5} {
            \foreach \j in {1,...,5}
                \node[glsite] at (\j,-\i) {$x_{\i,\j}$};
        }
    \end{tikzpicture}
\caption{A $5\times5$ schematic of the two-dimensional Ginzburg--Landau lattice. Each circle represents a site variable $x_{i,j}$, and edges connect nearest neighbors.}
    \label{fig:GL2d_lattice}
\end{figure}

The connected-cluster basis is defined on any interaction graph and, unlike chain-like tensor networks, does not require a one-dimensional ordering of the sites. As an example, we consider the two-dimensional Ginzburg--Landau model on a $d_1 \times d_1$ nearest-neighbor 2D lattice, with potential 
\begin{align*}
    U_{\rm 2D}(x) = \frac{\lambda}{2} \left[ \sum_{i=0}^{d_1}\sum_{j=1}^{d_1} \left(\frac{x_{i+1,j}-x_{i,j}}{h}\right)^2 + \sum_{i=1}^{d_1}\sum_{j=0}^{d_1} \left(\frac{x_{i,j+1}-x_{i,j}}{h}\right)^2 \right] + \frac{1}{4\lambda} \sum_{i=1}^{d_1}\sum_{j=1}^{d_1} (1-x_{i,j}^2)^2, 
\end{align*}
defined on the hypercube $[-1.5,1.5]^d$, where $d=d_1^2$, $h=1/(d_1+1)$, and the boundary values are fixed as
\begin{equation*}
    x_{0,j}=x_{d_1+1,j}=1, \quad 1\leq j\leq d_1, \qquad x_{i,0}=x_{i,d_1+1}=-1, \quad 1\leq i\leq d_1. 
\end{equation*}
Figure~\ref{fig:GL2d_lattice} illustrates the nearest-neighbor lattice using a $5\times5$ schematic.

\begin{figure}[!htbp]
    \centering
\includegraphics[width=0.75\textwidth]{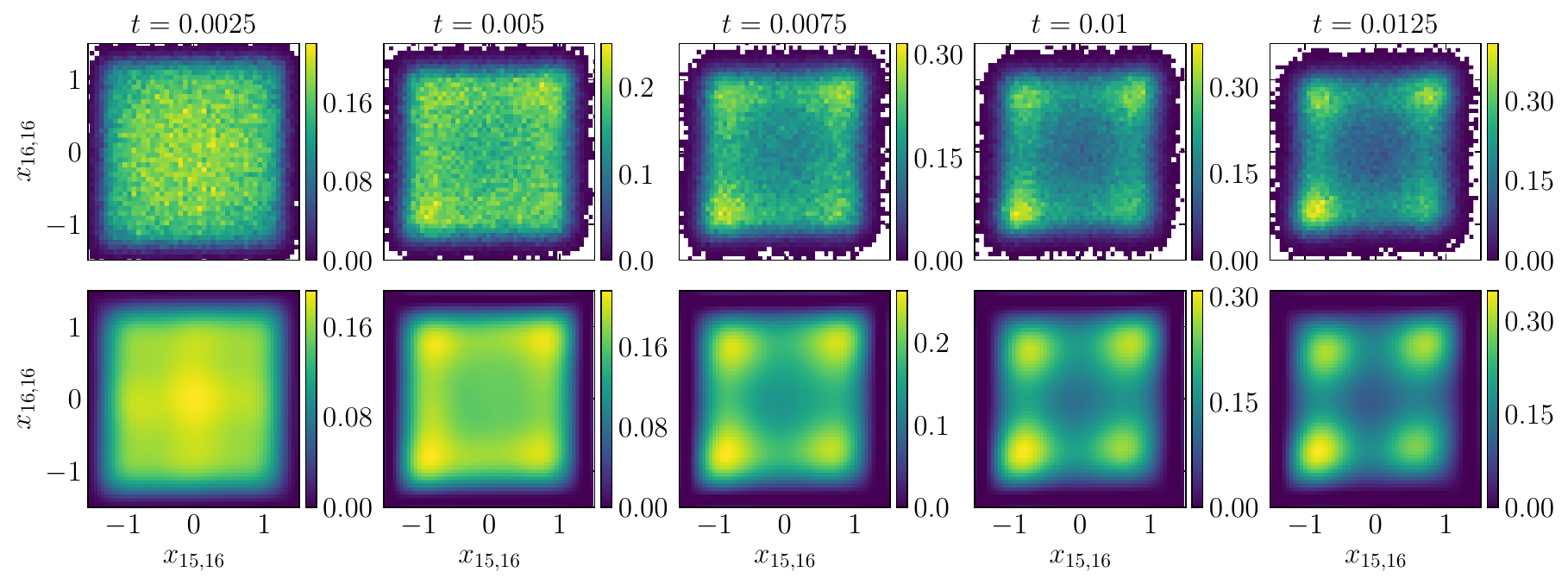} \caption{Evolution of the adjacent two-site marginal $S=\{(15,16),(16,16)\}$ in the $16\times16$ 2D Ginzburg--Landau model. Top: empirical reference (histograms); bottom: prediction obtained by iteratively applying the estimated transition moment matrix.}
    \label{fig:GL2d_d256_near_new}

\end{figure}

For such a 2D lattice system, representing the solution by a chain-like tensor network such as a tensor train is not straightforward. We take $d_1=16$, so that $d=256$, and set $\lambda=0.02$ and $\beta=1/4$. As in the 1D experiment, we predict a two-site marginal from an initial density $\rho_0$ given by the product of standard Gaussians truncated to $[-1.5,1.5]$. The basis is constructed as in the 1D experiment with $\xi=1$, total cluster order $k=2$, and 10 Legendre functions per site. The transition moment matrices are estimated from $N_{\rm src}=2\times10^4$ source states and $N_{\rm traj}=50$ trajectories per source state over a {time period} ${\tau}=2.5\times10^{-3}$, using an Euler--Maruyama step of $\Delta t={\tau}/20$. We again choose the uniform distribution on $[-1.5,1.5]^d$ as the reference measure $\rho_{\mathrm{ref}}$. An independent ensemble of $2\times10^5$ trajectories provides the empirical reference. Figure~\ref{fig:GL2d_d256_near_new} shows the evolution of the adjacent $\{(15,16),(16,16)\}$-marginal. The reconstructed marginal closely follows the empirical reference.

\subsection{Glauber dynamics for classical spin systems}
\label{sec:glauber_results}
We consider the Glauber dynamics of spin systems in 1D (Section~\ref{sec:1dising}) and 2D (Section~\ref{sec:glauber2d_results}).

\subsubsection{1D Ising chain}
\label{sec:1dising}
We next consider the Glauber dynamics for classical spin systems and consider the one-dimensional Ising chain with open boundaries and energy
\begin{equation*}
    U_{\text{1D}}(s) =-J\sum_{i=1}^{d-1}s_i s_{i+1} -h\sum_{i=1}^d s_i, \qquad s_i\in\{-1,1\}. 
\end{equation*}
We set $J=0.6$, $h=-0.2$, and $\beta=1$ in the heat-bath rates \eqref{eq:glauber_flip_rate}. For the moment prediction experiment, we take $d=100$ and initialize all spins at $+1$. We estimate the empirical Gram and transition moment matrices from $N_{\rm src}=4\times10^5$ independent source states sampled from the stationary Gibbs reference measure $\rho_{\mathrm{ref}}=\pi$ with $\pi(s)\propto\exp(-\beta U_{\text{1D}}(s))$, with $N_{\rm traj}=200$ short-time trajectories per source state over a time period $\tau=1$. We apply twenty one-step propagations up to the final time $T=20$. All trajectories evolve the full chain using the exact Gillespie algorithm, and an independent ensemble of $2\times10^5$ trajectories provides the empirical reference.

\begin{figure}[!htbp]
    \centering
\includegraphics[width=\textwidth]{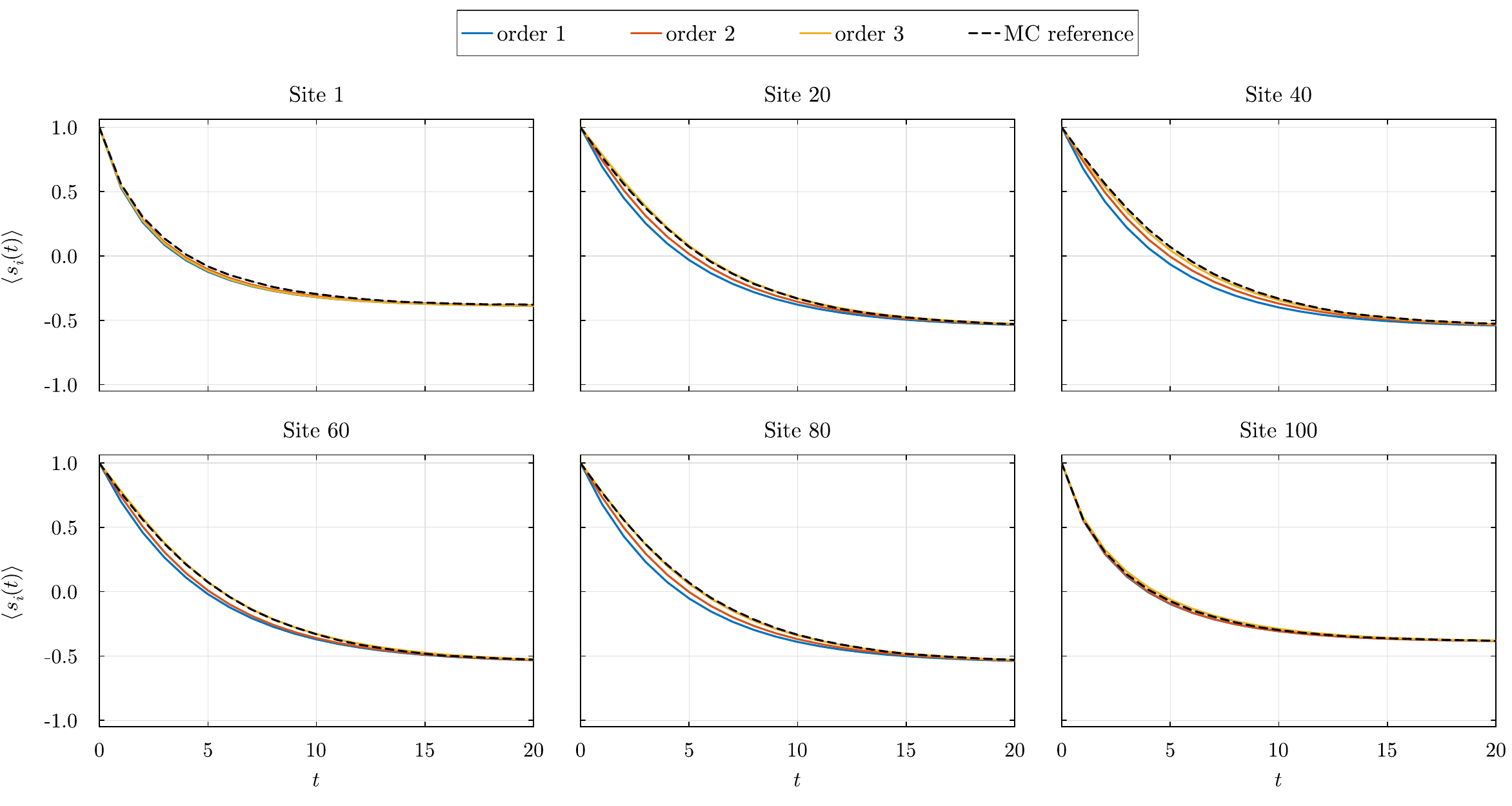}
\caption{Simultaneous compressed predictions of the single-spin means $\mu_i(t)$ at $i=1,20,40,60,80,100$ in the $d=100$ Glauber--Ising chain.}
    \label{fig:glauber_single_compressed}
\end{figure}

We first predict the single-spin means $\mu_i(t)=\E_{\rho_t}[s_i]$ simultaneously at sites $i=1,20,40,60,80,100$. For each cluster order $k=1,2,3$, a single compressed transition moment matrix on the connected-cluster basis $\mathcal B_k(V)$ supplies all six predictions. The corresponding numbers of basis functions are $n=101,200,396$. We compress both the empirical Gram and transition moment matrices using Algorithm~\ref{alg:matrix_free_compression}, with near-field radius $\delta=5$, rank bound $r=10$, and $\nu_{\max}=20$ alternating iterations. Figure~\ref{fig:glauber_single_compressed} compares the resulting predictions with the empirical reference. The predictions with $k=3$ closely follow the relaxation at all six sites. The relative time-series error averaged over the six sites decreases from $11.58\%$ at $k=1$ to $6.68\%$ at $k=2$ and $2.69\%$ at $k=3$.

\begin{figure}[!htbp]
    \centering
\includegraphics[width=0.75\textwidth]{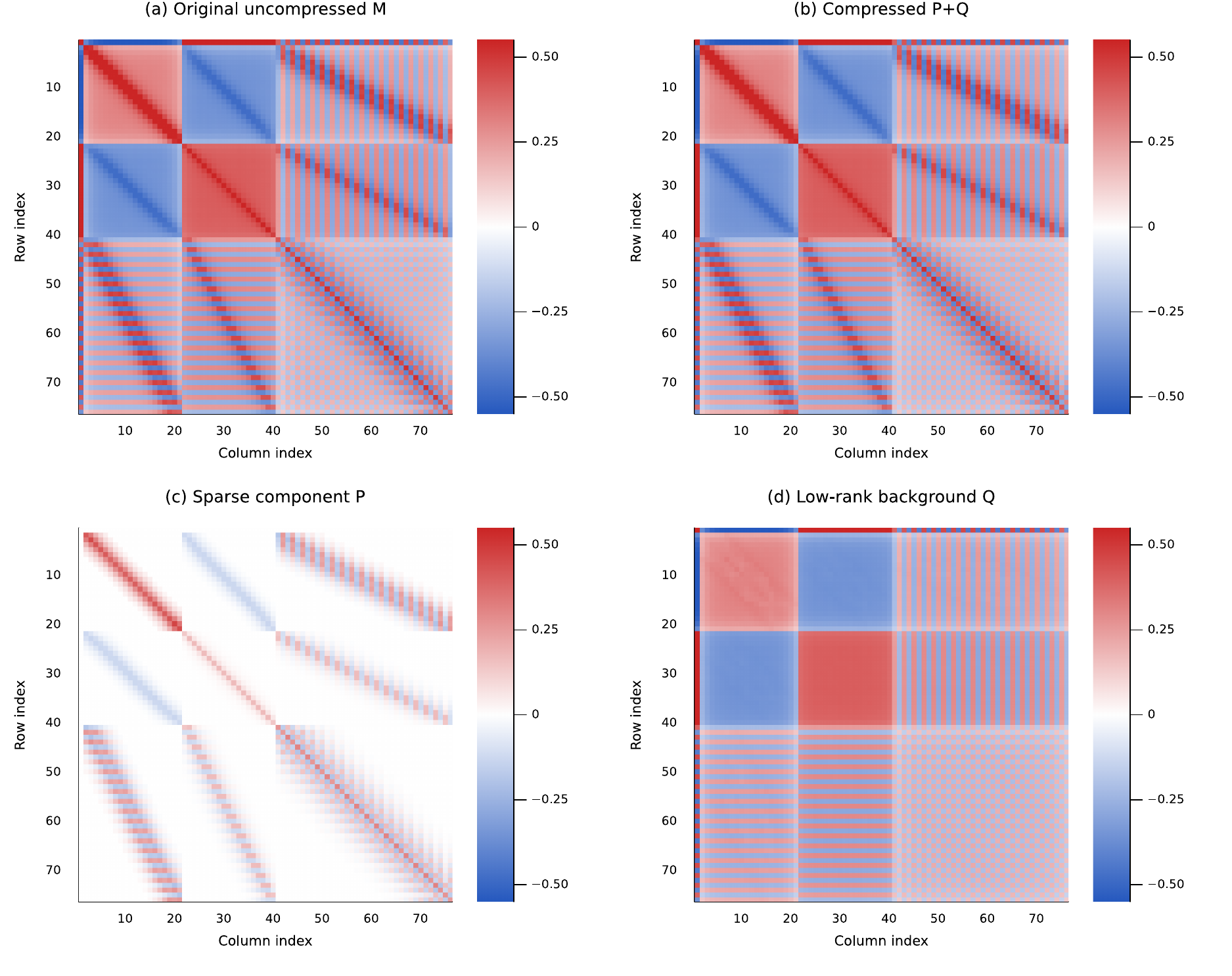}
\caption{Compression of the $76\times76$ empirical transition moment matrix for the $d=20$ Glauber--Ising chain with $k=3$. (a) Original matrix $M$; (b) compressed matrix $P+Q$; (c) sparse near-field component $P$; (d) low-rank component $Q$, with $\operatorname{rank}(Q)\leq10$.}
    \label{fig:glauber_compression_quality}
\end{figure}

Under the same model and compression settings, we next examine the compression quality for a smaller chain with $d=20$ and cluster order $k=3$. We use $N_{\rm src}=2\times10^5$ source states. The uncompressed empirical transition moment matrix $M:=\widehat M_{\rho_{\mathrm{ref}}}^{\tau}$ has size $76\times76$. Figure~\ref{fig:glauber_compression_quality} compares $M$ with its approximation $P+Q$ and displays the two components separately. The sparse component $P$ corrects the near-field entries, while $Q$ is stored in factored form and has rank at most ten. The compressed matrix reproduces the visible structure of $M$, with relative Frobenius error $\|M-(P+Q)\|_F/\|M\|_F\approx1.03\times10^{-3}$, demonstrating the accuracy of our proposed compression algorithm.

\begin{figure}[!htbp]
    \centering
\includegraphics[width=0.75\textwidth]{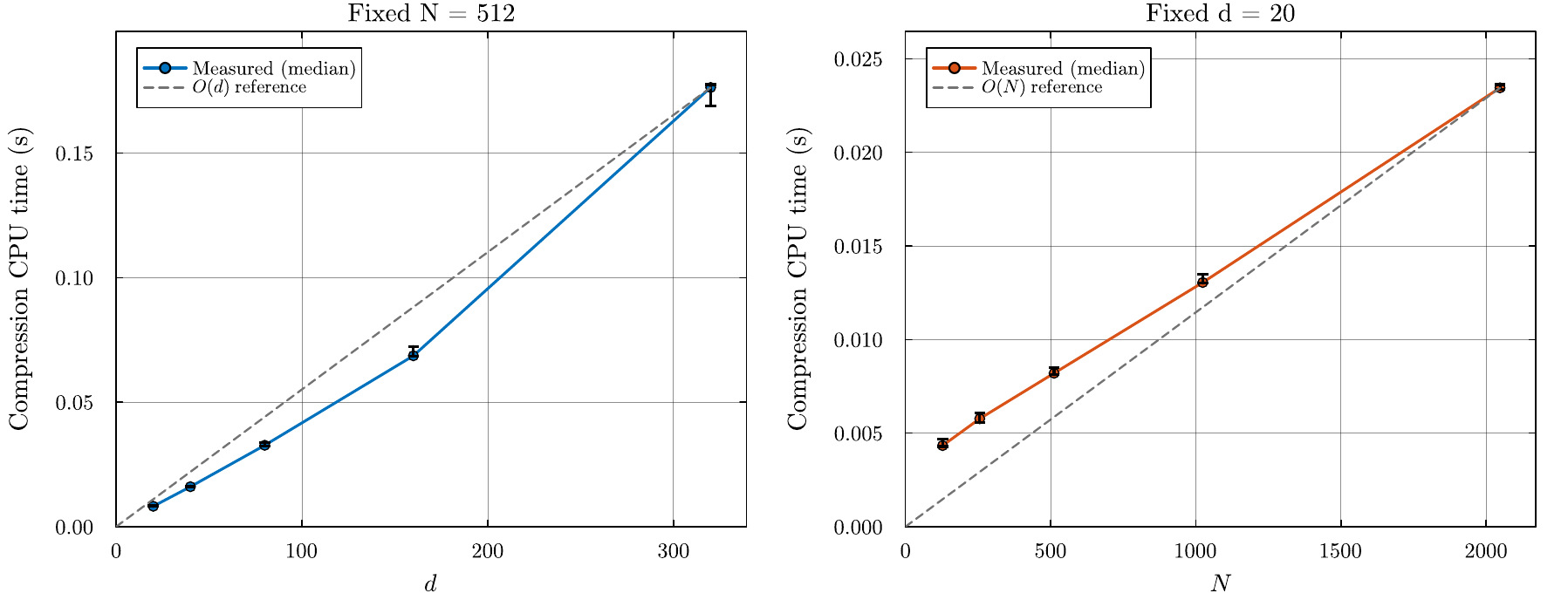}
\caption{CPU time for matrix-free compression. Left: varying chain length $d$ at fixed $N_{\rm src}=512$. Right: varying the number of source states $N_{\rm src}$ at fixed $d=20$ (the horizontal axis denotes $N=N_{\rm src}$). Points show medians of seven repetitions, and error bars span the minimum and maximum. Dashed lines  indicate linear growth in $d$ and in $N$.}
    \label{fig:glauber_cpu_scaling}
\end{figure}

Finally, we test the computational cost of constructing the compressed matrix once the trajectories are available, as discussed in Section~\ref{sec:compression}. We fix $k=3$, $\delta=5$, $r=10$, and $\nu_{\max}=20$, and use the same model parameters and $N_{\rm traj}=200$. Figure~\ref{fig:glauber_cpu_scaling} shows CPU time as $d$ varies over $20,40,80,160,320$ at fixed $N_{\rm src}=512$, and as $N_{\rm src}$ varies over $128,256,512,1024,2048$ at fixed $d=20$.
Each point is the median of seven repetitions after warmup. The observed growth is approximately linear in each variable, consistent with the $\mathcal O(dN_{\rm src})$ cost of this stage for fixed cluster order, near-field radius, rank, and iteration count.
The CPU timings are measured on an NVIDIA DGX Spark with a 20-core Arm CPU (NVIDIA GB10; 10 Cortex-X925 and 10 Cortex-A725 cores) using Julia 1.12.6.

\subsubsection{2D spin lattice}
\label{sec:glauber2d_results}
We next consider the Ising model on a $10\times10$ square lattice (so that $d=100$) with open boundaries, as in \Cref{fig:GL2d_lattice}. 
Identifying the vertex set $V$ with the lattice coordinates $\{1,\ldots,10\}^2$, we write the energy as
\begin{equation*}
    U_{\text{2D}}(s)=-J\sum_{\{i,j\}\in E}s_i s_j-h\sum_{i\in V}s_i,
\end{equation*}
where $E$ contains the horizontal and vertical nearest-neighbor edges. We set $J=0.3$, $h=-0.5$, and $\beta=1$ in~\eqref{eq:glauber_flip_rate}. Initially, the spins are independent and unbiased, giving $\rho_0(s)=2^{-100}$. We use the uniform reference measure $\rho_{\mathrm{ref}}(s)=2^{-100}$.
In all experiments on this lattice, we use a time period $\tau=0.25$ and final time $T=6$.

We predict the two-spin moments $\mu_{(1,1),(2,1)}(t)$ and $\mu_{(1,1),(10,10)}(t)$. The first pair consists of adjacent sites $(1,1)$ and $(2,1)$, and the second consists of the opposite corners $(1,1)$ and $(10,10)$. For the adjacent pair, we use a joint connected-cluster basis $\mathcal B_k(\mathcal N_3(\{(1,1),(2,1)\}))$. For the distant pair, we predict separate moment dynamics, using the bases $\mathcal B_k(\mathcal N_3(\{(1,1)\}))$ and $\mathcal B_k(\mathcal N_3(\{(10,10)\}))$, and approximate $\mu_{(1,1),(10,10)}(t)$ by $\mu_{(1,1)}(t)\mu_{(10,10)}(t)$ as in Remark~\ref{rem:disconnected_targets}. 

\begin{figure}[!htbp]
    \centering
    \includegraphics[width=0.75\textwidth]{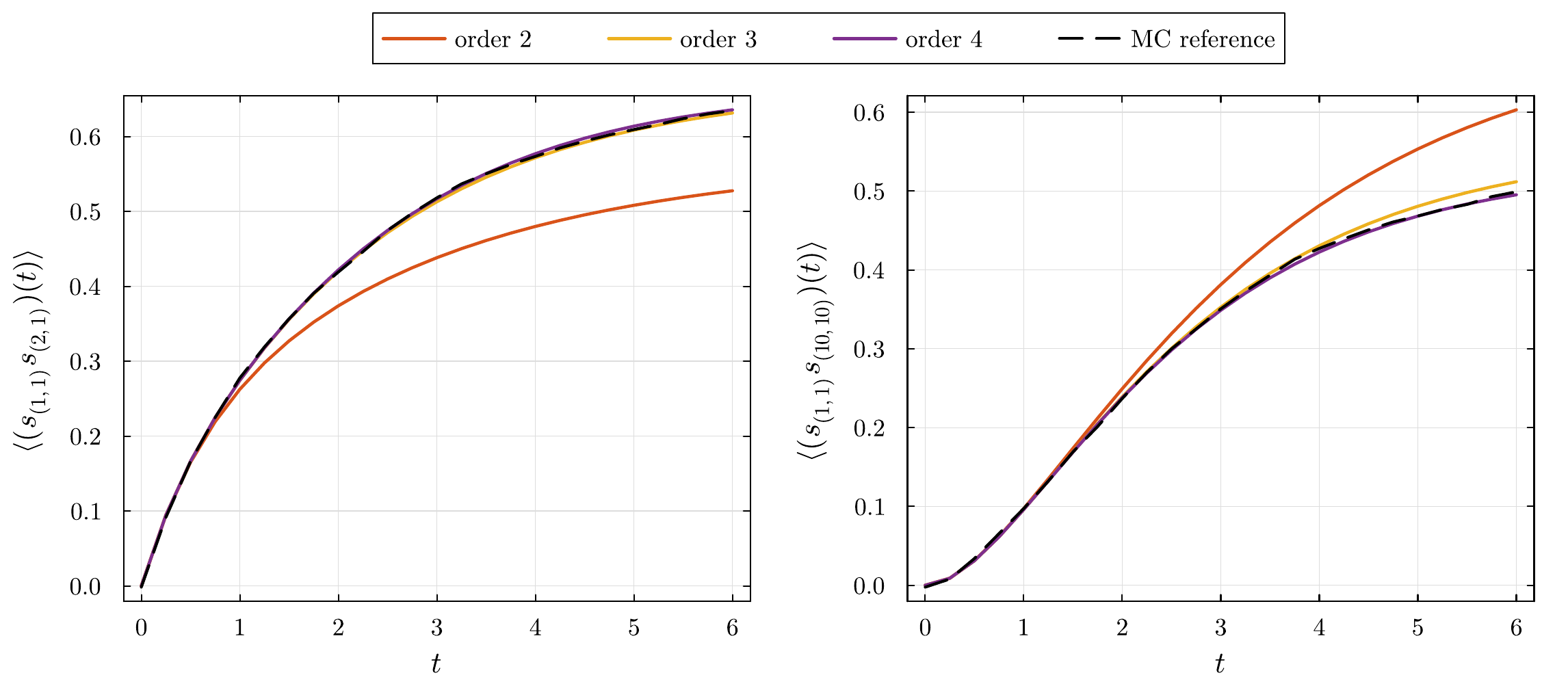}
    \caption{Two-spin moment predictions on the $10\times10$ Glauber--Ising lattice using repeated one-step propagation with cluster orders $k=2,3,4$. Left: joint prediction for the adjacent sites $(1,1)$ and $(2,1)$. Right: the opposite corners $(1,1)$ and $(10,10)$, predicted by multiplying the two predicted single-spin means. Dashed curves show  the empirical references for the true joint moments.}
    \label{fig:glauber2d_pairs_order}
\end{figure}

We first check the convergence with respect to cluster order, by varying the cluster order $k=2,3,4$ and applying repeated one-step propagation without TTM. All three use the same $N_{\rm src}=2\times10^5$ independent source states from the uniform reference measure and $N_{\rm traj}=200$ trajectories per source over the short time period $\tau$. The  bases for the adjacent pair contain $n=34,98,284$ functions, while each of  the two bases for the distant pair contains  $n=23,59,142$ functions. An independent ensemble of $2\times10^5$ trajectories up to the final time provides the empirical reference. Figure~\ref{fig:glauber2d_pairs_order} shows that increasing the cluster order from $2$ through $4$ reduces the relative time-series error over the full interval from $15.41\%$ to $0.51\%$ for the adjacent pair and from $15.45\%$ to $0.76\%$ for the distant pair. The agreement with the reference for the joint moment also supports the weak-correlation approximation for these distant sites over the tested interval.

\begin{figure}[!htbp]
    \centering
    \includegraphics[width=0.75\textwidth]{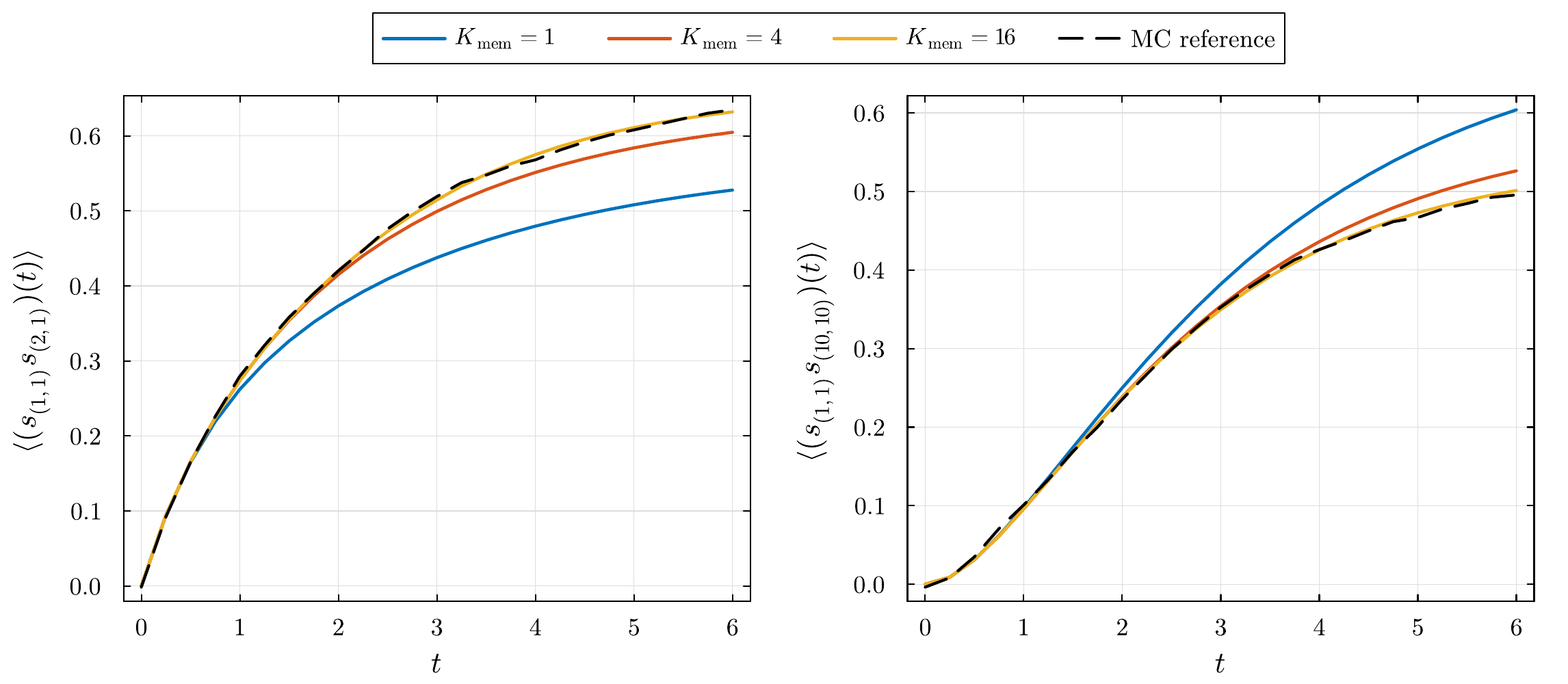}
    \caption{Two-spin moment predictions on the $10\times10$ Glauber--Ising lattice using TTM with memory lengths $K_{\rm mem}=1,4,16$ at fixed cluster order $k=2$. Left: joint prediction for the adjacent sites $(1,1)$ and $(2,1)$. Right: the opposite corners $(1,1)$ and $(10,10)$, predicted by multiplying the two predicted single-spin means. Dashed curves show the empirical references for the  joint moments.}
    \label{fig:glauber2d_pairs_memory}
\end{figure}

We next fix the cluster order at $k=2$ and vary the memory length $K_{\rm mem}=1,4,16$ to check the convergence {with respect to} the memory length. We estimate the dynamical maps $\mathcal E_m$ for $m=1,\ldots,16$ using $N_{\rm src}=4\times10^4$ uniform source states and $N_{\rm traj}=100$ trajectories per source state. An independent ensemble of $10^5$ trajectories provides the empirical reference. Figure~\ref{fig:glauber2d_pairs_memory} shows that increasing the memory length reduces the late-time error in both panels, while the cluster order and the matrix sizes are unchanged. In particular, the prediction with $K_{\rm mem}=16$ closely follows the empirical reference. For $t>4$, the relative time-series errors at $K_{\rm mem}=16$ are $0.49\%$ for the adjacent pair and $0.82\%$ for the distant pair. These results illustrate that the memory correction can improve the prediction at a low cluster order, without increasing it.

\subsection{Lindblad dynamics for open quantum systems}
\label{sec:quantum_results_new}
Finally, we consider the Lindblad dynamics of Section~\ref{sec:lindblad_application} and predict one-site moments. In all examples below, the number of quantum spins is $d=50$, and thus we use TJM to generate MPS trajectories to avoid {the} curse of dimensionality.

\subsubsection{Transverse-field Ising chain}
\begin{figure}[!htbp]
    \centering
\includegraphics[width=\textwidth] {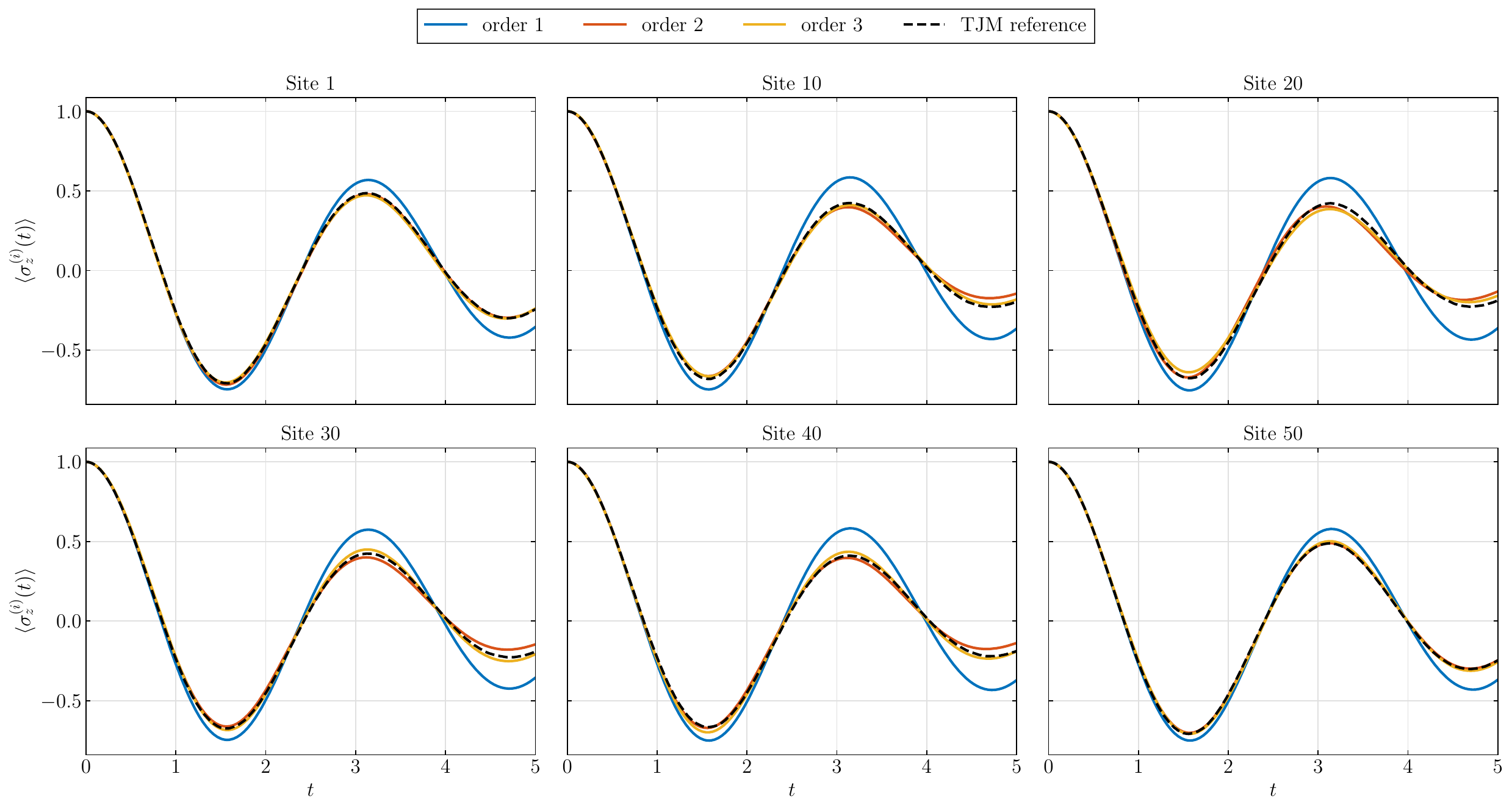} \caption{Compressed  predictions of $\langle\sigma_z^{(i)}\rangle$ at $i=1,10,20,30,40,50$ for the $d=50$ dissipative Ising chain. We use $h_x=1$, $J=0.2$, $\gamma_-=\gamma_z=0.1, \tau=0.02$, and $T=5$. We use the sparse-plus-rank-one initialization. Cluster orders $k=1,2,3$ are compared with an empirical TJM reference of $10^4$ trajectories.}
    \label{fig:quantum_d50_new}
\end{figure}

We consider the dissipative transverse-field Ising chain. For this model, the master equation in~\eqref{eqn:linblad} takes the form
\begin{equation*}
    \frac{\dd\rho_t}{\dd t} =-\mathrm{i}[H,\rho_t] +\sum_{i=1}^{2d}\left( L_i\rho_tL_i^* -\frac12\{L_i^*L_i,\rho_t\} \right), 
\end{equation*}
with Hamiltonian and jump operators
\begin{equation*}
\begin{aligned}
    H &= -J\sum_{i=1}^{d-1}\sigma_z^{(i)}\sigma_z^{(i+1)} -h_x\sum_{i=1}^{d}\sigma_x^{(i)}, \\
    L_i &=\begin{cases}
    \sqrt{\gamma_-}\,\sigma_-^{(i)}, & 1\le i\le d, \\
    \sqrt{\gamma_z}\,\sigma_z^{(i-d)}, & d+1\le i\le 2d.
    \end{cases}
\end{aligned}
\end{equation*}
Here $\sigma_-:=(\sigma_x+\mathrm{i}\sigma_y)/2.$ The first $d$ jump operators describe local relaxation, and the remaining $d$ describe local dephasing. All experiments on the Ising chain use $d=50$ and the  initial state
\begin{equation*}
  \rho_0=\bigotimes_{i=1}^{d}  \begin{pmatrix}
    1 & 0 \\
    0 & 0
  \end{pmatrix}{.}
\end{equation*}
We first use the connected-cluster basis $\mathcal{B}_k(V)$ of Pauli strings with compression to predict all one-site moments $\langle \sigma_z^{(i)}\rangle:=\mathrm{Tr}(\sigma_z^{(i)} \rho_t )$ simultaneously, with $h_x=1$, $J=0.2$, $\gamma_-=\gamma_z=0.1$, {time period} ${\tau}=0.02$, and final time $T=5$. We estimate the transition moment matrices from $N_{\rm src}=4\times10^4$ random product source states and $N_{\rm traj}=200$ TJM trajectories per source state. The empirical reference averages $10^4$ independent TJM trajectories. 
We compare {cluster orders} $k=1,2,3$, with near-field radius $\delta=2$ and the initial sparse-plus-rank-one compression as {discussed at the end of} \Cref{sec:compression}. Figure~\ref{fig:quantum_d50_new} shows that the predictions with $k=2,3$ remain close to the empirical TJM reference, while $k=1$ gives a larger amplitude error during the second oscillation.

\begin{figure}[!htb]
    \centering
    \begin{minipage}[t]{0.49\textwidth}
        \centering
        \includegraphics[height=0.76\linewidth]{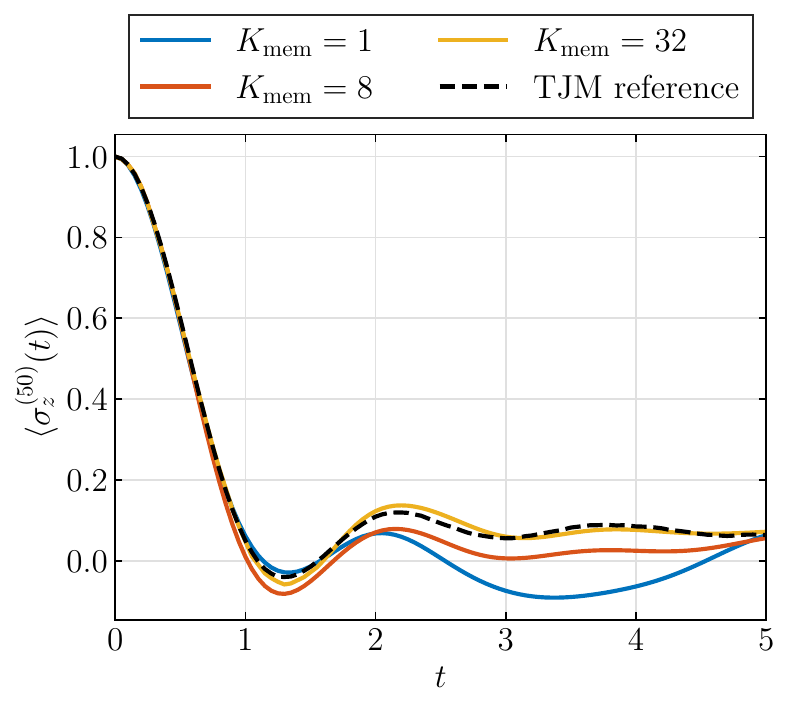}
        \par\smallskip {\small (a) Varying memory length}
    \end{minipage}\hfill
    \begin{minipage}[t]{0.49\textwidth}
        \centering
        \includegraphics[height=0.76\linewidth]{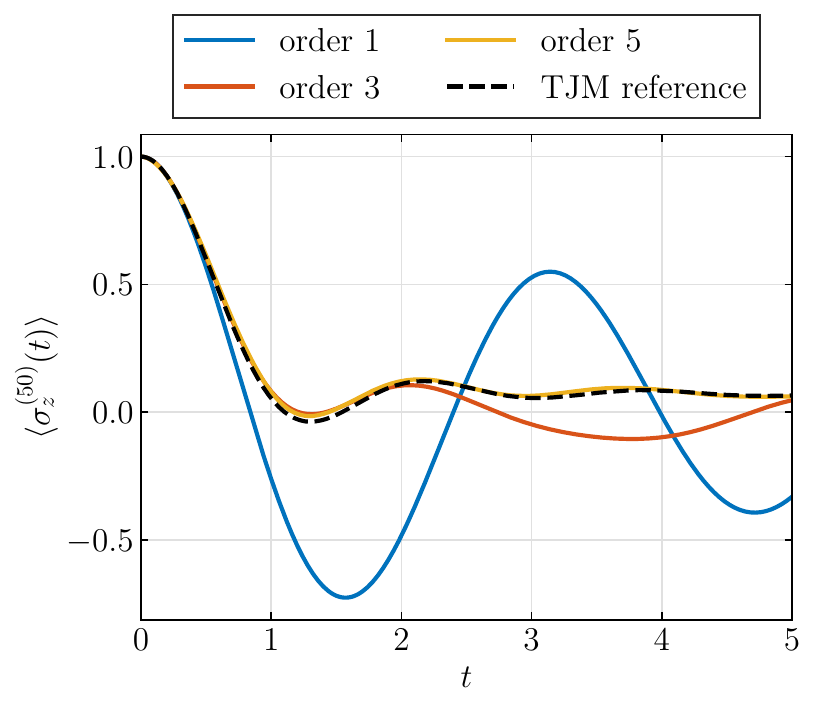}
        \par\smallskip {\small (b) Varying cluster order}
    \end{minipage}
    \caption{Predictions of $\langle\sigma_z^{(50)}\rangle$ for the $d=50$ dissipative Ising chain with $h_x=J=1$, $\gamma_-=\gamma_z=0.1$, and final time $T=5$, using connected-cluster bases on $\{40,\ldots,50\}$.
    (a) TTM predictions at fixed cluster order $k=3$, with time period $\tau=0.05$ and memory lengths $K_{\rm mem}=1,8,32$.
    (b) Repeated one-step propagation with time period $\tau=0.02$ and cluster orders $k=1,3,5$.
    Dashed curves show empirical TJM references averaging $5000$ trajectories in (a) and $8\times10^4$ trajectories in (b).}
    \label{fig:quantum_ttm_d50_J1}
\end{figure}

We next increase the coupling to $J=1$ and focus on the boundary moment $\langle\sigma_z^{(50)}\rangle$, keeping $h_x=1$, $\gamma_-=\gamma_z=0.1$, and $T=5$. The following two experiments use uncompressed transition moment matrices on the connected-cluster basis of Pauli strings on the neighborhood $\mathcal N_{\xi}(\{50\})=\{40,\ldots,50\}$ with $\xi=10$.
To assess TTM at cluster order $k=3$, we estimate the dynamical maps $\mathcal E_m$ {for successive multiples of the time period} ${\tau}=0.05$ from trajectories up to time $1.6$, using $N_{\rm src}=5000$ random product source states and $N_{\rm traj}=50$ TJM trajectories per source state. The internal TJM time step is $\Delta t=0.005$, with maximum MPS bond dimension $16$. Figure~\ref{fig:quantum_ttm_d50_J1}(a) compares memory lengths $K_{\rm mem}=1,8,32$ with an empirical reference of $5000$ TJM trajectories. Repeated one-step propagation ($K_{\rm mem}=1$) develops a substantial late-time error. Increasing the memory length improves the prediction, and $K_{\rm mem}=32$ closely follows the empirical reference through $T=5$.

Finally, we examine repeated one-step iteration with cluster orders $k=1,3,5$ in the same neighborhood. For this experiment, we use $N_{\rm src}=8\times10^4$ random product source states and $N_{\rm traj}=200$ short TJM trajectories per source state. The one-step {time period} is ${\tau}=0.02$, resolved by ten TJM steps of size $\Delta t=0.002$, with a maximum MPS bond dimension of $16$. The empirical reference averages $8\times10^4$ independent TJM trajectories. Figure~\ref{fig:quantum_ttm_d50_J1}(b) shows that the prediction with $k=5$ closely follows the empirical reference over the prediction interval, while lower orders have larger errors. These predictions use repeated one-step propagation, without compression or TTM.

\subsubsection{Heisenberg chain}

\begin{figure}[!htb]
    \centering
\includegraphics[width=0.72\textwidth] {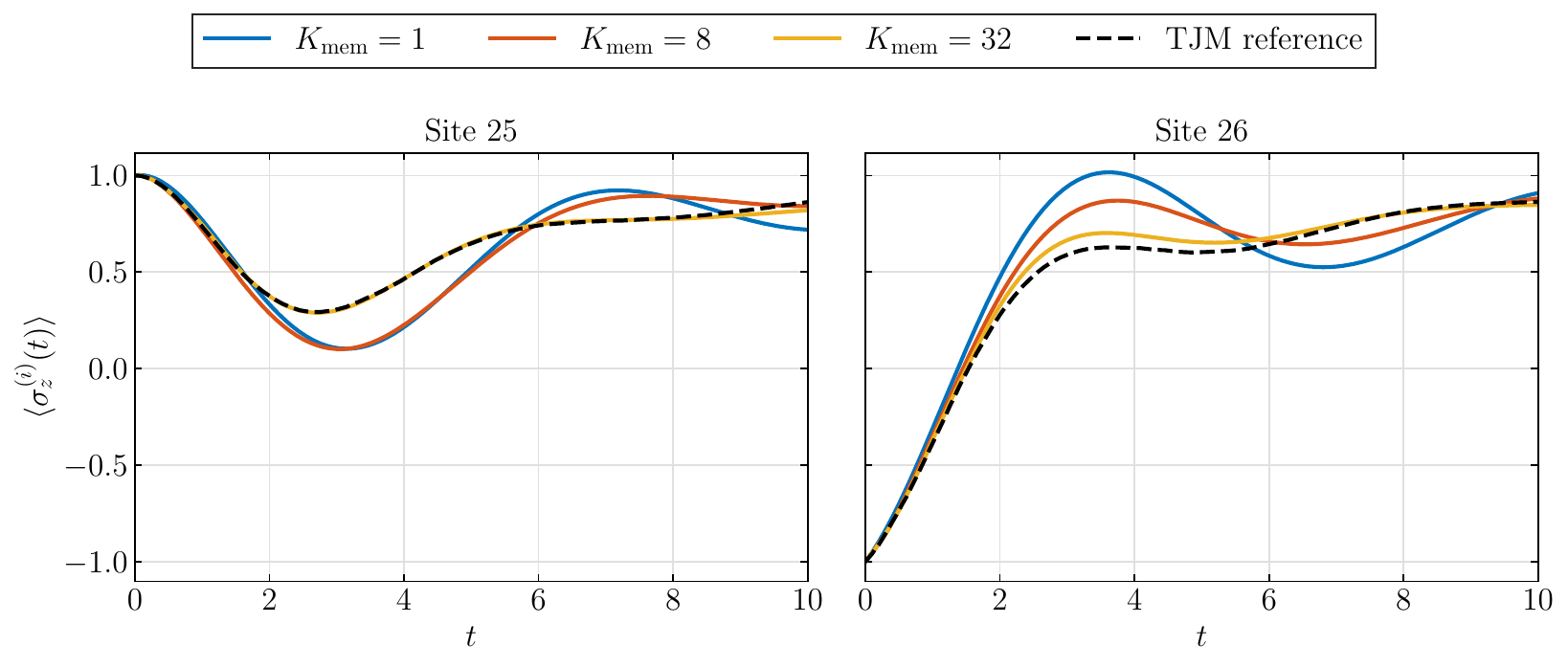} \caption{ TTM predictions of the moments $\langle \sigma_z^{(25)} \rangle$ and $\langle \sigma_z^{(26)} \rangle$ in the $d=50$ dissipative Heisenberg XXX chain with memory lengths $K_\mathrm{mem}=1,8,32$ compared with the empirical TJM reference. We use $J=0.15,h=0.1,\gamma_-=0.2,\tau=0.05,T=10$, and the connected-cluster basis with $k=2$ on the neighborhood $\mathcal N_3(\{25,26\})=\{22,\ldots,29\}$.}
    \label{fig:heisenberg_ttm_d50_new}
\end{figure}

We consider the  $d$-site Heisenberg XXX chain with
\begin{equation*}
\begin{split}
    H & = -J\sum_{i=1}^{d-1} \left( \sigma_x^{(i)}\sigma_x^{(i+1)} +\sigma_y^{(i)}\sigma_y^{(i+1)}+ \sigma_z^{(i)}\sigma_z^{(i+1)} \right)-h\sum_{i=1}^d \sigma_z^{(i)},\\
    L_i& =\sqrt{\gamma_-}\; I^{\otimes(i-1)} \otimes \begin{pmatrix} 0&1\\ 0&0 \end{pmatrix} \otimes I^{\otimes(d-i)}. 
\end{split}
\end{equation*}
We use $d=50, J=0.15,h=0.1,\gamma_-=0.2$, and use the Néel product state as the initial state, in which the odd and even sites are in the $+1$ and $-1$ eigenstates of $\sigma_z$, respectively, so that $ \langle \sigma_z^{(i)}\rangle=(-1)^{i+1}$ at $t=0.$ We use $N_\mathrm{src}=5000$ and $N_\mathrm{traj}=50$. The {dynamical maps} are estimated from trajectories up to time $1.6$ with time period $\tau=0.05$, and TTM predicts the moments up to $T=10$. We use the connected-cluster basis $\mathcal B_2(\mathcal N_3(\{25,26\}))$ of Pauli strings  to predict  $\langle \sigma_z^{(25)}\rangle$ and $\langle \sigma_z^{(26)}\rangle$, giving $n=88$. The trajectories used for estimation use a TJM time step of $\Delta t=0.00625$ and a maximum MPS bond dimension of $8$. The empirical TJM reference averages $4000$ trajectories with $\Delta t_{\mathrm{ref}}=0.003125$ and maximum bond dimension $16$. In Figure~\ref{fig:heisenberg_ttm_d50_new}, we compare the TTM predictions at the two central sites with the empirical TJM reference.

\FloatBarrier
\section{Conclusion}
\label{sec:conclusion}

We have developed a framework for predicting local observable dynamics $\mathcal P_t g_C$ in classical and quantum many-body systems from short-time trajectories.
A Galerkin projection of the backward Markov operator onto connected-cluster basis functions yields a reduced transition moment matrix, which is compressed in sparse-plus-low-rank form to enable efficient prediction of many moments.
We further use the transfer tensor method to account for memory effects arising from the projection. Numerical experiments demonstrate the framework for Langevin and Glauber dynamics, as well as Lindblad dynamics through stochastic unravelings.

In addition, for reversible local binary-spin dynamics, we establish $L^2(\pi)$ error bounds for connected-cluster approximation and repeated one-step propagation using exact moment matrices under the stationary law $\pi$.
Under the locality and spectral-gap assumptions, a neighborhood radius of $\mathcal O(\log(1/\varepsilon))$ and a cluster order of $|C|+\mathcal O(\log^2(1/\varepsilon))$ suffice to approximate $\mathcal P_t g_C$ to accuracy $\varepsilon$, uniformly in time.
We also bound the additional observable propagation error introduced by sparse-plus-rank-one compression of the Gram and transition moment matrices. Extending these guarantees to other reference measures, Monte Carlo estimation, subsequent low-rank fitting, and finite-memory TTM remains open.

\medskip

\noindent\textbf{Use of AI tools.}
The authors designed the proposed algorithm without AI assistance, including the use of local-cluster bases to approximate observable dynamics, sparse-plus-low-rank compression of transition moment matrices for efficient computation, and the transfer tensor method to approximate memory effects. During manuscript preparation, the authors used ChatGPT-6 (OpenAI) to help explore mathematical arguments, check derivations, and refine the exposition. They also used Claude Opus 5.5 (Anthropic) to edit the text and assist with the implementation of selected computational subroutines. The authors reviewed all AI-assisted outputs and take full responsibility for the mathematical results, numerical computations, and final content of the manuscript.

\appendix

\section{Proofs of Section~\ref{sec:glauber_guarantees}}
\label{app:gb-proofs}

\subsection{Local approximation and propagation error}

We give the proof of Theorem~\ref{thm:st-log-radius} in this appendix. To this end, we first introduce site oscillations and use them to bound the error of freezing the exterior in Appendix~\ref{app:st-site-freezing}, and then construct a local polynomial approximation in Appendix~\ref{app:st-polynomial}. These estimates yield the connected-cluster approximation bound in Appendix~\ref{app:st-local-approximation} and the repeated one-step propagation bound in Appendix~\ref{app:st-exact-propagation}. 

\subsubsection{Site oscillations and exterior freezing}
\label{app:st-site-freezing}

For a real observable $h$, define its site oscillations by
\[
 \omega_i(h)=\max_{\substack{s,y\in\Omega\\s_j=y_j\ \forall j\ne i}}|h(s)-h(y)|.
\]
We recall the model \eqref{eq:st-generator}, and introduce the total oscillation of the target and the propagation speed:
\begin{equation}
 A_g=\sum_{i\in C}\omega_i(g_C),\qquad v=e(R_{\mathrm{loc}}-1)J_{\mathrm{loc}}.
 \label{eq:st-local-propagation-constants}
\end{equation}

Site oscillations quantify how an observable depends on individual sites and are a standard tool in the analysis of interacting particle systems~\cite{liggett1985interacting}.
For dynamics that allow a single jump to update several sites, Doukhan et al.~\cite[Proposition~8.1]{doukhan2008functional} bound the site-oscillation vector of the evolved observable componentwise by applying the exponential of the jump-rate influence matrix to the initial site-oscillation vector.
Wu~\cite[Proposition~2.5 and Theorem~3.1]{wu2006poincare} gives Dobrushin-type oscillation bounds for single-site heat-bath dynamics and, more generally, for single-site jump dynamics.
We give a direct proof using a symmetric comparison matrix whose off-diagonal entries sum the maximal exit rates of the blocks containing both sites. This construction gives explicit bounds on the interaction range and operator norm of the matrix, leading to the exponential spatial decay estimate in Proposition~\ref{prop:st-spatial-locality}.

\begin{lemma}[Site comparison]
\label{lem:st-site-comparison}
For each block generator $\mathcal{L}_A$, let
\[
\gamma_A:=\frac12\|\cL_A\|_{\infty\to\infty},
\]
and define the site matrix
\[
\Gamma_{ij}:=\mathbf 1_{\{i\ne j\}} \sum_{A:i,j\in A}\gamma_A.
\]
Then, for every real-valued observable $h$ and $t\ge0$,
\begin{equation*}
\omega(\cP_t h)\le e^{t\Gamma}\omega(h) \quad\text{entrywise}.
\end{equation*}
Moreover, $\Gamma$ is symmetric and entrywise nonnegative, with
\[
\Gamma_{ij}=0\quad\text{if }\dist_G(i,j)>a, \qquad \|\Gamma\|_{1\to1} =\|\Gamma\|_{\infty\to\infty} \le (R_{\mathrm{loc}}-1)J_{\mathrm{loc}}.
\]
\end{lemma}

\begin{proof}
We first establish the comparison for a single block update $\cL_A$.
For each nonzero block $A$, define $K_A=I+\cL_A/\gamma_A$. 
By the definition of $\gamma_A$, the operator $K_A$ is stochastic and updates only the spins in $A$. 
For each configuration $x$, the block update acts as 
\[
(K_Ah)(x) =\sum_{z_A\in\{-1,1\}^A} k_A(x_A,z_A)h(z_A,x_{A^c}),
\]
where $k_A$ is a stochastic matrix. If $x$ and $y$ differ only at $i\notin A$, then $x_A=y_A$, and the configurations $(z_A,x_{A^c})$ and $(z_A,y_{A^c})$ also differ only at $i$. We hence have 
\[
\begin{aligned}
|(K_Ah)(x)-(K_Ah)(y)| \le \sum_{z_A}k_A(x_A,z_A) |h(z_A,x_{A^c})-h(z_A,y_{A^c})|\le \omega_i(h),
\end{aligned}
\]
where the last inequality follows from $\sum_{z_A}k_A(x_A,z_A)=1$.
Taking the supremum over such $x,y$ gives $\omega_i(K_Ah)\le\omega_i(h)$.
If $i\in A$ and $x,y$ differ only at $i$, then $x_{A^c}=y_{A^c}=:w$.
Since $k_A$ is stochastic, both $(K_Ah)(x)$ and  $(K_Ah)(y)$ are convex combinations of $\{h(z_A,w):z_A\in\{-1,1\}^A\}$. This gives 
\[
\begin{aligned}
|(K_Ah)(x)-(K_Ah)(y)| \le \max_{z_A}h(z_A,w)-\min_{z_A}h(z_A,w) \le \sum_{j\in A}\omega_j(h),
\end{aligned}
\]
where the last inequality follows by changing the spins in $A$ one at a time.
Taking the supremum gives $\omega_i(K_Ah)\le\sum_{j\in A}\omega_j(h)$.
Define the site matrix $(B_A)_{ij}=\mathbf 1_{\{i\ne j,\ i,j\in A\}}$.
Combining the two cases yields 
\begin{equation} \label{auxeq1}
    \omega(K_Ah)\le(I+B_A)\omega(h), \qquad \Gamma=\sum_A\gamma_AB_A.
\end{equation}
Since $I+B_A$ is entrywise nonnegative, this comparison extends by induction to arbitrary ordered products of block updates.

We consider general $\cL= \sum_A \cL_A = \sum_A\gamma_AK_A-\gamma I$ with $\gamma=\sum_A\gamma_A$. We expand the semigroup as
\begin{equation*}
\cP_t=e^{-\gamma t} \sum_{n=0}^\infty\frac{t^n}{n!} \sum_{A_1,\ldots,A_n} \left(\prod_{j=1}^n\gamma_{A_j}\right) K_{A_n}\cdots K_{A_1},
\end{equation*}
where the empty product is the identity. By subadditivity of oscillations, nonnegativity of the coefficients, and the single-block comparison \eqref{auxeq1}, we obtain
\[
\begin{aligned}
\omega(\cP_t h) & \le e^{-\gamma t} \exp\big(t\sum_A\gamma_A(I+B_A)\big)\omega(h)\\
& = e^{-\gamma t}e^{t(\gamma I+\Gamma)}\omega(h) = e^{t\Gamma}\omega(h).
\end{aligned}
\]

Finally, $\Gamma_{ij}=0$ whenever $\dist_G(i,j)>a$, since no block contains both $i,j$.
For each column $j$,
\[
\sum_i\Gamma_{ij} =\sum_{A:j\in A}(|A|-1)\gamma_A \le (R_{\mathrm{loc}}-1) \sum_{A:j\in A}\|\cL_A\|_{\infty\to\infty}.
\]
Taking the maximum over $j$ gives
\[
\|\Gamma\|_{1\to1} \le (R_{\mathrm{loc}}-1)J_{\mathrm{loc}}.
\]
Since $\Gamma$ is symmetric and entrywise nonnegative, $\|\Gamma\|_{\infty\to\infty}=\|\Gamma\|_{1\to1}$, which completes the proof.
\end{proof}

Let $h$ depend only on the spins in a nonempty set $S\subseteq V$, and let $q\ge0$ be an integer.
Lemma~\ref{lem:st-site-comparison} gives
\begin{equation}
\left\|\bigl(\omega_i(\cP_t h)\bigr)_{\dist_G(i,S)>(q-1)a}\right\|_{\ell^p}\le\|\omega(h)\|_{\ell^p}e^{vt-q},\qquad p\in\{1,\infty\},\quad t\ge0.
\label{eq:st-oscillation-tail}
\end{equation}
Indeed, $\Gamma^n\omega(h)$ is supported in $\cN_{na}(S)$, so the terms of order $n<q$ vanish at the sites on the left-hand side.
The operator norm bound on $\Gamma$ controls the remaining terms by $\|\omega(h)\|_{\ell^p}\sum_{n\ge q}x^n/n!$, where $x=(R_{\mathrm{loc}}-1)J_{\mathrm{loc}}t$.
Since $e^{n-q}\ge1$ for $n\ge q$, this series is at most $e^{ex-q}=e^{vt-q}$.

We next show that freezing the spins outside $C_\ell=\cN_{\ell a}(C)$ yields a localized evolution depending only on the spins in $B=\cN_{(\ell+1)a}(C)$ that approximates $\cP_tg_C$ with exponentially small error in the neighborhood radius. This allows us to construct the Krylov approximation within the connected-cluster space $W_k(B)$.

\begin{lemma}[Freezing the exterior]
\label{lem:st-frozen-generator}
Let $R_{\mathrm{loc}}\ge2$ and $\ell\ge1$ be an integer.
For each integer $j\ge0$, set $C_j=\cN_{ja}(C)$ and let $B=C_{\ell+1}$.
Define the frozen generator $\cL^{(\ell)}$ by suppressing all transitions that change spins outside $C_\ell$:
\begin{equation}
\begin{aligned}
\cL^{(\ell)}(s,y) &=\cL(s,y)\mathbf 1_{\{s_{C_\ell^c} =y_{C_\ell^c}\}}, \qquad s\ne y,\\
\cL^{(\ell)}(s,s) &=-\sum_{y\ne s}\cL^{(\ell)}(s,y).
\end{aligned}
\label{eq:st-frozen-generator}
\end{equation}
Then $\cL^{(\ell)}$ and $\mathcal H_\ell:=\cL-\cL^{(\ell)}$ are Markov generators reversible with respect to $\pi$.
Moreover, $\cL^{(\ell)}$ admits a local block decomposition with block norms no larger than those of $\cL$, and every nonzero block is contained in $B$.
Thus, $\cP_t^{(\ell)}g_C$ and $(\cL^{(\ell)})^j g_C$, where $\cP_t^{(\ell)}=e^{t\cL^{(\ell)}}$, depend only on the spins in $B$.

For every $t\ge0$, the localization errors satisfy
\begin{equation}
\begin{aligned}
\int_0^t \|\mathcal H_\ell\cP_s^{(\ell)}g_C\|_\infty\,\dd s &\le A_g e^{vt-\ell},\\
\|\cP_tg_C-\cP_t^{(\ell)}g_C\|_\infty &\le A_g e^{vt-\ell},
\end{aligned}
\label{eq:st-freezing-error}
\end{equation}
where $A_g$ is the total oscillation of $g_C$ in \eqref{eq:st-local-propagation-constants}. 
Both estimates also hold in $L^2(\pi)$.
Furthermore, for every observable $h$ depending only on the spins in $B$,
\begin{equation}
\|\mathcal H_\ell h\|_\pi \le J_{\mathrm{loc}}|B|\|h\|_\pi.
\label{eq:st-localized-boundary-norm}
\end{equation}
\end{lemma}

\begin{proof}
We first show that the frozen generator is reversible and preserves observables supported in $B$.
The restriction in \eqref{eq:st-frozen-generator} is symmetric in $s$ and $y$, so detailed balance for $\cL$ implies detailed balance for both $\cL^{(\ell)}$ and $\mathcal H_\ell$.
Applying the restriction to each block gives $\cL^{(\ell)}=\sum_A\cL_A^{(\ell)}$, where  $\cL_A^{(\ell)}$ is still a Markov generator on $A$ and satisfies
\[
\|\cL_A^{(\ell)}\|_{\infty\to\infty} \le \|\cL_A\|_{\infty\to\infty}.
\]
Every nonzero restricted block $A$ contains a site in $C_\ell$ and is therefore contained in $B=C_{\ell+1}$ by the diameter bound.
Thus $\cL^{(\ell)}$ preserves functions of $s_B$, as claimed.

The frozen generator satisfies the bounds on block size, block diameter, and local strength in \eqref{eq:st-generator}.
Applying \eqref{eq:st-oscillation-tail} with $S=C$, $q=\ell$, and $p=1$ gives
\begin{equation}
\sum_{i\notin C_{\ell-1}}\omega_i(\cP_s^{(\ell)}g_C)\le A_g e^{vs-\ell}.
\label{eq:st-frozen-oscillation-tail}
\end{equation}

Now, note that every transition removed by the freezing procedure changes at least one spin outside $C_\ell$, so its associated block $A$ intersects $C_\ell^c$. Since $\operatorname{diam}_G(A)\le a$, such a block is disjoint from $C_{\ell-1}$.
Since a jump on $A$ changes an observable $h$ by at most $\sum_{i\in A}\omega_i(h)$, we have
\begin{equation*}
\begin{aligned}
\|\mathcal H_\ell h\|_\infty &\le \sum_{A\not\subseteq C_\ell} \gamma_A\sum_{i\in A}\omega_i(h) \le J_{\mathrm{loc}} \sum_{i\notin C_{\ell-1}}\omega_i(h).
\end{aligned}
\end{equation*}
Combining this with \eqref{eq:st-frozen-oscillation-tail} yields
\[
\begin{aligned}
\int_0^t \|\mathcal H_\ell\cP_s^{(\ell)}g_C\|_\infty\,\dd s &\le J_{\mathrm{loc}}A_ge^{-\ell} \int_0^t e^{vs}\,\dd s \le A_ge^{vt-\ell},
\end{aligned}
\]
where we used $J_{\mathrm{loc}}/v = 1/[e(R_{\mathrm{loc}}-1)]\le1$.
Moreover, Duhamel's formula gives
\[
\cP_tg_C-\cP_t^{(\ell)}g_C =\int_0^t \cP_{t-s}\mathcal H_\ell \cP_s^{(\ell)}g_C\,\dd s.
\]
The sup-norm contraction of $\cP_{t-s}$ proves the second inequality in \eqref{eq:st-freezing-error}.
Both estimates also hold in $L^2(\pi)$ since $\pi$ is a probability measure.

Finally, we prove \eqref{eq:st-localized-boundary-norm}.
Let $\mathcal H_{\ell,B}$ be the generator obtained from $\mathcal H_\ell$ by keeping only jumps that change at least one spin in $B$ and adjusting the diagonal to preserve zero row sums.
This restriction is symmetric in the initial and final configurations, so $\mathcal H_{\ell,B}$ is reversible with respect to $\pi$.
Moreover, for every function $h$ of $s_B$, we have
\[
\mathcal H_\ell h=\mathcal H_{\ell,B}h.
\]
Note that the maximal exit rate of $\mathcal H_{\ell,B}$ satisfies
\[
\begin{aligned}
q_{\max} &:=\max_s[-\mathcal H_{\ell,B}(s,s)] \le\sum_{A:A\cap B\ne\varnothing}\gamma_A \le \frac12 J_{\mathrm{loc}}|B|.
\end{aligned}
\]
For any jump generator $\mathcal G$ reversible with respect to $\pi$, detailed balance gives
\begin{equation}
-\langle h,\mathcal Gq\rangle_\pi = \frac12\sum_{s,y}\pi(s)\mathcal G(s,y)(h(y)-h(s))(q(y)-q(s)).
\label{eq:st-dirichlet-jump}
\end{equation}
Applying \eqref{eq:st-dirichlet-jump} with $\mathcal G=\mathcal H_{\ell,B}$ and $h=q=z$, and using $(z(y)-z(s))^2\le 2|z(y)|^2+2|z(s)|^2$, gives
\[
0\le \langle z,-\mathcal H_{\ell,B}z\rangle_\pi \le2q_{\max}\|z\|_\pi^2.
\]
Since $\mathcal H_{\ell,B}$ is self-adjoint and nonpositive on $L^2(\pi)$,
\[
\|\mathcal H_\ell h\|_\pi =\|\mathcal H_{\ell,B}h\|_\pi \le2q_{\max}\|h\|_\pi \le J_{\mathrm{loc}}|B|\|h\|_\pi,
\]
which completes the proof.
\end{proof}

\subsubsection{Polynomial approximation}
\label{app:st-polynomial}

Krylov subspace approximations of the matrix exponential are classical~\cite{saad1992analysis,hochbruck1997krylov}. Botchev et al.~\cite[Section~4]{botchev2013residual} use variation of constants to bound the approximation error in terms of the residual.
In the lemma below, we construct a Krylov approximation to $e^{t\mathcal G}g_C$ and give explicit bounds on both the approximation error and the time-integrated residual. We will use this residual bound to control the error of repeated one-step propagation.
Here $\mathcal G$ can be either the full generator $\cL$ or its frozen version $\cL^{(\ell)}$.
For simplicity, we write
\begin{equation*}
 \sigma=\bigl(J_{\mathrm{loc}} \max\{|C|,R_{\mathrm{loc}}-1\}\bigr)^{-1}, \qquad M_g=2\|g_C-\pi(g_C)\|_\infty, \qquad w(t)=\frac{t}{t+\sigma}.
\end{equation*}

\begin{lemma}[Local polynomial approximation]
\label{lem:st-polynomial-residual}
Let $\mathcal G=\sum_{A\in\mathcal I}\mathcal G_A$ be reversible with respect to $\pi$, where each $\mathcal G_A$ is a local Markov generator on a block $A$ satisfying \eqref{eq:st-generator}, with
\[
 \max_i\sum_{A:i\in A} \|\mathcal G_A\|_{\infty\to\infty} \le J_{\mathrm{loc}}.
\]
For an integer $N\ge1$, define the Krylov space
\[
 \mathcal K_N(\mathcal G) =\operatorname{span} \{\mathcal G^j(g_C-\pi(g_C)):0\le j<N\},
\]
and let $\Pi_N(\mathcal G)$ denote the $L^2(\pi)$-orthogonal projection onto this space. Then, every Walsh monomial appearing in $\mathcal K_N(\mathcal G)$ has degree at most $|C|+(N-1)(R_{\mathrm{loc}}-1)$ and is supported in $\cN_{(N-1)a}(C)$.
If $C$ and all generator blocks are connected, each such support is contained in a connected set of at most $|C|+(N-1)(R_{\mathrm{loc}}-1)$ sites.
In particular, if $C\subseteq B$ and every nonzero block is contained in $B$, then
\[
 \mathcal K_N(\mathcal G)\subseteq W_k(B), \qquad k\ge |C|+(N-1)(R_{\mathrm{loc}}-1).
\]

Moreover, the $L^2$-approximation error satisfies
\begin{equation}
 \|(I-\Pi_N(\mathcal G)) e^{t\mathcal G}(g_C-\pi(g_C))\|_\pi \le M_gw(t)^N, \qquad t\ge0,
 \label{eq:st-sharp-krylov-error}
\end{equation}
and there exists a continuously differentiable curve $q_N:[0,\infty)\to\mathcal K_N(\mathcal G)$ with
\begin{equation*}
 q_N(0)=g_C-\pi(g_C),
\end{equation*}
such that, for every $t\ge0$,
\begin{align}
 \|e^{t\mathcal G}(g_C-\pi(g_C))-q_N(t)\|_\pi &\le M_g\frac{w(t)^N}{1-w(t)},
 \label{eq:st-exact-curve-error}\\
 \int_0^t\|q_N'(s)-\mathcal Gq_N(s)\|_\pi\,\dd s &\le 2M_g\frac{w(t)^N}{(1-w(t))^2}.
 \label{eq:st-exact-integrated-residual}
\end{align}
\end{lemma}

\begin{proof}
We first establish the support and degree bounds for the Krylov space.
Write $g^\circ=g_C-\pi(g_C)$.
For each nonzero product $\mathcal G_{A_j}\cdots\mathcal G_{A_1}g^\circ$, define $U_0=C$ and $U_i=U_{i-1}\cup A_i$.
Since the partial product through step $i-1$ depends only on $s_{U_{i-1}}$, the next block $A_i$ must intersect $U_{i-1}$ for the product to be nonzero.
Thus each block adds at most $R_{\mathrm{loc}}-1$ sites, all within distance $a$ of $U_{i-1}$, giving
\[
|U_j|\le |C|+j(R_{\mathrm{loc}}-1), \qquad U_j\subseteq\cN_{ja}(C).
\]
If $C$ and the blocks are connected, every $U_j$ is connected.
Moreover, if $C$ and all nonzero blocks lie in $B$, then $U_j\subseteq B$.
Since the Walsh expansion of each block product involves only supports contained in $U_j$, these observations give the asserted degree, support, and connected-cluster bounds.

We next bound the growth of the Krylov vectors.
For any intermediate support $U_i$, the sum of the norms of blocks that can act nontrivially is bounded by
\[
\sum_{A:A\cap U_i\ne\varnothing} \|\mathcal G_A\|_{\infty\to\infty} \le \sum_{x\in U_i}\sum_{A:x\in A} \|\mathcal G_A\|_{\infty\to\infty} \le J_{\mathrm{loc}}|U_i|.
\]
Summing over successive block products therefore yields
\begin{equation*}
\begin{aligned}
\|\mathcal G^j g^\circ\|_\pi &\le \|\mathcal G^j g^\circ\|_\infty\\
&\le \|g^\circ\|_\infty J_{\mathrm{loc}}^j \prod_{i=0}^{j-1} \bigl(|C|+i(R_{\mathrm{loc}}-1)\bigr)\\
&\le \|g^\circ\|_\infty\sigma^{-j}j!,
\end{aligned}
\end{equation*}
where the last inequality follows from
\[
|C|+i(R_{\mathrm{loc}}-1) \le(i+1)\max\{|C|,R_{\mathrm{loc}}-1\}.
\]
Consequently, the exponential series gives
\[
\|e^{-\sigma\mathcal G/2}g^\circ\|_\pi \le \sum_{j=0}^\infty \frac{(\sigma/2)^j}{j!}\|\mathcal G^jg^\circ\|_\pi \le 2\|g^\circ\|_\infty=M_g.
\]

We now use analyticity to obtain the approximation bounds.
Extend $L^2(\pi)$ to complex-valued functions, equipped with the Hermitian inner product.
Since $\mathcal G$ is self-adjoint and nonpositive,
\[
\|e^{z\mathcal G}\|_{\pi\to\pi}\le1, \qquad \operatorname{Re}z\ge0.
\]
For $\operatorname{Re}z>-\sigma/2$, the factorization
\[
e^{z\mathcal G}g^\circ =e^{(z+\sigma/2)\mathcal G} e^{-\sigma\mathcal G/2}g^\circ
\]
therefore gives $\|e^{z\mathcal G}g^\circ\|_\pi\le M_g$.
The change of variables $z=\sigma w/(1-w)$ maps the unit disk onto the half-plane $\operatorname{Re}z>-\sigma/2$.
In particular, the nonnegative real time axis corresponds to $0\le w(t)<1$.
Hence, the analytic function 
\begin{equation*}
\begin{gathered}
F(w):=e^{\sigma w\mathcal G/(1-w)}g^\circ =\sum_{j=0}^\infty a_jw^j,
\end{gathered}
\end{equation*}
satisfies
\begin{equation*}
    \|F(w)\|_\pi\le M_g\quad(|w|<1), \qquad \|a_j\|_\pi\le M_g.
\end{equation*}
The coefficient bound follows from Cauchy's integral formula by letting the circle radius approach one.
Since $(\sigma w/(1-w))^i$ starts at order $w^i$, expanding the exponential gives $a_0=g^\circ$ and
\[
a_j\in \operatorname{span}\{\mathcal G^ig^\circ:0\le i\le j\}.
\]

To prove the $L^2$-approximation estimate, consider, for any $\|h\|_\pi=1$, the scalar analytic function
\[
\langle h,(I-\Pi_N(\mathcal G))F(w)\rangle_\pi.
\]
It is bounded by $M_g$ and vanishes to order at least $N$ at $w=0$, since $a_j\in\mathcal K_N(\mathcal G)$ for $j<N$.
The maximum modulus principle, applied after dividing by $w^N$, gives
\[
|\langle h,(I-\Pi_N(\mathcal G))F(w)\rangle_\pi| \le M_g|w|^N.
\]
Taking the supremum over $h$ and setting $w=w(t)$ proves \eqref{eq:st-sharp-krylov-error}.

Finally, we construct a Krylov curve with a controlled integrated residual.
Truncate the expansion of $F$ at order $N-1$:
\begin{equation*}
q_N(t)=\sum_{j=0}^{N-1}a_jw(t)^j.
\end{equation*}
The coefficients are real and belong to $\mathcal K_N(\mathcal G)$ for $j<N$.
Thus $q_N(t)\in\mathcal K_N(\mathcal G)$, $q_N(0)=g^\circ$, and summing the omitted terms gives
\[
\|e^{t\mathcal G}g^\circ-q_N(t)\|_\pi \le M_g\frac{w(t)^N}{1-w(t)},
\]
which proves \eqref{eq:st-exact-curve-error}.

To control the residual, differentiate $F$ to obtain
\[
\frac{(1-w)^2}{\sigma}F'(w)=\mathcal GF(w).
\]
Comparing coefficients and using $w'(t)=(1-w(t))^2/\sigma$ gives the exact identity
\begin{equation*}
q_N'(t)-\mathcal Gq_N(t) =-\frac{N}{\sigma}a_Nw(t)^{N-1} +\frac{N-1}{\sigma}a_{N-1}w(t)^N.
\end{equation*}
Since $\|a_j\|_\pi\le M_g$ and $\dd t=\sigma\,\dd w/(1-w)^2$, we conclude that
\[
\begin{aligned}
\int_0^t \|q_N'(s)-\mathcal Gq_N(s)\|_\pi\,\dd s &\le 2NM_g \int_0^{w(t)} \frac{w^{N-1}}{(1-w)^2}\,\dd w\\
&\le 2M_g\frac{w(t)^N}{(1-w(t))^2}.
\end{aligned}
\]
This proves \eqref{eq:st-exact-integrated-residual} and completes the proof.
\end{proof}

\begin{remark}
The polynomial approximation argument in \Cref{lem:st-polynomial-residual} extends to more general local generator decompositions, yielding uniform-in-time $L^2(\pi)$ approximation with both neighborhood radius and Walsh degree of order $\mathcal O(\log^2(1/\varepsilon))$.
Indeed, the argument only requires the stated block size, range, and norm bounds, together with $\mathcal G_A\chi_S=0$ whenever $A\cap S=\varnothing$, and $\mathcal G_A\chi_S$ depending only on $s_{S\cup A}$ otherwise.
The individual summands need not be Markov generators, provided their sum $\mathcal G$ is a reversible Markov generator with a spectral gap.
Applying the polynomial argument to $\mathcal G=\cL$ with $T$ and $N$ from \eqref{eq:st-local-best-cutoffs} gives error at most $\varepsilon$ for all $t\ge0$, using Walsh monomials of degree at most $|C|+(N-1)(R_{\mathrm{loc}}-1)$ supported in $\cN_{(N-1)a}(C)$.
The sharper logarithmic radius $\mathcal O(\log(1/\varepsilon))$ in \Cref{prop:st-local-approximation} below and the subsequent compression estimates rely on the stronger local Markov block structure through Lemma~\ref{lem:st-site-comparison}.
\end{remark}

\subsubsection{Spatial locality and connected-cluster approximation}
\label{app:st-local-approximation}
\leavevmode\par

\smallskip
\noindent\textbf{Spatial truncation.}
For $B=\cN_\xi(C)$, define the conditional expectation with the spins on $B$ pinned:
\[
  \mathsf{Q}_\xi h = \E_\pi[h \mid s_B],
\]
which is the $L^2(\pi)$-orthogonal projection onto the subspace of functions depending only on $s_B$.

\begin{proposition}[Spatial locality]
\label{prop:st-spatial-locality}
For $\xi\ge0$, let $q=\lfloor\xi/a\rfloor+1$.
If $s$ and $y$ agree on $B=\cN_\xi(C)$, then
\begin{equation}
\begin{aligned}
|(\cP_tg_C)(s)-(\cP_tg_C)(y)|
&\le A_g e^{vt-q}, \qquad t\ge0.
\end{aligned}
\label{eq:st-spatial-pairwise}
\end{equation}
It follows that 
\begin{align}
\|(I-\mathsf Q_\xi)\cP_tg_C\|_\pi &\le \min\left\{ A_g e^{vt-q}, \|g_C-\pi(g_C)\|_\pi e^{-\lambda t} \right\},
\label{eq:st-spatial-projection}\\
\sup_{t\ge0}\|(I-\mathsf Q_\xi)\cP_tg_C\|_\pi &\le \max\left\{A_g,\|g_C-\pi(g_C)\|_\pi\right\} \exp\left(-\frac{\lambda q}{\lambda+v}\right).
\label{eq:st-spatial-uniform}
\end{align}
\end{proposition}

\begin{proof}
Every site outside $B$ is at distance greater than $(q-1)a$ from $C$, so \eqref{eq:st-oscillation-tail} with $S=C$ and $p=1$ gives $\sum_{i\notin B}\omega_i(\cP_tg_C)\le A_g e^{vt-q}$.
If $s_B=y_B$, changing the spins in $B^c$ one at a time gives
\[
|(\cP_tg_C)(s)-(\cP_tg_C)(y)| \le \sum_{i\notin B}\omega_i(\cP_tg_C),
\]
which proves \eqref{eq:st-spatial-pairwise}.

For $h=\cP_tg_C$, averaging \eqref{eq:st-spatial-pairwise} over $y$ with law $\pi(\cdot\mid y_B=s_B)$ gives $$|h(s)-\mathsf Q_\xi h(s)|\le A_g e^{vt-q}$$ for every $s$.
Hence $\|(I-\mathsf Q_\xi)\cP_tg_C\|_\pi\le A_g e^{vt-q}$.
On the other hand, the constant $\pi(g_C)$ belongs to the range of $\mathsf Q_\xi$.
The spectral gap therefore gives
\[
\begin{aligned}
\|(I-\mathsf Q_\xi)\cP_tg_C\|_\pi &\le \|\cP_tg_C-\pi(g_C)\|_\pi \le e^{-\lambda t} \|g_C-\pi(g_C)\|_\pi.
\end{aligned}
\]
Combining these estimates proves \eqref{eq:st-spatial-projection}.

Finally, we obtain the uniform-in-time bound.
For any $T\ge0$, use the spatial estimate on $[0,T]$ and the $L^2(\pi)$ relaxation of $\mathcal{P}_t$ on $[T,\infty)$ to obtain
\[
\begin{aligned}
\sup_{t\ge0}\|(I-\mathsf Q_\xi)\cP_tg_C\|_\pi \le \max\bigl\{ A_g e^{vT-q}, \|g_C-\pi(g_C)\|_\pi e^{-\lambda T} \bigr\}.
\end{aligned}
\]
Choosing $T=q/(v+\lambda)$ makes both exponents equal to $-\lambda q/(v+\lambda)$, which proves \eqref{eq:st-spatial-uniform}.
\end{proof}

\noindent\textbf{Spatial and cluster cutoffs.}
We now show that $\cP_tg_C$ admits a uniform-in-time approximation in a connected-cluster space with logarithmic neighborhood radius and quadratic-logarithmic cluster order.
To control both cutoffs, we combine exterior freezing with Krylov approximation.
Indeed, conditioning a low-degree polynomial on the spins in a neighborhood may increase its Walsh degree.
For example, under $\pi(s)=2^{-3}(1+\rho s_1s_2s_3)$ on three spins, where $0<|\rho|<1$, we have $\E_\pi[s_3\mid s_1,s_2]=\rho s_1s_2$.

\begin{proposition}[Spatially local connected-cluster approximation]
\label{prop:st-local-approximation}
Assume that $C$ and the generator blocks are connected, and let $R_{\mathrm{loc}}\ge2$.
For integers $\ell,N\ge1$, set $B=\cN_{(\ell+1)a}(C)$ and choose $k\ge |C|+(N-1)(R_{\mathrm{loc}}-1)$.
Let $\Pi$ denote the $L^2(\pi)$-orthogonal projection onto $W_k(B)$. 
Then, for every $t\ge0$, there holds 
\begin{equation}
\begin{aligned}
\|(I-\Pi)\cP_tg_C\|_\pi \le \min\bigl\{ &\|g_C-\pi(g_C)\|_\pi e^{-\lambda t}, A_g e^{vt-\ell}+M_g w(t)^N \bigr\}.
\end{aligned}
\label{eq:st-local-best-error}
\end{equation}
In particular, let $H_g=\max\{A_g,M_g\}$ and $0<\varepsilon<H_g$.
Choose
\begin{equation}
T=\lambda^{-1}\log\frac{2H_g}{\varepsilon},\qquad \ell=\left\lceil vT+\log\frac{2H_g}{\varepsilon}\right\rceil,\qquad N=\left\lceil\frac{\log(2H_g/\varepsilon)}{\log(1+\sigma/T)}\right\rceil.
\label{eq:st-local-best-cutoffs}
\end{equation}
Here $T$ is the time up to which we use the Krylov approximation, $\ell a$ is the radius beyond which the spins are frozen, and $N$ is the number of Krylov terms involved. 
Take the neighborhood radius $\xi=(\ell+1)a$ and the cluster order $k=|C|+(N-1)(R_{\mathrm{loc}}-1)$. Then
\[
\sup_{t\ge0}\|(I-\Pi)\cP_tg_C\|_\pi \le\varepsilon,
\]
with
\begin{equation}
\xi=\mathcal O(\log(1/\varepsilon)), \qquad k=|C|+\mathcal O(\log^2(1/\varepsilon)).
\label{eq:st-local-best-orders}
\end{equation}
\end{proposition}

\begin{remark}
We emphasize that neither this proposition nor Proposition~\ref{prop:st-spatial-locality} requires a uniform bound on the maximum degree of $G$ assumed in Section~\ref{sec:glauber_guarantees}. 
\end{remark}

\begin{proof}
We first establish \eqref{eq:st-local-best-error} by combining exterior freezing with Krylov approximation.
Let $g=g_C-\pi(g_C)$ and let $\cL^{(\ell)}$ be the frozen generator from Lemma~\ref{lem:st-frozen-generator}, with $\cP_t^{(\ell)}=e^{t\cL^{(\ell)}}$.
Since $\cL^{(\ell)}$ is reversible and all its nonzero blocks are contained in $B$, Lemma~\ref{lem:st-polynomial-residual} gives
\[
\mathcal K_N(\cL^{(\ell)})\subseteq W_k(B).
\]
Let $\Pi_N^{(\ell)}$ be the orthogonal projection onto this Krylov space.
The same lemma yields
\[
\|(I-\Pi_N^{(\ell)}) \cP_t^{(\ell)}g\|_\pi \le M_gw(t)^N.
\]
Combining this estimate with the freezing error \eqref{eq:st-freezing-error}, we obtain
\[
\begin{aligned}
\|(I-\Pi)\cP_tg_C\|_\pi &\le \|\cP_tg_C-\cP_t^{(\ell)}g_C\|_\pi +\|(I-\Pi_N^{(\ell)}) \cP_t^{(\ell)}g\|_\pi\\
&\le A_g e^{vt-\ell}+M_gw(t)^N.
\end{aligned}
\]
Here we used the fact that $\pi(g_C)+\Pi_N^{(\ell)}\cP_t^{(\ell)}g$ belongs to $W_k(B)$.
Alternatively, since the constant $\pi(g_C)$ also belongs to $W_k(B)$, the spectral gap gives
\[
\|(I-\Pi)\cP_tg_C\|_\pi \le \|\cP_tg_C-\pi(g_C)\|_\pi \le e^{-\lambda t}\|g\|_\pi.
\]
Taking the smaller of these bounds proves \eqref{eq:st-local-best-error}.

We next give the uniform-in-time estimate.
For $0\le t\le T$, both terms in the second bound of \eqref{eq:st-local-best-error} are increasing in $t$.
The choices of $\ell$ and $N$ in \eqref{eq:st-local-best-cutoffs} ensure that
\[
A_g e^{vT-\ell}\le\frac{\varepsilon}{2}, \qquad M_g w(T)^N\le\frac{\varepsilon}{2}.
\]
For $t\ge T$, the $L^2(\pi)$ relaxation bound \eqref{eq:st-gap} gives
\[
\|(I-\Pi)\cP_tg_C\|_\pi \le \|g\|_\pi e^{-\lambda T} \le\frac{\varepsilon}{2}.
\]
Thus the approximation error is at most $\varepsilon$ for all $t\ge0$. 
Finally, we characterize the cutoff scalings.
Writing $L=\log(2H_g/\varepsilon)$ and using $\log(1+x)\ge x/(1+x)$ for $x>0$, we obtain
\[
\begin{aligned}
\xi \le a\left[2+\left(1+\frac v\lambda\right)L\right], \qquad N \le 1+L+\frac{L^2}{\lambda\sigma}.
\end{aligned}
\]
These estimates imply \eqref{eq:st-local-best-orders}, completing the proof. 
\end{proof}

\subsubsection{Propagation with exact moment matrices}
\label{app:st-exact-propagation}

\begin{proof}[Proof of Theorem~\ref{thm:st-log-radius}]
We prove the uniform propagation bound by combining a finite-time Krylov approximation with long-time $L^2(\pi)$ relaxation. If $g_C$ is constant or $R_{\mathrm{loc}}=1$, take $B=C$ and $k=|C|$. In the first case, constants are preserved by $\cP_t$; in the second, the space of functions of $s_C$ is invariant under $\cP_t$.
Thus the projected iteration is exact in both cases.
For $\varepsilon\ge M_g$, the same choice of $B$ and $k$ suffices, since both the exact and projected evolutions preserve $\pi(g_C)$ and are contractions in $L^2(\pi)$. Consequently, $
\|\cP_{m\tau}g_C-u_m\|_\pi
\le 2\|g_C-\pi(g_C)\|_\pi
\le M_g\le\varepsilon$.

It remains to consider $R_{\mathrm{loc}}\ge2$ and $0<\varepsilon<M_g$.
Fix integers $\ell,N\ge1$ and choose the window $B$ and cluster order $k$ so that
\begin{equation*}
B=\cN_{(\ell+1)a}(C), \qquad k\ge |C|+(N-1)(R_{\mathrm{loc}}-1).
\end{equation*}
For any time horizon $T>0$, set $X_T=1+T/\sigma$ and $w_T=w(T)$.

We first control the propagation error for $m\tau\le T$.
Let $\cL^{(\ell)}$ and $\mathcal H_\ell=\cL-\cL^{(\ell)}$ be the frozen and discarded generators from Lemma~\ref{lem:st-frozen-generator}.
The frozen generator is reversible, and its nonzero blocks lie in $B$ with the same local strength bound.
By Lemma~\ref{lem:st-polynomial-residual}, there exists a curve $q_N(t)\in W_k(B)$ satisfying
\begin{align}
\|\cP_t^{(\ell)}g_C-\pi(g_C)-q_N(t)\|_\pi &\le M_gX_Tw_T^N,
\label{eq:st-frozen-polynomial-error}\\
\int_0^t \|q_N'(s)-\cL^{(\ell)}q_N(s)\|_\pi\,\dd s &\le 2M_gX_T^2w_T^N,
\label{eq:st-frozen-polynomial-residual}
\end{align}
for every $0\le t\le T$. 
To control the residual with respect to the full generator, write
\[
\begin{aligned}
q_N'(s)-\cL q_N(s) ={}&q_N'(s)-\cL^{(\ell)}q_N(s) -\mathcal H_\ell\cP_s^{(\ell)}g_C -\mathcal H_\ell \bigl(q_N(s)-[\cP_s^{(\ell)}g_C-\pi(g_C)]\bigr),
\end{aligned}
\]
where we used $\mathcal H_\ell 1=0$.
The first term is controlled by \eqref{eq:st-frozen-polynomial-residual}, and the second by the integrated freezing estimate \eqref{eq:st-freezing-error}.
The difference in the last term depends only on $s_B$, so \eqref{eq:st-localized-boundary-norm} and \eqref{eq:st-frozen-polynomial-error} apply.
Combining these estimates gives, for $0\le t\le T$,
\begin{equation}
\begin{aligned}
\int_0^t \|q_N'(s)-\cL q_N(s)\|_\pi\,\dd s \le {} A_g e^{vT-\ell} +M_g\bigl(2X_T^2 +J_{\mathrm{loc}}T|B|X_T\bigr)w_T^N,
\end{aligned}
\label{eq:st-frozen-full-residual}
\end{equation}

We now use this residual bound to control repeated one-step propagation.
Let
\[
u_m^\circ=(\Pi\cP_\tau)^m (g_C-\pi(g_C))=u_m-\pi(g_C).
\]
Since $q_N(t)\in W_k(B)$, applying $\Pi$ to Duhamel's formula yields
\begin{equation*}
\begin{aligned}
&q_N((m+1)\tau) -\Pi\cP_\tau q_N(m\tau) = \Pi\int_{m\tau}^{(m+1)\tau} \cP_{(m+1)\tau-s} \bigl(q_N'(s)-\cL q_N(s)\bigr)\,\dd s.
\end{aligned}
\end{equation*}
Starting from $q_N(0)=u_0^\circ=g_C-\pi(g_C)$, we iterate this identity.
The contractivity of $\Pi\cP_\tau$ and $\cP_t$ then gives
\begin{equation}
\|q_N(m\tau)-u_m^\circ\|_\pi \le \int_0^{m\tau} \|q_N'(s)-\cL q_N(s)\|_\pi\,\dd s.
\label{eq:st-projected-duhamel}
\end{equation}

For $m\tau\le T$, the triangle inequality bounds the error by the freezing error \eqref{eq:st-freezing-error}, the Krylov error \eqref{eq:st-frozen-polynomial-error}, and the integral in \eqref{eq:st-projected-duhamel}.
We bound the integral by \eqref{eq:st-frozen-full-residual} and use $X_T+2X_T^2\le3X_T^2$.
For $m\tau\ge T$, the spectral gap controls both the exact and projected evolutions.
Since $1\in W_k(B)$, the projection $\Pi$ preserves the stationary mean.
For every centered $h\in W_k(B)$, we have 
\[
\|\Pi\cP_\tau h\|_\pi \le e^{-\lambda\tau}\|h\|_\pi.
\]
Thus
\begin{equation*}
\begin{aligned}
\|\cP_{m\tau}g_C-\pi(g_C)-u_m^\circ\|_\pi &\le \|\cP_{m\tau}g_C-\pi(g_C)\|_\pi +\|u_m^\circ\|_\pi\\
&\le 2e^{-\lambda m\tau} \|g_C-\pi(g_C)\|_\pi\\
&\le 2e^{-\lambda T} \|g_C-\pi(g_C)\|_\pi.
\end{aligned}
\end{equation*}
Combining the short- and long-time estimates gives
\begin{equation}
\begin{aligned}
\sup_{\tau>0,\ m\in\mathbb N_0}\|\cP_{m\tau}g_C-u_m\|_\pi \le \max\Bigl\{&2e^{-\lambda T}\|g_C-\pi(g_C)\|_\pi,\\
&2A_g e^{vT-\ell}+M_g\bigl(3X_T^2+J_{\mathrm{loc}}T|B|X_T\bigr)w_T^N\Bigr\}.
\end{aligned}
\label{eq:st-log-radius-bound}
\end{equation}

Finally, we choose the cutoffs to achieve accuracy $\varepsilon$.
Since every vertex has degree at most $\Delta$,
\[
|B|\le\mathcal V_\ell :=|C|\sum_{j=0}^{\lceil(\ell+1)a\rceil}\Delta^j.
\]
Set $T=\lambda^{-1}\log(2M_g/\varepsilon)$ and $\ell=\max\{1,\lceil vT+\log(4A_g/\varepsilon)\rceil\}$.
Then take $$N=\bigl\lceil \log\!\bigl(2M_g[3X_T^2+J_{\mathrm{loc}}T\mathcal V_\ell X_T]/\varepsilon\bigr)/\log(1+\sigma/T)\bigr\rceil.$$
The choice of $T$ bounds the long-time term in \eqref{eq:st-log-radius-bound} by $\varepsilon/2$, since $2\|g_C-\pi(g_C)\|_\pi\le M_g$.
The choices of $\ell$ and $N$ bound each of the two short-time terms by $\varepsilon/2$.
These parameters are well defined: $g_C-\pi(g_C)\ne0$ implies $A_g>0$, while $0<\varepsilon<M_g$ ensures $T>0$ and $N\ge1$.
Hence \eqref{eq:st-log-radius-bound} gives \eqref{eq:st-propagation-accuracy}.

Finally, note that for fixed local parameters,
\[
T,\ell=\mathcal O(\log(1/\varepsilon)), \qquad \log\mathcal V_\ell=\mathcal O(\ell+1).
\]
Using $\log(1+\sigma/T)\ge\sigma/(T+\sigma)$, we further obtain $N=\mathcal O(\log^2(1/\varepsilon))$.
Taking $\xi=(\ell+1)a$ and $k=|C|+(N-1)(R_{\mathrm{loc}}-1)$ proves \eqref{eq:st-log-radius-orders}. 
\end{proof}

\subsection{Compression of moment matrices}

We prove Theorem~\ref{thm:st-matrix-compression} and give sufficient conditions for exponential decay of the spatial tail $E_\delta$.

\subsubsection{Dirichlet kernel and matrix compression}
\label{app:st-matrix-compression}

For $S\in\cR_+$, let $\phi_S=\chi_S-\pi(\chi_S)$ be the centered observable. The covariance block defined in Section~\ref{sec:gb-compression} has entries $C^t(S,T)=\langle\phi_S,\cP_t\phi_T\rangle_\pi$ for $S,T\in\cR_+$.
In the basis consisting of $1$ and the centered observables $\phi_S$, the Gram and transition moment matrices have the block forms $\operatorname{diag}(1,G_c)$ and $\operatorname{diag}(1,C^\tau)$, respectively.
Since compression preserves the constant row and column, the compressed matrices have the same block structure, with centered blocks $\widetilde G_c$ and $\widetilde C^\tau$ given by \eqref{eq:st-common-defect}.
In particular, both the exact and compressed coefficient maps preserve constants and stationary means whenever defined.

With $v$ as in \eqref{eq:st-local-propagation-constants}, set $\eta=\lambda/[2a(v+\lambda)]$ and $K=eR_{\mathrm{loc}}kJ_{\mathrm{loc}}$. Define the spatial tail by
\begin{equation*}
S_\delta(\eta)=\sup_{S\in\cR_+} \sum_{\substack{T\in\cR_+\\ \dist_G(S,T)\ge\delta}} e^{-\eta\dist_G(S,T)}.
\end{equation*}
The constant $E_\delta$ in Theorem~\ref{thm:st-matrix-compression} is
\begin{equation}
E_\delta = \frac{2K}{\lambda}S_\delta(\eta).
\label{eq:st-matrix-tail-bound}
\end{equation}
If $\cR_+=\varnothing$, we set $S_\delta(\eta)=E_\delta=0$.

\begin{lemma}[Dirichlet kernel]
\label{lem:st-kernel}
Under \eqref{eq:st-generator}--\eqref{eq:st-gap}, the Dirichlet kernel satisfies
\begin{equation}
|\mathcal D(\phi_S,\cP_t\phi_T)| \le K e^{-\eta\dist_G(S,T)}e^{-\lambda t/2}, \qquad S,T\in\cR_+,\quad t\ge0,
\label{eq:st-kernel-bound}
\end{equation}
where $K$ and $\eta$ are defined above.
\end{lemma}

\begin{proof}
For $q=\lceil\dist_G(i,T)/a\rceil$, we have $\dist_G(i,T)>(q-1)a$ and $q\ge\dist_G(i,T)/a$.
Applying \eqref{eq:st-oscillation-tail} with $h=\chi_T$, $S=T$, and $p=\infty$, and using $\|\omega(\chi_T)\|_{\ell^\infty}=2$, gives
\begin{equation}
\omega_i(\cP_t\chi_T)\le 2e^{vt-\dist_G(i,T)/a}.
\label{eq:st-site-bound}
\end{equation}
By \eqref{eq:st-dirichlet-jump} with $\mathcal G=\cL$, only blocks intersecting $S$ contribute to $\mathcal D(\phi_S,\cP_t\phi_T)$.
If $A\cap S\ne\varnothing$, then every $i\in A$ satisfies
\[
\dist_G(i,T)\ge\dist_G(S,T)-a.
\]
Moreover, a transition on $A$ changes $\cP_t\chi_T$ by at most $\sum_{i\in A}\omega_i(\cP_t\chi_T)$.
Using \eqref{eq:st-site-bound}, $|\chi_S(y)-\chi_S(s)|\le2$, and
\[
\sum_{A:A\cap S\ne\varnothing}\gamma_A \le\frac{|S|J_{\mathrm{loc}}}{2},
\]
we obtain
\begin{equation}
|\mathcal D(\phi_S,\cP_t\phi_T)| \le eR_{\mathrm{loc}}kJ_{\mathrm{loc}}\, e^{vt-\dist_G(S,T)/a}.
\label{eq:st-kernel-local}
\end{equation}

We next establish temporal decay using the spectral gap.
Since
\[
\|\cL\chi_S\|_\pi \le\|\cL\chi_S\|_\infty \le kJ_{\mathrm{loc}}, \qquad \|\phi_T\|_\pi\le1,
\]
reversibility and $L^2(\pi)$ relaxation give
\begin{equation}
\begin{aligned}
|\mathcal D(\phi_S,\cP_t\phi_T)| &=|\langle-\cL\chi_S,\cP_t\phi_T\rangle_\pi| \le kJ_{\mathrm{loc}}e^{-\lambda t}.
\end{aligned}
\label{eq:st-kernel-gap}
\end{equation}

Finally, take the weighted geometric mean of \eqref{eq:st-kernel-local} and \eqref{eq:st-kernel-gap}, with weights $\lambda/[2(v+\lambda)]$ and $1-\lambda/[2(v+\lambda)]$, respectively.
The time exponent becomes $-\lambda t/2$, while the spatial exponent becomes $-\eta\dist_G(S,T)$.
Since $K=eR_{\mathrm{loc}}kJ_{\mathrm{loc}}
\ge kJ_{\mathrm{loc}}$, the resulting prefactor
is at most $K$.
This proves \eqref{eq:st-kernel-bound}.
\end{proof}

\begin{proof}[Proof of Theorem~\ref{thm:st-matrix-compression}]
We obtain the two compression bounds by integrating the Dirichlet kernel estimate.
Since the spectral gap implies $C^t\to0$ as $t\to\infty$, and
\[
\frac{\dd}{\dd t}C^t(S,T) =-\mathcal D(\phi_S,\cP_t\phi_T),
\]
we have
\begin{equation*}
\begin{aligned}
C^t(S,T) &=\int_t^\infty \mathcal D(\phi_S,\cP_s\phi_T)\,\dd s,\\
D_\tau(S,T) &=\int_0^\tau \mathcal D(\phi_S,\cP_s\phi_T)\,\dd s.
\end{aligned}
\end{equation*}
By Lemma~\ref{lem:st-kernel},
\[
\begin{aligned}
|C^t(S,T)| &\le \frac{2K}{\lambda} e^{-\eta\dist_G(S,T)}e^{-\lambda t/2},\\
|D_\tau(S,T)| &\le \frac{2K}{\lambda} e^{-\eta\dist_G(S,T)} (1-e^{-\lambda\tau/2}).
\end{aligned}
\]
The compression removes only entries with $\dist_G(S,T)\ge\delta$.
Both truncation errors are symmetric, and the constant row and column of $\widetilde M^t-M^t$ vanish.
Their operator norms are therefore bounded by their maximum absolute row sums.
Summing the preceding estimates over omitted supports and using \eqref{eq:st-matrix-tail-bound} gives \eqref{eq:st-matrix-error} and \eqref{eq:st-defect-error}.
\end{proof}

\subsubsection{Spatial-tail bounds}

\begin{proposition}[A sufficient growth condition]
\label{prop:st-growth}
Let $\cR$ consist of the supports in \eqref{eq:st-dictionary} or a subcollection.
Assume that each graph has maximum degree at most $\Delta$ and satisfies the polynomial volume-growth bound
\begin{equation}
|\cN_\ell(i)| \le V_0(1+\ell)^\nu, \qquad i\in V,\quad\ell\ge0,
\label{eq:st-polynomial-growth}
\end{equation}
with common constants $\Delta$, $V_0$, and $\nu$ across the family of graphs.
For fixed $k$, if the local generator bounds in \eqref{eq:st-generator} are uniform and the spectral gap has a uniform positive lower bound, then $$E_\delta\lesssim  e^{-b\delta}$$ for every integer $\delta\ge1$, with constants $E,b>0$ independent of graph volume and window size.
\end{proposition}

\begin{proof}
We first bound the number of retained supports containing a given vertex $i$.
Every connected set of $j+1$ vertices containing $i$ admits a spanning tree rooted at $i$.
A depth-first traversal of this tree gives a walk of length $2j$, with at most $\Delta^{2j}$ possibilities.
Each such set contains at most $2^j$ subsets containing $i$.
Consequently, the number of involved supports containing $i$ is at most
\[
N_{k,\Delta} :=\sum_{j=0}^{k-1}(2\Delta^2)^j.
\]

We next estimate the spatial tail.
If $\dist_G(S,T)=\ell$, then $T$ contains a vertex at distance $\ell$ from $S$.
Since $|S|\le k$, the polynomial growth bound gives at most $kV_0(1+\ell)^\nu$ possible vertices.
Each belongs to at most $N_{k,\Delta}$ supports.
Therefore,
\[
S_\delta(\eta) \le kN_{k,\Delta}V_0 \sum_{\ell\ge\delta} (1+\ell)^\nu e^{-\eta\ell}.
\]
Using $e^{-\eta\ell/2}\le e^{-\eta\delta/2}$ for $\ell\ge\delta$, we obtain
\begin{equation*}
\begin{aligned}
S_\delta(\eta) &\le S_0e^{-\eta\delta/2}, \qquad
S_0 :=kN_{k,\Delta}V_0 \sum_{\ell=0}^\infty (1+\ell)^\nu e^{-\eta\ell/2}.
\end{aligned}
\end{equation*}
Combining this with \eqref{eq:st-matrix-tail-bound} proves the exponential decay of $E_\delta$.
For fixed $k$, uniform local bounds and a uniform positive spectral gap keep $\eta$ bounded away from zero and $2K/\lambda$ uniformly bounded.
Thus the constants are independent of graph volume and window size.
\end{proof}

The connected-cluster structure also ensures sparsity of the compressed covariance matrix. 
For each nonempty support $S$, the number of retained supports $T$ with $\dist_G(S,T)<\delta$ is at most
\[
N_{k,\Delta}|\cN_{\delta-1}(S)|.
\]
Since $|S|\le k$, this number is uniformly bounded for fixed $k,\delta$, and $\Delta$. Consequently, the sparse covariance component has $\mathcal O(|\cR|)$ nonzero entries.

\subsection{Uniform compressed propagation}

We prove Theorem~\ref{thm:st-compression} by first deriving a matrix perturbation bound and then applying it to the exact and compressed one-step maps.

\subsubsection{Relative perturbations of the Gram and dissipation matrices}

\begin{lemma}[Uniform perturbation of powers]
\label{lem:st-joint-perturbation}
Let $G,M,\widehat G,\widehat M$ be real symmetric matrices, with $G$ positive definite and $0\preceq M\preceq G$.
Set $D=G-M$ and $\widehat D=\widehat G-\widehat M$.
Suppose that
\begin{equation*}
(1-\alpha)G\preceq\widehat G\preceq(1+\alpha)G,\qquad (1-\beta)D\preceq\widehat D\preceq(1+\beta)D,
\end{equation*}
for $\alpha,\beta\ge0$ with $2\alpha+\beta<1$. 
Define
\[
T=G^{-1}M,\qquad \widehat T=\widehat G^{-1}\widehat M, \qquad \|A\|_G=\|G^{1/2}AG^{-1/2}\|_2.
\]
Then
\begin{equation}
\sup_{m\ge0}\|\widehat T^m-T^m\|_G \le 2\alpha\left(\frac{1}{1-\alpha}+\frac{1}{1-2\alpha-\beta}\right)+\frac{\beta}{\sqrt{(1-\beta)(1-2\alpha-\beta)}}.
\label{eq:st-uniform-powers-bound}
\end{equation}
In particular, if $\alpha,\beta\le1/8$, the right-hand side is bounded by $6\alpha+2\beta$.
\end{lemma}

\begin{proof}
We first establish energy estimates for the exact and perturbed propagators.
A simultaneous congruence by $G^{-1/2}$ reduces the proof to $G=I$, with the $G$-operator norm becoming the Euclidean operator norm.
We retain the notation $T$ and $\widehat T$ for the propagators in these coordinates and set
\[
H=\widehat G,\qquad U=H-I,\qquad V=\widehat D-D.
\]
Then $T=I-D$ and $\widehat T=I-H^{-1}(D+V)$, with $0\preceq D\preceq I$ and
\[
\|U\|_2\le\alpha, \qquad -\beta D\preceq V\preceq\beta D.
\]
The latter inequalities imply
\begin{equation}
|x^\top Vy| \le\beta\|D^{1/2}x\|_2\|D^{1/2}y\|_2.
\label{eq:st-relative-defect-polarization}
\end{equation}
Indeed, the positive semidefinite matrices $\beta D\pm V$ vanish on $\ker D$.
On $(\ker D)^\perp$, conjugation by $D^{-1/2}$ gives the stated bound.

The matrix $\widehat T$ is self-adjoint in the $H$-inner product because $H^{1/2}\widehat T H^{-1/2}=I-H^{-1/2}(D+V)H^{-1/2}$ is symmetric.
The bounds $0\preceq D+V\preceq(1+\beta)I$ and $H\succeq(1-\alpha)I$ give
\[
(\kappa-1)I\preceq H^{1/2}\widehat T H^{-1/2}\preceq I,\qquad \kappa=\frac{1-2\alpha-\beta}{1-\alpha}\in(0,1].
\]
Thus the spectrum of $\widehat T$ lies in $[\kappa-1,1]$, and $\|\widehat T^j\|_H\le1$ for all $j\ge0$.

By the spectral decomposition, we claim
\begin{equation}
\sum_{j=0}^\infty \|D^{1/2}T^jv\|_2^2\le\|v\|_2^2,\qquad \sum_{j=0}^\infty \|D^{1/2}\widehat T^jw\|_2^2\le\frac{\|w\|_H^2}{(1-\beta)\kappa}.
\label{eq:st-power-energy-estimates}
\end{equation}
Indeed, for the first estimate, we use an orthonormal eigenbasis of $D$. Since $T=I-D$, a mode with eigenvalue $d\in(0,1]$ has weight
\[
d\sum_{j=0}^\infty(1-d)^{2j}=\frac{1}{2-d}\le1,
\]
and modes with $d=0$ contribute zero.
For the second estimate, $(1-\beta)D\preceq D+V=H(I-\widehat T)$ gives
\[
\|D^{1/2}\widehat T^jw\|_2^2\le\frac{1}{1-\beta}\langle \widehat T^jw,(I-\widehat T)\widehat T^jw\rangle_H.
\]
Expanding $w$ in an $H$-orthonormal eigenbasis of $\widehat T$ and summing over $j$ gives the weight
\[
\frac{1}{1-\beta}\sum_{j=0}^\infty(1-s)s^{2j}=\frac{1}{(1-\beta)(1+s)}\le\frac{1}{(1-\beta)\kappa}
\]
for each mode with eigenvalue $s\in[\kappa-1,1)$. 

Since $\widehat T-T=H^{-1}(UD-V)$, telescoping gives
\begin{equation} \label{auxeqqq}
\widehat T^m-T^m=\sum_{j=0}^{m-1} \widehat T^{m-1-j}H^{-1}UDT^j-\sum_{j=0}^{m-1} \widehat T^{m-1-j}H^{-1}VT^j.
\end{equation}
By self-adjointness of $\widehat T$ in the $H$-inner product,
\[
u^\top \widehat T^{m-1-j}H^{-1}VT^jv = (\widehat T^{m-1-j}H^{-1}u)^\top VT^jv.
\]
Applying \eqref{eq:st-relative-defect-polarization} to each summand and then using Cauchy--Schwarz in $j$ and \eqref{eq:st-power-energy-estimates} gives
\[
\left|u^\top\sum_{j=0}^{m-1} \widehat T^{m-1-j}H^{-1}VT^jv\right|\le\frac{\beta}{\sqrt{(1-\beta)\kappa}}\|H^{-1}u\|_H\|v\|_2.
\]
Since $\|H^{-1}u\|_H=\|H^{-1/2}u\|_2\le\|u\|_2/\sqrt{1-\alpha}$, taking the supremum over unit vectors $u,v$ yields
\begin{equation}
\left\|\sum_{j=0}^{m-1} \widehat T^{m-1-j}H^{-1}VT^j\right\|_2\le\frac{\beta}{\sqrt{(1-\alpha)(1-\beta)\kappa}}.
\label{eq:st-defect-power-sum}
\end{equation}

To bound the first sum in \eqref{auxeqqq}, set $K=H^{-1}U$. Since $\|Kx\|_H=\|H^{-1/2}Ux\|_2$, we have $\|K\|_{2\to H}\le\alpha/\sqrt{1-\alpha}$.
The discrete differences of the two propagators satisfy
\begin{equation}
\|DT^j\|_2\le\frac{1}{j+1},\qquad \|(I-\widehat T)\widehat T^j\|_H\le\frac{2-\kappa}{\kappa(j+1)}.
\label{eq:st-power-derivatives}
\end{equation}
These estimates follow by maximizing $(1-s)|s|^j$ over the spectrum of each propagator. For $s\in[0,1]$, the geometric sum gives $(j+1)(1-s)s^j\le1-s^{j+1}\le1$. For a negative eigenvalue $s=-x$ of $\widehat T$, we have $0<x\le1-\kappa$, so
\[
(1+x)x^j\le(2-\kappa)(1-\kappa)^j\le\frac{2-\kappa}{\kappa(j+1)},
\]
where the last inequality uses the same geometric-sum bound at $s=1-\kappa$. 
For $m\ge2$, split the sum at $q=\lfloor m/2\rfloor$. On the first part, summation by parts transfers the difference $DT^j=T^j-T^{j+1}$ from powers of $T$ to powers of $\widehat T$.
\begin{equation*}
\begin{aligned}
\sum_{j=0}^{q-1}\widehat T^{m-1-j}KDT^j ={}&\widehat T^{m-1}K-\widehat T^{m-q}KT^q +\sum_{j=1}^{q-1}(I-\widehat T)\widehat T^{m-1-j}KT^j.
\end{aligned}
\end{equation*}
Contractivity bounds the norm of each boundary term by $\|K\|_{2\to H}$. By \eqref{eq:st-power-derivatives}, the remaining sum above has norm at most $(2-\kappa)\|K\|_{2\to H}/\kappa$, and the part with $q\le j<m$ has norm at most $\|K\|_{2\to H}$, since $\sum_{j=1}^{q-1}(m-j)^{-1}\le1$ and $\sum_{j=q}^{m-1}(j+1)^{-1}\le1$.
Thus
\[
\left\|\sum_{j=0}^{m-1}\widehat T^{m-1-j}KDT^j\right\|_{2\to H}
\le\left(3+\frac{2-\kappa}{\kappa}\right)\|K\|_{2\to H}
=2(1+\kappa^{-1})\|K\|_{2\to H}.
\]
The bound also holds for $m=1$, when the sum is $KD$ and $\|D\|_2\le1$, and for $m=0$, when the sum is empty.
Using $\|z\|_2\le(1-\alpha)^{-1/2}\|z\|_H$ and the bound on $\|K\|_{2\to H}$ gives
\begin{equation}
\left\|\sum_{j=0}^{m-1}\widehat T^{m-1-j}H^{-1}UDT^j\right\|_2\le\frac{2\alpha}{1-\alpha}(1+\kappa^{-1}).
\label{eq:st-gram-power-sum}
\end{equation}

Combining \eqref{eq:st-defect-power-sum} and \eqref{eq:st-gram-power-sum}, and substituting $\kappa=(1-2\alpha-\beta)/(1-\alpha)$, proves \eqref{eq:st-uniform-powers-bound}.
For $\alpha,\beta\le1/8$, the coefficient of $\alpha$ is at most $2(8/7+8/5)<6$, while that of $\beta$ is at most $8/\sqrt{35}<2$.
\end{proof}

\subsubsection{Proof of the compressed propagation theorem}
\label{app:st-compressed-propagation}

\begin{proof}[Proof of Theorem~\ref{thm:st-compression}]
If the basis contains only the constant, then $\widetilde M^0=[1]$ and both iterations fix $f$.
Otherwise, set $\zeta_\delta=E_\delta/g_*$.
The cutoff condition \eqref{eq:st-compression-cutoff} gives $\zeta_\delta\le\varepsilon/8\le1/8$.

For any $\tau>0$, reversibility and the spectral gap give $0\preceq\cP_\tau\preceq e^{-\lambda\tau}I$ on centered functions.
In the centered basis this yields
\begin{equation*}
0\preceq C^\tau\preceq e^{-\lambda\tau}G_c,\qquad (1-e^{-\lambda\tau})G_c\preceq D_\tau\preceq G_c.
\end{equation*}
Using $G_c\succeq g_*I$ and $D_\tau\succeq(1-e^{-\lambda\tau})g_*I$, the compression bounds \eqref{eq:st-matrix-error} at $t=0$ and \eqref{eq:st-defect-error} give
\begin{equation*}
(1-\alpha)G_c\preceq\widetilde G_c\preceq(1+\alpha)G_c,\qquad (1-\beta)D_\tau\preceq\widetilde D_\tau\preceq(1+\beta)D_\tau,
\end{equation*}
where $\alpha=\zeta_\delta$ and $\beta=\zeta_\delta(1-e^{-\lambda\tau/2})/(1-e^{-\lambda\tau})=\zeta_\delta/(1+e^{-\lambda\tau/2})\le\zeta_\delta$.
Since $\alpha<1$, the first bound gives $\widetilde G_c\succ0$.
The centered block form in Appendix~\ref{app:st-matrix-compression} then gives $\widetilde M^0\succ0$.

Take $(G,M)=(G_c,C^\tau)$ and $(\widehat G,\widehat M)=(\widetilde G_c,\widetilde C^\tau)$.
Since $\alpha,\beta\le1/8$, Lemma~\ref{lem:st-joint-perturbation} yields
\begin{equation}
\sup_{m\ge0}\left\|(\widetilde G_c^{-1}\widetilde C^\tau)^m-(G_c^{-1}C^\tau)^m\right\|_{G_c}\le6\alpha+2\beta\le8\zeta_\delta.
\label{eq:st-centered-power-bound}
\end{equation}
Both iterations fix the constant component $\pi(f)$.
A centered coefficient vector $z$ represents $\sum_{S\in\cR_+}z_S\phi_S$, whose squared $L^2(\pi)$ norm is $z^\top G_cz$.
Using \eqref{eq:st-centered-power-bound} and $8\zeta_\delta\le\varepsilon$ gives
\begin{equation*}
\sup_{\tau>0,\ m\in\mathbb N_0}\|\widetilde u_m-u_m\|_\pi\le8\zeta_\delta\|f-\pi(f)\|_\pi\le\varepsilon\|f-\pi(f)\|_\pi,
\end{equation*}
which proves \eqref{eq:st-compression-propagation}.
\end{proof}

\printbibliography

@article{glauber1963time,
  author  = {Glauber, Roy J.},
  title   = {Time-dependent statistics of the {Ising} model},
  journal = {Journal of Mathematical Physics},
  volume  = {4},
  number  = {2},
  year    = {1963},
  pages   = {294--307},
  doi     = {10.1063/1.1703954},
}

@article{lindblad1976generators,
  author  = {Lindblad, G{\"o}ran},
  title   = {On the generators of quantum dynamical semigroups},
  journal = {Communications in Mathematical Physics},
  volume  = {48},
  number  = {2},
  year    = {1976},
  pages   = {119--130},
  doi     = {10.1007/BF01608499},
}

@article{PhysRevA.70.012106,
  author  = {Breuer, Heinz-Peter},
  title   = {Genuine quantum trajectories for non-{Markovian} processes},
  journal = {Physical Review A},
  volume  = {70},
  number  = {1},
  year    = {2004},
  eid     = {012106},
  doi     = {10.1103/PhysRevA.70.012106},
}

@article{wild2023classical,
  author  = {Wild, Dominik S. and Alhambra, {\'A}lvaro M.},
  title   = {Classical simulation of short-time quantum dynamics},
  journal = {PRX Quantum},
  volume  = {4},
  number  = {2},
  year    = {2023},
  eid     = {020340},
  doi     = {10.1103/PRXQuantum.4.020340},
}

@article{barthel2012quasilocality,
  author  = {Barthel, Thomas and Kliesch, Martin},
  title   = {Quasilocality and efficient simulation of {Markovian} quantum dynamics},
  journal = {Physical Review Letters},
  volume  = {108},
  number  = {23},
  year    = {2012},
  eid     = {230504},
  doi     = {10.1103/PhysRevLett.108.230504},
}

@book{Pavliotis2014,
  author    = {Pavliotis, Grigorios A.},
  title     = {Stochastic Processes and Applications: Diffusion Processes, the {Fokker--Planck} and
               {Langevin} Equations},
  series    = {Texts in Applied Mathematics},
  number    = {60},
  year      = {2014},
  publisher = {Springer},
  address   = {New York, NY},
  doi       = {10.1007/978-1-4939-1323-7},
}

@book{hoffmann2012GL,
  author    = {Hoffmann, Karl-Heinz and Tang, Qi},
  title     = {{Ginzburg--Landau} Phase Transition Theory and Superconductivity},
  series    = {International Series of Numerical Mathematics},
  number    = {134},
  year      = {2012},
  publisher = {Birkh{\"a}user},
  address   = {Basel},
  doi       = {10.1007/978-3-0348-8274-3},
}

@article{doi:10.1137/090752286,
  author  = {Oseledets, I. V.},
  title   = {Tensor-train decomposition},
  journal = {SIAM Journal on Scientific Computing},
  volume  = {33},
  number  = {5},
  year    = {2011},
  pages   = {2295--2317},
  doi     = {10.1137/090752286},
}

@article{TANG2024113110,
  author  = {Tang, Xun and Ying, Lexing},
  title   = {Solving high-dimensional {Fokker--Planck} equation with functional hierarchical
             tensor},
  journal = {Journal of Computational Physics},
  volume  = {511},
  year    = {2024},
  eid     = {113110},
  doi     = {10.1016/j.jcp.2024.113110},
}

@article{doi:10.1137/24M1654075,
  author  = {Tang, Xun and Collis, Leah and Ying, Lexing},
  title   = {Solving high-dimensional {Kolmogorov} backward equations with functional hierarchical
             tensor operators},
  journal = {SIAM Journal on Scientific Computing},
  volume  = {47},
  number  = {2},
  year    = {2025},
  pages   = {A613--A632},
  doi     = {10.1137/24M1654075},
}

@article{halko2011finding,
  author  = {Halko, Nathan and Martinsson, Per-Gunnar and Tropp, Joel A.},
  title   = {Finding structure with randomness: Probabilistic algorithms for constructing
             approximate matrix decompositions},
  journal = {SIAM Review},
  volume  = {53},
  number  = {2},
  year    = {2011},
  pages   = {217--288},
  doi     = {10.1137/090771806},
}

@article{thiede2019galerkin,
  author  = {Thiede, Erik H. and Giannakis, Dimitrios and Dinner, Aaron R. and Weare, Jonathan},
  title   = {{Galerkin} approximation of dynamical quantities using trajectory data},
  journal = {The Journal of Chemical Physics},
  volume  = {150},
  number  = {24},
  year    = {2019},
  eid     = {244111},
  doi     = {10.1063/1.5063730},
}

@article{drautz2019atomic,
  author  = {Drautz, Ralf},
  title   = {Atomic cluster expansion for accurate and transferable interatomic potentials},
  journal = {Physical Review B},
  volume  = {99},
  number  = {1},
  year    = {2019},
  eid     = {014104},
  doi     = {10.1103/PhysRevB.99.014104},
}

@article{dusson2022atomic,
  author  = {Dusson, Genevi{\`e}ve and Bachmayr, Markus and Cs{\'a}nyi, G{\'a}bor and Drautz, Ralf
             and Etter, Simon and van der Oord, Cas and Ortner, Christoph},
  title   = {Atomic cluster expansion: Completeness, efficiency and stability},
  journal = {Journal of Computational Physics},
  volume  = {454},
  year    = {2022},
  eid     = {110946},
  doi     = {10.1016/j.jcp.2022.110946},
}

@article{meneghelli2014mayer,
  author  = {Meneghelli, Carlo and Yang, Gang},
  title   = {Mayer-cluster expansion of instanton partition functions and thermodynamic {Bethe}
             ansatz},
  journal = {Journal of High Energy Physics},
  volume  = {2014},
  number  = {5},
  year    = {2014},
  eid     = {112},
  doi     = {10.1007/JHEP05(2014)112},
}

@incollection{brydges1984short,
  author    = {Brydges, David C.},
  title     = {A short course on cluster expansions},
  booktitle = {Critical Phenomena, Random Systems, Gauge Theories, Part I},
  editor    = {Osterwalder, Konrad and Stora, Raymond},
  year      = {1986},
  pages     = {129--183},
  note      = {Les Houches, Session XLIII, 1984},
  publisher = {North-Holland},
  address   = {Amsterdam},
  url       = {https://personal.math.ubc.ca/~db5d/Seminars/les_houches_84.pdf},
}

@article{chen2023speed,
  author  = {Chen, Chi-Fang Anthony and Lucas, Andrew and Yin, Chao},
  title   = {Speed limits and locality in many-body quantum dynamics},
  journal = {Reports on Progress in Physics},
  volume  = {86},
  number  = {11},
  year    = {2023},
  eid     = {116001},
  doi     = {10.1088/1361-6633/acfaae},
}

@article{Lucke2022tgEDMD,
  author  = {L{\"u}cke, Marvin and N{\"u}ske, Feliks},
  title   = {{tgEDMD}: Approximation of the {Kolmogorov} operator in tensor train format},
  journal = {Journal of Nonlinear Science},
  volume  = {32},
  number  = {4},
  year    = {2022},
  eid     = {44},
  doi     = {10.1007/s00332-022-09801-0},
}

@article{osti_2281727,
  author  = {Lu, Lu and Jin, Pengzhan and Pang, Guofei and Zhang, Zhongqiang and Karniadakis, George
             Em},
  title   = {Learning nonlinear operators via {DeepONet} based on the universal approximation
             theorem of operators},
  journal = {Nature Machine Intelligence},
  volume  = {3},
  number  = {3},
  year    = {2021},
  pages   = {218--229},
  doi     = {10.1038/s42256-021-00302-5},
}

@inproceedings{li2020fourier,
  author    = {Li, Zongyi and Kovachki, Nikola and Azizzadenesheli, Kamyar and Liu, Burigede and
               Bhattacharya, Kaushik and Stuart, Andrew and Anandkumar, Anima},
  title     = {{Fourier} neural operator for parametric partial differential equations},
  booktitle = {International Conference on Learning Representations},
  year      = {2021},
  url       = {https://iclr.cc/virtual/2021/poster/3281},
}

@article{cerrillo2014non,
  author  = {Cerrillo, Javier and Cao, Jianshu},
  title   = {Non-{Markovian} dynamical maps: Numerical processing of open quantum trajectories},
  journal = {Physical Review Letters},
  volume  = {112},
  number  = {11},
  year    = {2014},
  eid     = {110401},
  doi     = {10.1103/PhysRevLett.112.110401},
}

@article{pollock2018tomographically,
  author  = {Pollock, Felix A. and Modi, Kavan},
  title   = {Tomographically reconstructed master equations for any open quantum dynamics},
  journal = {Quantum},
  volume  = {2},
  year    = {2018},
  eid     = {76},
  doi     = {10.22331/q-2018-07-11-76},
}

@article{nakajima1958quantum,
  author  = {Nakajima, Sadao},
  title   = {On quantum theory of transport phenomena: Steady diffusion},
  journal = {Progress of Theoretical Physics},
  volume  = {20},
  number  = {6},
  year    = {1958},
  pages   = {948--959},
  doi     = {10.1143/PTP.20.948},
}

@article{zwanzig1960ensemble,
  author  = {Zwanzig, Robert},
  title   = {Ensemble method in the theory of irreversibility},
  journal = {The Journal of Chemical Physics},
  volume  = {33},
  number  = {5},
  year    = {1960},
  pages   = {1338--1341},
  doi     = {10.1063/1.1731409},
}

@article{PhysRevLett.104.190401,
  author  = {Poulin, David},
  title   = {{Lieb--Robinson} bound and locality for general {Markovian} quantum dynamics},
  journal = {Physical Review Letters},
  volume  = {104},
  number  = {19},
  year    = {2010},
  eid     = {190401},
  doi     = {10.1103/PhysRevLett.104.190401},
}

@article{JMLR:v24:21-1524,
  author  = {Kovachki, Nikola and Li, Zongyi and Liu, Burigede and Azizzadenesheli, Kamyar and
             Bhattacharya, Kaushik and Stuart, Andrew and Anandkumar, Anima},
  title   = {Neural operator: Learning maps between function spaces with applications to {PDEs}},
  journal = {Journal of Machine Learning Research},
  volume  = {24},
  number  = {89},
  year    = {2023},
  pages   = {1--97},
  url     = {https://www.jmlr.org/papers/v24/21-1524.html},
}

@misc{khoo2025fastoperator,
  author      = {Khoo, Yuehaw and Wang, Yuguan and Yang, Siyao},
  title       = {Fast operator learning for mapping correlations},
  year        = {2025},
  eprint      = {2512.09286},
  eprinttype  = {arxiv},
  eprintclass = {math.NA},
}

@article{sander2025tjm,
  author  = {Sander, Aaron and Fr{\"o}hlich, Maximilian and Eigel, Martin and Eisert, Jens and
             Ge{\ss}, Patrick and Hinterm{\"u}ller, Michael and Milbradt, Richard M. and Wille,
             Robert and Mendl, Christian B.},
  title   = {Large-scale stochastic simulation of open quantum systems},
  journal = {Nature Communications},
  volume  = {16},
  year    = {2025},
  eid     = {11074},
  doi     = {10.1038/s41467-025-66846-x},
}

@article{RevModPhys.77.259,
  author  = {Schollw{\"o}ck, Ulrich},
  title   = {The density-matrix renormalization group},
  journal = {Reviews of Modern Physics},
  volume  = {77},
  number  = {1},
  year    = {2005},
  pages   = {259--315},
  doi     = {10.1103/RevModPhys.77.259},
}

@article{PhysRevLett.68.580,
  author  = {Dalibard, Jean and Castin, Yvan and M{\o}lmer, Klaus},
  title   = {Wave-function approach to dissipative processes in quantum optics},
  journal = {Physical Review Letters},
  volume  = {68},
  number  = {5},
  year    = {1992},
  pages   = {580--583},
  doi     = {10.1103/PhysRevLett.68.580},
}

@article{RevModPhys.70.101,
  author  = {Plenio, M. B. and Knight, P. L.},
  title   = {The quantum-jump approach to dissipative dynamics in quantum optics},
  journal = {Reviews of Modern Physics},
  volume  = {70},
  number  = {1},
  year    = {1998},
  pages   = {101--144},
  doi     = {10.1103/RevModPhys.70.101},
}

@article{gillespie1977exact,
  author  = {Gillespie, Daniel T.},
  title   = {Exact stochastic simulation of coupled chemical reactions},
  journal = {The Journal of Physical Chemistry},
  volume  = {81},
  number  = {25},
  year    = {1977},
  pages   = {2340--2361},
  doi     = {10.1021/j100540a008},
}

@article{williams2015data,
  author  = {Williams, Matthew O. and Kevrekidis, Ioannis G. and Rowley, Clarence W.},
  title   = {A data-driven approximation of the {Koopman} operator: Extending dynamic mode
             decomposition},
  journal = {Journal of Nonlinear Science},
  volume  = {25},
  number  = {6},
  year    = {2015},
  pages   = {1307--1346},
  doi     = {10.1007/s00332-015-9258-5},
}

@misc{facelli2026fast,
  author      = {Facelli, Giorgio and Fawzi, Hamza and Fawzi, Omar and Rouz{\'e}, Cambyse and
                 Slezak, Sam and Stilck Fran{\c{c}}a, Daniel},
  title       = {Fast convergence of propagation algorithms for open quantum systems},
  year        = {2026},
  eprint      = {2610.06769},
  eprinttype  = {arxiv},
  eprintclass = {quant-ph},
}

@article{GREENGARD1987325,
  author  = {Greengard, Leslie and Rokhlin, Vladimir},
  title   = {A fast algorithm for particle simulations},
  journal = {Journal of Computational Physics},
  volume  = {73},
  number  = {2},
  year    = {1987},
  pages   = {325--348},
  doi     = {10.1016/0021-9991(87)90140-9},
}

@article{hackbusch1999sparse,
  author  = {Hackbusch, Wolfgang},
  title   = {A sparse matrix arithmetic based on {$\mathcal{H}$}-matrices. {Part I}: Introduction to
             {$\mathcal{H}$}-matrices},
  journal = {Computing},
  volume  = {62},
  number  = {2},
  year    = {1999},
  pages   = {89--108},
  doi     = {10.1007/s006070050015},
}

@book{liggett1985interacting,
  author    = {Liggett, Thomas M.},
  title     = {Interacting Particle Systems},
  series    = {Grundlehren der mathematischen Wissenschaften},
  number    = {276},
  year      = {1985},
  publisher = {Springer-Verlag},
  address   = {New York},
  doi       = {10.1007/978-1-4613-8542-4},
}

@article{doukhan2008functional,
  author  = {Doukhan, Paul and Lang, Gabriel and Louhichi, Sana and Ycart, Bernard},
  title   = {A functional central limit theorem for interacting particle systems on transitive
             graphs},
  journal = {Markov Processes and Related Fields},
  volume  = {14},
  number  = {1},
  year    = {2008},
  pages   = {79--114},
  url     = {https://math-mprf.org/journal/articles/id1145/},
}

@article{wu2006poincare,
  author  = {Wu, Liming},
  title   = {Poincar{\'e} and transportation inequalities for {Gibbs} measures under the {Dobrushin}
             uniqueness condition},
  journal = {The Annals of Probability},
  volume  = {34},
  number  = {5},
  year    = {2006},
  pages   = {1960--1989},
  doi     = {10.1214/009117906000000368},
}

@inproceedings{zhou2011godec,
  author    = {Zhou, Tianyi and Tao, Dacheng},
  title     = {{GoDec}: Randomized low-rank \& sparse matrix decomposition in noisy case},
  booktitle = {Proceedings of the 28th International Conference on Machine Learning},
  year      = {2011},
  pages     = {33--40},
  url       = {https://icml.cc/2011/papers/41_icmlpaper.pdf},
}

@article{gorini1976completely,
  author  = {Gorini, Vittorio and Kossakowski, Andrzej and Sudarshan, E. C. G.},
  title   = {Completely positive dynamical semigroups of {$N$}-level systems},
  journal = {Journal of Mathematical Physics},
  volume  = {17},
  number  = {5},
  year    = {1976},
  pages   = {821--825},
  doi     = {10.1063/1.522979},
}

@article{saad1992analysis,
  author  = {Saad, Yousef},
  title   = {Analysis of some {Krylov} subspace approximations to the matrix exponential operator},
  journal = {SIAM Journal on Numerical Analysis},
  volume  = {29},
  number  = {1},
  year    = {1992},
  pages   = {209--228},
  doi     = {10.1137/0729014},
}

@article{hochbruck1997krylov,
  author  = {Hochbruck, Marlis and Lubich, Christian},
  title   = {On {Krylov} subspace approximations to the matrix exponential operator},
  journal = {SIAM Journal on Numerical Analysis},
  volume  = {34},
  number  = {5},
  year    = {1997},
  pages   = {1911--1925},
  doi     = {10.1137/S0036142995280572},
}

@article{botchev2013residual,
  author  = {Botchev, Mike A. and Grimm, Volker and Hochbruck, Marlis},
  title   = {Residual, restarting, and {Richardson} iteration for the matrix exponential},
  journal = {SIAM Journal on Scientific Computing},
  volume  = {35},
  number  = {3},
  year    = {2013},
  pages   = {A1376--A1397},
  doi     = {10.1137/110820191},
}

\end{document}